\documentclass[10pt]{iopart}
\usepackage{graphicx}
\usepackage[colorlinks=true, allcolors=blue]{hyperref}
\usepackage{upgreek}
\usepackage[numbers,square,sort&compress]{natbib}

\usepackage{ulem}
\usepackage{comment}
\usepackage{braket}

\usepackage{tabularx}
\usepackage{xtab,afterpage}
\usepackage{multirow}
\usepackage{booktabs}

\usepackage{xfrac}

\begin{document}

\topical[Many-body methods for high-precision studies of isotope shifts]{Recent advancements in atomic many-body methods for high-precision studies of isotope shifts}

\author{
B~K~Sahoo$^{1}$,
S~Blundell$^2$, 
A~V~Oleynichenko$^{3,4}$,
R~F~Garcia~Ruiz$^5$, 
L~V~Skripnikov$^{3,6}$, and
B~Ohayon$^7$ 
}

\address{$^1$ Atomic, Molecular and Optical Physics Division, Physical Research Laboratory, Navrangpura, Ahmedabad 380058, Gujarat, India}
\ead{bijaya@prl.res.in}

\address{$^2$ Univ.\ Grenoble Alpes, CEA, CNRS, IRIG, SyMMES, 38000 Grenoble, France}
\ead{steven.blundell@cea.fr}

\address{$^3$ Petersburg Nuclear Physics Institute named by B.P.\ Konstantinov of National Research Center ``Kurchatov Institute''  (NRC ``Kurchatov Institute'' -- PNPI), 1 Orlova roscha, Gatchina, 188300 Leningrad region,~Russia }

\address{$^4$ Moscow Center for Advanced Studies, 20 Kulakova Str., 123592 Moscow, Russia}

\ead{oleynichenko\_av@pnpi.nrcki.ru}

\address{$^5$  Department of Physics, Massachusetts Institute of Technology, Cambridge Massachusetts 02139, USA}
\ead{rgarciar@mit.edu}

\address{$^6$ Saint Petersburg State University, 7/9 Universitetskaya nab., St. Petersburg, 199034, Russia }
\ead{skripnikov\_lv@pnpi.nrcki.ru}

\address{$^7$ The Helen Diller Quantum Center, Department of Physics, Technion-Israel Institute of Technology, Haifa, 3200003, Israel}
\ead{bohayon@technion.ac.il}

\begin{abstract}
The development of atomic many-body methods, capable of incorporating electron correlation effects accurately, is required for isotope shift (IS) studies.
In combination with precise measurements, such calculations help to extract nuclear charge radii differences, and to probe for signatures of physics beyond the Standard Model of particle physics.
We review here a few recently-developed methods in the relativistic many-body perturbation theory (RMBPT) and relativistic coupled-cluster (RCC) theory frameworks for calculations of IS factors in the highly charged ions (HCIs), and neutral or singly-charged ions, respectively. The results are presented for a wide range of atomic systems in order to demonstrate the interplay between quantum electrodynamics (QED) and electron correlation effects.
In view of this, we start our discussions with the RMBPT calculations for a few HCIs by rigorously treating QED effects; then we outline methods to calculate IS factors in the one-valence atomic systems using two formulations of the RCC approach. Then we present calculations for two valence atomic systems, by employing the Fock-space RCC methods. For completeness, we briefly discuss theoretical input required for the upcoming experiments, their possibilities to probe nuclear properties and implications to fundamental physics studies. 
\end{abstract}
\vspace{-2 mm}


\submitto{\jpb}
\maketitle

\ioptwocol

\section{Introduction}

Adding or removing neutrons to an atomic nucleus results in variations of its nuclear density, which in turn influences the energy levels of its atomic electrons. The change in an energy level due to this effect is termed isotope shift (IS). Measurements of ISs can be used for exploring diverse nuclear and particle physics phenomena (see e.g.~\cite{Geb15,2017-HiggsLike,2021-Ag,Ber18,Bar21,2023-Review,2024-Struct}).

In the leading order, IS measurements are sensitive to changes in mean-squared nuclear charge radii and nuclear masses~\cite{king,Mil19,Gar16,Reinhard2020,Allehabi:20}.
As nuclear masses can be measured with sufficient precision by other means~\cite{Blaum_2011,2013-Blaum,2018-Mass,2022-Mass}, ISs are predominantly used to measure changes in radii. 
This approach has been established as one of the main tools to study the evolution of nuclear size far away from the stable region of the nuclear chart~\cite{2023-Review}. The extension of these measurements towards regions of extreme proton-to-neutron ratios is of growing interest for nuclear structure studies, as the knowledge of the nuclear size is essential to our understanding of the atomic nucleus and nuclear matter at extreme conditions~\cite{Gar16,Gor19,2020-Cu,kar24,kon24}. On the other hand, if the nuclear properties are known or constrained, precise measurements of ISs can be a highly sensitive probe of the existence of possible new electron-nucleon interactions~\cite{2017-Mik,2017-Yotam,2017-HiggsLike,2017-Fuchs,Ber18,Sta18,2020-KP}. This increasing interest in the use of ISs in fundamental physics has stimulated considerable developments in high-precision experimental techniques, which are now able to achieve sub-kHz precision~\cite{Man19,2020-CaKP,2020-YbKP,2020-YbBud,Hur22,Ono22,2023-CaDoret,2024-free,2024-YbKP}. Such experimental progress is not only promising for searches of new physics beyond the Standard Model of particle physics, but it could also offer means to access elusive nuclear observables such as the higher-order radial moment, $\langle r^4 \rangle$ \cite{Reinhard2020}, and the nuclear dipole polarizability~\cite{Pap16,Flam18}. 
Precise measurements of these nuclear properties would have a marked impact on our knowledge of nuclear structure and nuclear matter~\cite{Reinhard2020,Kau20,Sky21,kon24,2024-Struct}.

As electronic structure calculations are needed in most cases in order to translate measured ISs to useful fundamental observables, theoretical developments must evolve concurrently with experimental capabilities to enable an accurate interpretation of IS data. Thus, to capitalize on the potential of IS measurements, advancements in atomic theory are of critical importance. Such calculations are highly demanding, lying at the forefront of atomic theory.

Although much attention has been paid to review the ever-evolving experimental methods, see e.g.~\cite{Blaum_2011,2013-Blaum,nortershauser2014nuclear,2015-revExp,2022-ReviewExp,2023-Review,2024-Struct}, the literature focusing on the theoretical aspects is scarce. To our knowledge, the latest review highlighting the importance of atomic theory calculations is more than a decade old~\cite{2012-Co}. That review emphasized the limitations of the atomic theory for interpreting IS measurements, pointing that the accuracy of current methods restricts the determinations of radii differences in most atomic systems. 

In recent years, the adaptation of methods from computational quantum many-body methods combined with the possibilities of modern powerful computers has largely changed this situation, making an increasing set of many-electron systems amenable to accurate calculations.
In this review, we focus on the impact of the theoretical methods of the next generation for radii extractions from IS measurements in atoms. We limit the discussion to so-called ``many-electron'' systems, which we loosely refer to as those with more than six electrons, for which atomic calculations with arbitrarily high numerical accuracy are not feasible.
Nevertheless, it is of great importance to test the performance of many-electron methods by calculating observables in few-electron systems and comparing them with ones calculated with few-electron methods (see e.g.~\cite{1995-LiDrake,1998-LiDrake,2002-LiDrake,2008-LiBe,2009-Lit,2010-C2,2011-Li,2014-BeLikeQED,2014-1sBe,2014-LiLike,2014-BeIS,2015-QEDLi,2015-Shab,2015-HeIS,2015-BIS,2019-C,2019-QEDSMS,2019-Be,2020-CII,2020-QEDNMS,2021-BoronECG,2021-BenchBe,2023-C,2023-CII,2024-Be,2024-CIS}).

\begin{figure*}\label{Fig:overview}
    \includegraphics[trim=60 30 40 60,clip,width=0.96\textwidth]{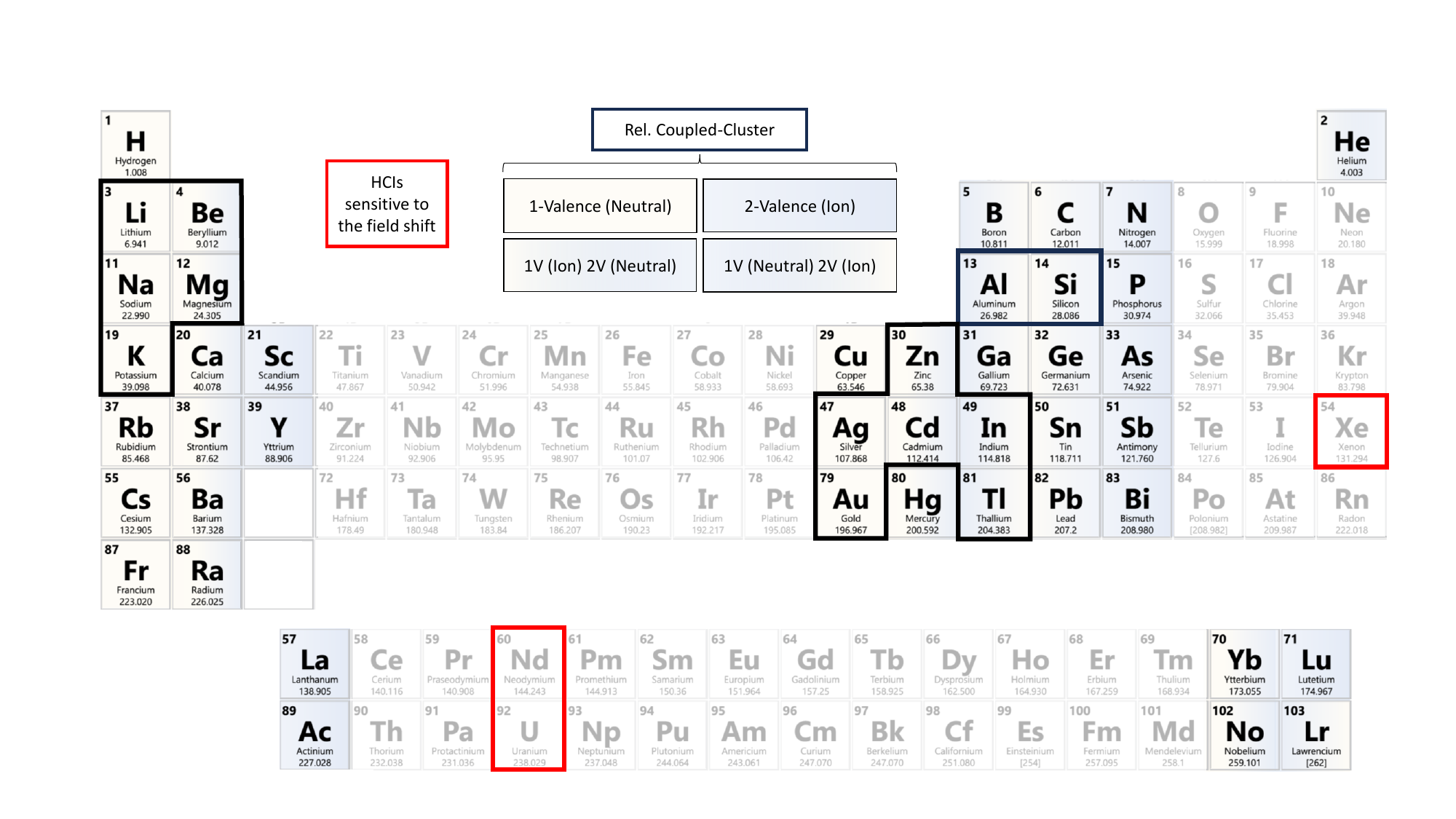}
    \caption{The atomic systems whose isotope-shift factor calculations are discussed in this review.}
    \label{fig:Overview}
\end{figure*}

In this review we focus on the theory input needed for interpreting ISs in many-electron systems in terms of nuclear charge radii differences. 
We first briefly consider many-electron highly-charged-ions (HCIs).
In these systems, electron correlations can often be treated perturbatively and the treatment of relativistic and quantum electrodynamics (QED) effects must be as rigorous as possible. As the landscape of HCIs is very rich, we will limit the discussion to the handful of cases where radii differences could be determined. We refer the reader to the comprehensive review of Indelicato~\cite{2019-QEDtest} for more information on QED tests with HCIs which we will not cover here.
The main topic covered in this review is the many-body calculations of IS factors in neutral or singly charged systems. Here electron correlations can be strong and must be treated non-perturbatively (``to all-orders''), while QED corrections can be safely presumed to be small, making them amenable to approximate methods. We will limit the discussion to the recently-developed state-of-the-art methods of calculating IS factors in many-electron systems, i.e., the relativistic coupled-cluster (RCC) method, including its version for strongly multi-reference cases.
The relevant atomic systems mentioned in this review are highlighted in Fig.~\ref{fig:Overview}.

We refer the interested reader to reviews on other methods successfully applied in the field which will not be discussed here: Relativistic configuration interaction (RCI)~\cite{1993-RCI,1994-RCI,1995-RCI,2003-RelIS,2007-Boronlike,2008-RCI,2010-RCI,2012-RCI,2014-RCI,2015-RCI,2017-RCI,2020-CAR}, combined configuration interaction (CI) and many-body perturbation theory (MBPT) (denoted as CI-MBPT)~\cite{1996-CIMBPT, 1998.CIMBPT, 2003-CI, 2006-CIMBPT,Kozlov:15,2017-CI,2019-CIMBPT,2019-ambit,2022-CIPT}, combined CI and Dirac-Fock-Sturm method (denoted as CI-DFS)~\cite{Ulrich:2003,2005-CIDFS,kozhedub-10-qed,2014-LiLike,2014-BeLikeQED,Zubova:2016,2018-CIDFS}, combined CI and coupled-cluster theory (denoted as CI+all-order)~\cite{Blundell:91,Safronova:99,Kozlov:04,Pal:07,Safronova:08,Safronova:09,2015-LCC,Safronova:Th:18,Raeder:18,Cheung:21},
the multiconfiguration Dirac-Hartree-Fock (MCDHF) method~\cite{1997-Per,2007-grasp,2009-MCDHF,2012-LiLike,jonsson-13-is, NAZE20132187, 2019-grasp} and the multiconfiguration Dirac-Fock General Matrix Elements (MCDFGME) method~\cite{2005-MCDFGME, 2007-GME,2013-PaulQED}. 

The structure of this review is as follows. 
We first briefly review the relevant theoretical background needed to discuss the first-order IS effects due to finite-size, field shift (FS), and finite-mass, mass shift (MS), of the nucleus.
We then remind the reader of the form of the relevant operators, followed by a discussion of the QED effects. The quantum-mechanical operators used in the field are discussed in details, followed by the discussion of the QED effects.
Then, the finite-field (FF) approach to calculate expectation values of these operators is briefly discussed, with other approaches mentioned in later sections.
We then discuss some electronic structure theory methods of solving the many-electron problem in the context of the IS studies.
We focus on the application of MBPT in single-valence HCIs.
Then we consider the implementation of both single- and multi-reference RCC methods as well as different approaches to calculate the IS factors within these methods.
In section~\ref{sec:res}, we present some selected published results which demonstrate the applicability of the discussed theoretical methods.
Finally the theory input needed for upcoming experiments is discussed and a brief outlook on the development of new calculation methods is given.

\section{Theory}

\subsection{Background}

Estimating ISs as first-order perturbations is a good enough approximation for extracting valuable information from most measurements.
Within this approximation, contributions to the IS are divided into FS and MS, given by~\cite{bohr1922difference,1922-Ehr}
\begin{equation}
\delta \nu_{A,A'} \simeq 
K\mu_{A,A'} + F \delta r^2_{A,A'},
\label{eq:IS}
\end{equation}
where $\nu_{A,A'}\equiv \nu_{A'}-\nu_A$, $\mu_{A,A'}=1/M_{A}-1/M_{A'}$, and $\delta r^2_{A,A'}=\langle r_{A'}^2 \rangle -\langle r_{A}^2 \rangle$.
The MS can further be divided into the normal mass shift (NMS) and the specific mass shift (SMS)~\cite{1930-HE}. From theoretical point of view, these shifts are estimated by determining their respective IS factors $K$ and $F$, known as the MS and FS factors, respectively~\cite{Martensson-Pendrill_1992}.

To apply perturbation theory to the IS, electronic wave functions of the non-perturbed multi-electron systems are needed. These wave functions can be obtained in at least two steps; first, a mean-field approach is adopted, and then, the neglected electron correlation effects are included systematically~\cite{bartlett,lindgren}. There are different methods to include electronic correlations, befitting the behavior of electron correlation effects and configurations of atomic systems. 
The overlap between different theoretical techniques allows one method to test another.

\subsection{The Atomic Hamiltonian}

An accurate determination of IS factors demands that the atomic wave functions are calculated using proficient many-body methods, as discussed in Sec.~\ref{sec:many}. For this purpose, it is also necessary to consider the atomic Hamiltonian in the relativistic framework. Calculations of wave functions are performed mainly in the infinite nuclear mass limit using the Dirac-Coulomb (DC) Hamiltonian. In atomic units (a.~u.) it is given by the expression~\cite{Dirac}
\begin{eqnarray}\label{eq:DC}
H^{\text{DC}} &=& 
\Lambda_{+} ( \sum_i \left [c \vec\alpha_i^D \cdot \vec p_i+(\beta_i-1)c^2+V_n(r_i)\right] 
\nonumber \\ 
&& + \frac{1}{2}\sum_{i\neq j}\frac{1}{r_{ij}} ) \Lambda_{+}, 
\end{eqnarray}
where $c$ is the speed of light, $\vec\alpha_i^D$ and $\beta_i$ are the Dirac matrices corresponding to electron $i$, $\vec p_i$ is the single particle momentum operator, $\frac{1}{r_{ij}}$ represents the instantaneous Coulomb potential between the electrons located at the $i^{th}$ and $j^{th}$ positions, and $\Lambda_{+}$ is the projector to the positive-energy states. The form of the nuclear potential $V_n(r)$ depends on the nuclear model employed and is discussed in the latter part of this section.

Contributions from the Breit interactions between electrons beyond the DC Hamiltonian, $H^{\text{DC}}$, can be estimated by adding the following potential term~\cite{breit}
\begin{eqnarray}\label{eq:DHB}
V^{\text{B}} (r_{ij}) &=& -  \frac{1}{2}\sum_{i\neq j} \frac{\vec \alpha_i^D \cdot \vec \alpha_j^D +
(\vec\alpha_i^D \cdot \hat r_{ij})(\vec\alpha_j^D \cdot \hat r_{ij})}{2r_{ij}} , \ \ \
\end{eqnarray}
where $\hat r_{ij}$ is the unit vector along $\vec r_{ij}$.
This is referred to as the Dirac-Coulomb-Breit (DCB) Hamiltonian.

\subsection{Mass-shift operators}

The Schr\"odinger Hamiltonian, describing a many-electron system with a point-like nucleus in the non-relativistic framework, can be written, in a.u., with an exact dependence on the nuclear mass as
\begin{equation}
    H^{\text{SC}}_A=\sum_i \left(\frac{\vec p_i^{~2}}{2\mu_A}-\frac{Z}{r_i}\right)+
    \frac{1}{2}\sum_{i\neq j} \left(\frac{\vec p_i\cdot\vec p_j}{M_A}-\frac{1}{r_{ij}}\right),
    \label{eq:SC}
\end{equation}
with $\mu_A\equiv M_A/(1+M_A)$ is the reduced mass.
A derivation of Eq.~(\ref{eq:SC}) is given, e.g., in Sec.~13.2.3 of Ref.~\cite{2022-QEDbook}. When changing the nuclear mass, the difference in the Hamiltonians for two masses reads
\begin{equation}\label{eq:dSC}
    H^{\text{SC}}_{A'}-H^{\text{SC}}_{A}=
    \frac{\mu_{A,A'}}{2}
    \left(
    \sum_i {\vec p}_i^{~2}
    +
    \sum_{i\neq j} {\vec p}_i\cdot{\vec p}_j
    \right). 
\end{equation}
We can thus write down the NMS operator, corresponding to the one-body part of Eq.~(\ref{eq:dSC}), as
\begin{equation}\label{eq:NRNMS}
O^{\text{NMS}} = \frac{1}{2}\sum_i {\vec p}_i^{~2} 
\end{equation}
and the SMS operator, corresponding to the two-body part of Eq.~(\ref{eq:dSC}), as
\begin{equation}\label{eq:NRSMS}
O^{\text{SMS}} = \frac{1}{2} \sum_{i\ne j} {\vec p}_i \cdot {\vec p}_j.
\end{equation}

In the nonrelativistic approximation, the virial theorem~\cite{virial}, relating the kinetic and total energies of an atomic system predicts that, up to corrections of order $1/M_A$, $\langle O^{\text{NMS}}\rangle_i \equiv K^{\text{NMS}}_i\approx - \Delta E_i$, with $i$ denoting an energy level (or interval), and $\Delta E_i$ is the ionization energy of that level (or the excitation energy of an interval).
To illustrate this point, we take the calculated nonrelativistic energy of the $2s-2p$ transition in Li~I from Table~III of~\cite{2008-Li}, to get $K^{\text{NMS}}_{2s-2p}\approx245.10~$GHz~u. This value is compared with a calculation derived from nonrelativistic values given in Table~I of Ref.~\cite{kozhedub-10-qed}, namely $K^{\text{NMS}}_{2s-2p}=K^{\text{MS}}_{2s-2p}-K^{\text{SMS}}_{2s-2p}=245.13~$GHz~u. The fractional deviation is of the expected order, namely $1/M_A\approx10^{-4}$.
In light many-electron systems, for which relativistic effects are small, the scaling procedure above may produce more accurate result for $K^{\text{NMS}}$ than a numerical calculation, which naturally has a finite ability to account for electron correlations. Nevertheless, there are strong cancellations between the neglected contributions in $K^{\text{NMS}}$ and $K^{\text{SMS}}$~\cite{1987-Palmer}, so that it is advisable to calculate both of them on equal footing.

Following the discussions in Refs.~\cite{shabaev1985mass, 1987-Palmer, 1988-Shab, 1994-Shab}, the relativistic forms of the NMS and SMS operators, up to the order of $(\alpha Z)^2$ relative to the non-relativistic forms of Eqs.~(\ref{eq:NRNMS}) and~(\ref{eq:NRSMS}), are given by
\begin{eqnarray}
O^{\text{NMS}} &=& \frac{1}{2}\sum_i \left ({\vec p}_i^{~2} - \frac{\alpha Z}{r_i} {\vec \alpha}_i^D \cdot {\vec p}_i 
\right. \nonumber \\ && \left.
- \frac{\alpha Z}{r_i} ({\vec \alpha}_i^D \cdot {\vec C}_i^1){\vec C}_i^1 \cdot {\vec p}_i \right ), 
\label{nmsexp}
\end{eqnarray}
and
\begin{eqnarray}
O^{\text{SMS}} &=&
\frac{1}{2} \sum_{i\ne j} \left ({\vec p}_i \cdot {\vec p}_j - \frac{\alpha Z}{r_i} {\vec \alpha}_i^D \cdot {\vec p}_j 
\right. \nonumber \\ &&
\left.  - \frac{\alpha Z}{r_i} ({\vec \alpha}_i^D \cdot {\vec C}_i^1) ({\vec p}_j \cdot {\vec C}_j^1) \right ),
\label{smsexp}
\end{eqnarray}
respectively, where $\alpha$ is the fine-structure constant. 
Eqs.~(\ref{nmsexp}) and (\ref{smsexp}) reduce to  Eqs.~(\ref{eq:NRNMS}) and~(\ref{eq:NRSMS}) in the non-relativistic limit.
%L.S. I'm not sure that it is correct to write $\alpha \rightarrow 0$. Usually one uses the ``infinite'' value of the spead of light for this. If necessary, we can write $c \rightarrow 0$.
%$\alpha \rightarrow 0$.

\subsection{Different forms of the field shift operator}

The definition of the FS operator follows directly from Eqs.~(\ref{eq:IS}) and~(\ref{eq:DC})~\cite{johnson2007atomic} %[{\bf Lecture Notes on Atomic Physics, W. R. Johnson}]
\begin{equation}
F(r) = - \frac{\partial V_n(r)}{ \partial \langle r^2 \rangle},
\label{Fdef1}
\end{equation}
where the nuclear potential depends on its assumed nuclear charge density distribution (its ``shape").
%\LS{With this definition the FS factor for the transition is determined by the FS factor for the lower state minus that in the upper state of the transition~\cite{Martensson-Pendrill_1992}.}[{\bf BKS: This is not a general statement, so it should not be mentioned here. It would be better to make this statement wherever results are presented by adopting this notion.}]
From Eq.~(\ref{Fdef1}), it is obvious that the choice of $V_n(r)$, which also depends on the nuclear radius, may be important for accurate evaluation of the FS factors in an atomic system.
Nevertheless, for valence transitions in natural systems, the dependence of $F$ on the radius of a specific isotope is negligible (see e.g.~\cite{2021-Yb,2022-CdHG,2024-Ag}), so that it can be considered an elemental property.

Assuming that charges are uniformly distributed within a spherical nucleus of radius $r_N$, the nuclear potential can be defined as
\begin{eqnarray}
V_n(r) =-Z \left\{\begin{array}{ll}
\frac{1}{2r_N} \left [ 3- \frac{r^2}{r_N^2} \right ]  & \mbox{for $r \leq r_N$}\\
\frac{1}{r}  & \mbox{for $r >r_N$} ,
\end{array}\right.   ,
\end{eqnarray}
where 
$r_N^2= \frac{5}{3}\langle r^2 \rangle$. 
It yields
\begin{eqnarray}
F(r) = - \left\{\begin{array}{ll}
%-\frac{5Z}{4r_N^3} \left [ 1- \frac{r^2}{r_N^2} \right ]  & \mbox{for $r \leq r_N$}\\
\frac{5Z}{4r_N^3} \left [ 1- \frac{r^2}{r_N^2} \right ]  & \mbox{for $r \leq r_N$}\\
0  & \mbox{for $r >r_N$} 
\end{array}\right.   . 
\end{eqnarray}

In the Gaussian nuclear charge distribution
model, nuclear potential observed by an electron as~\cite{visscher:1997dirac} 
\begin{eqnarray}
V_n (r) = - \frac{Z}{r} \text{erf}(\sqrt{\xi} r ) 
\end{eqnarray}
with $\xi = \frac{1.5}{\langle r^2 \rangle}$. From this expression, we can define~\cite{skripnikov2024isotopeQED}
\begin{eqnarray}
\label{Fgauss}
F(r) = - \frac{2Z \xi^{3/2}}{3 \sqrt{\pi}} e^{- \xi r^2} .
\end{eqnarray}

Except for the lightest nuclei, the two-parameter Fermi nuclear charge distribution 
\begin{equation}\label{eq:fermi}
    \rho_N(r)\equiv \frac{\rho_0}{1+\exp{(r-b)/a}},
\end{equation}
where $\rho_0$ is a normalization constant, $b$ is the half-charge density, and $a\approx 2.3/4\ln(3)$ is the skin thickness, is considered to be a more realistic approximation.
The corresponding nuclear electrostatic potential is expressed by~\cite{1985-Fermi}
\begin{eqnarray}
 V_n(r) = -\frac{Z}{\mathcal{N}r} \times  \nonumber\\
\left\{\begin{array}{rl}
\frac{1}{b}(\frac{3}{2}+\frac{a^2\pi^2}{2b^2}-\frac{r^2}{2b^2}+\frac{3a^2}{b^2}P_2^+\frac{6a^3}{b^2r}(S_3-P_3^+)) & \mbox{for~$r_i \leq b$}\\
\frac{1}{r_i}(1+\frac{a62\pi^2}{b^2}-\frac{3a^2r}{c^3}P_2^-+\frac{6a^3}{b^3}(S_3-P_3^-))                           & \mbox{for~$r_i >b$,}  
\end{array}\right. \nonumber 
\label{eq12}
\end{eqnarray}
where the factors are 
\begin{eqnarray}
\mathcal{N} &=& 1+ \frac{a^2\pi^2}{b^2} + \frac{6a^3}{b^3}S_3  \nonumber \\
\text{with} \ \ \ \ S_k &=& \sum_{l=1}^{\infty} \frac{(-1)^{l-1}}{l^k}e^{-lb/a} \ \ \  \nonumber \\
\text{and} \ \ \ \ P_k^{\pm} &=& \sum_{l=1}^{\infty} \frac{(-1)^{l-1}}{l^k}e^{\pm l(r-b)/a} .
\end{eqnarray}
Here $b$ is related to the (RMS) radius of a nucleus by
\begin{equation}
b\approx\sqrt{\frac{5}{3} \langle r^2 \rangle - \frac{7}{3} a^2 \pi^2},
\end{equation}
so that it may be estimated using the measured $\langle r^2 \rangle$,
or by using the empirical formula~\cite{1985-R}.
\begin{equation}\label{eq:rad}
\langle r^2 \rangle = 0.836 A^{1/3} + 0.57 ~ \text{fm} 
\end{equation}
The FS operator may now be calculated from the relation
\begin{equation}
F(r)= - \frac{\partial V_n(r)}{ \partial b} \frac{\partial b}{ \partial \langle r^2 \rangle}.
\label{Fdef4}
\end{equation}

There is another form of the FS factor used by the community. It can be obtained by evaluating 
the energy shift for electrons moving in a finite-size nucleus potential as compared to the ones moving in the Coulomb potential of the point-like nucleus~\cite{1997-Per}
%\LS{In Eqs.(20,21) there should be the minus sign in front of $Z/r'$, see Eqs. (1,20) in Ref.~\cite{1997-Per}.}
% Ben: corrected
%\LS{Corrected: we should calculate $V_n - V_point=V_n - (-Z/r)=V_n + Z/r$. Note that $V_n$ is not an electrostatic potential, it is potential energy which have a different sign.}
\begin{equation}
%E_{\text{FS}} = \int \left ( V_n(r') - \frac{Z}{r'} \right ) \rho_e(r') d^3r' ,
E_{\text{FS}} = \int \left ( V_n(r') + \frac{Z}{r'} \right ) \rho_e(r') d^3r' ,
\end{equation}
where $\rho_e(r)= \langle \Psi (r') | \delta (r - r') | \Psi(r') \rangle$ stands for the electron density. 
% L.S. We actually have the same definition of $F$
%Note that the sign of the expression is changed from Ref.~\cite{1997-Per} to be consistent with the present definition of $F$.
Using the identity $\nabla^2 r^2 =6$, 
and assuming that the electron charge distribution is nearly constant over the nuclear volume, we can approximately write down
\begin{eqnarray}
E_{\text{FS}} 
%&=& 
&\approx&
% \frac{1}{6} \rho_e(0) \int \left ( V_{e-N}(r') - \frac{Z}{r'} \right ) \nabla^2 {r'}^2 d^3r' \nonumber \\
%   &=&  \frac{1}{6} \rho_e(0) \int {r'}^2 \nabla^2 \left ( V_{e-N}(r') - \frac{Z}{r'} \right ) d^3r' ,
\frac{1}{6} \rho_e(0) \int \left ( V_{e-N}(r') + \frac{Z}{r'} \right ) \nabla^2 {r'}^2 d^3r' \nonumber \\
  &=&  \frac{1}{6} \rho_e(0) \int {r'}^2 \nabla^2 \left ( V_{e-N}(r') + \frac{Z}{r'} \right ) d^3r' ,
\end{eqnarray}
where $\rho_e(0)$ is the electron density at the origin. Using the identities, $\nabla^2 \left ( \frac{1} {r} \right )=- 4 \pi \delta(r)$ and 
%L.S.: V_{e-N} is not an electrostatic potential according to Eq. (1) 
%$\nabla^2 V_{e-N} (r) = - 4 \pi Z \rho_N(r) $,
$\nabla^2 V_{e-N} (r) = + 4 \pi Z \rho_N(r) $
together with the nuclear charge distribution density $\rho_N(r)$ normalized to unity, it yields
%\LS{Please, check Eq. (22). From Eq. (21) (with a corrected sign in front of $Z/r'$) one has $E_{\text{FS}} = \frac{1}{6} \rho_e(0) \int {r'}^2 ( -4\pi Z\rho_N(r') +4\pi \delta(r') ) d^3r'=-\frac{2}{3}\pi Z\rho_e(0)(\langle r^2 \rangle + 0)=-\frac{2}{3}\pi Z\rho_e(0)\langle r^2 \rangle$. This is a ``field shift'' of the finite nucleus system with respect to the point nucleus one as stated in the beginning of this paragraph (more correctly it is the finite size correction to the electronic energy~\cite{1997-Per}). This differs from Eq. (22) below which seems to be written for the FS of the system with one $<r^2>$ relative to other $<r^2>$, but this is not $E_{\text{FS}}$. Eq. (22) should coincide with Eq. (26) of Ref.~\cite{1997-Per} with an opposite sign of the FS operator used in the present paper.}
%\LS{Corrected (now this Eq. equals to Eq. (26) of Ref.~\cite{1997-Per}):}
\begin{equation}
%E_{\text{FS}} =-\frac{2}{3} \pi Z \rho_e(0) \delta \langle r^2 \rangle . 
E_{\text{FS}} =\frac{2}{3} \pi Z \rho_e(0) \langle r^2 \rangle . 
\end{equation}
From the above derivation, the FS factor can be deduced as
%\LS{Can we give any reference where the FS \textit{\textbf{operator}} (i.e. $F(r)$) is written in such a form? It seems that the operator $F(r)$ should be proportional to $\delta(r)$.}
% I removed the word "operator".
%\LS{Corrected (now we obtain the negative value of F for 2s-2p0.5 transition in Li as follows in Tables with results.):}
\begin{equation}\label{Fdef5}
%F = \frac{2 \pi}{3} Z |\langle \Psi (r) | \Psi(r)\rangle|_{r \rightarrow 0},
F = - \frac{2 \pi}{3} Z \rho_e(0).
\end{equation}
%where $|\Psi(r)\rangle$ is the electronic wave function. 

\subsection{Seltzer's moments}

In heavy atoms, the MS in the IS expression is usually much smaller than the FS, so that higher-order contributions due to the FS could be of importance.
The traditional approach to deal with this is to assume that the nuclear potential can be expressed as functions of second and higher-order radial moments. So, the change of the nuclear potential between isotopes $A$ and $A'$ can be expressed as
%L.S. Corrected to be consistent with Eq.~(\ref{Fdef1})}
\begin{equation}
%\delta V_n(r) = {\cal F} \lambda^{A,A'} ,
\delta V_n(r) = -{\cal F} \lambda^{A,A'} ,
\end{equation}
where ${\cal F}$ is the more generalized form of the FS factor and
\begin{eqnarray}\label{eq:Seltzer}
\lambda^{A,A'} =  \delta \langle r^2 \rangle^{A,A'} + \frac{C_2}{C_1}  \delta \langle r^4 \rangle^{A,A'}  + \frac{C_3}{C_1}  \delta \langle r^6 \rangle^{A,A'} + \cdots . \nonumber \\
\label{eq:Seltzer}
\end{eqnarray}
In the above expression, $C_n$s are known as the Seltzer's factors and $\langle r^n \rangle^{A,A'}$ are known as the Seltzer's moments with $n=2,4,6,\cdots$~\cite{1969-Seltzer,1987-Blund}. In the systems with heavy nuclei, contributions from the higher-moments are significant. They also become prominent in the context of probing new physics with IS studies.

The aforementioned approach has its drawbacks. Namely in a perturbative expansion of the energy levels of a hydrogenic system~\cite{Pachucki-3P}, other moments appear; the first one being Friar's~\cite{1979-Friar}.
Moreover, for high $Z$ the leading power is no longer $2$ but $2\gamma$, where $\gamma=\sqrt{1-(Z \alpha)^2}$~\cite{1993-Shab}.
However, in the present review, we mostly restrict our discussions to the leading FS effect.

\subsection{QED corrections from approximate to rigorous}

The methods of relativistic MBPT (RMBPT), relativistic CI (RCI), and RCC can be regarded as different approximations to the QED perturbation theory~\cite{sapirstein-87-qed,Shabaev:2024:94:inbook}. To carry out the calculations conveniently, the dominant contributions from the DC Hamiltonian (and Breit interactions) are estimated first. In the next step, higher-order effects arising from the QED theory are included. Highly accurate calculations including a rigorous treatment of QED effects are possible for HCIs. However, applying these methods to neutral or weakly charged many-electron atoms is extremely complicated. Therefore, the approximate \textit{model} approaches, considering QED effects as quantum-mechanical operators, are necessary~\cite{Shabaev:2024:94:inbook}.

\subsubsection{Model QED approach for the field shift}\label{QED-FS}

Recently, an approach has been proposed~\cite{skripnikov2024isotopeQED} to calculate the QED contribution to the field shift factor in many-electron systems, that can be combined with modern methods of electronic structure theory designed to accurately treat high-order electron correlation effects. Within this formalism one starts by considering some effective operator $H^{\rm QED}(\mathbf{r}, \langle r^2 \rangle)$ (different expressions can be considered), which describes the vacuum polarization (VP) and self-energy (SE) effects of QED. These operators explicitly depend on the RMS nuclear charge radius, $\langle r^2 \rangle$, and the electron coordinate $\mathbf{r}$. The QED contribution to the electronic energy in the first-order is given by the expectation value
\begin{equation}
E^{\rm QED}(\langle r^2 \rangle) = \langle \Psi^{\langle r^2 \rangle} | \sum_i H^{\rm QED}(\mathbf{r}_i, \langle r^2 \rangle) | \Psi^{\langle r^2 \rangle}\rangle,   
\label{EQED}
\end{equation}
where $|\Psi^{\langle r^2 \rangle} \rangle$ is the many-electron wave function, which depends on the RMS nuclear charge radius through the electron-nucleus interaction operator $V_{\rm n}(r)$ in Eq.~(\ref{eq:DC}). The QED contribution to the FS factor, $F^{\rm QED}$, can be calculated using the following relation:
\begin{equation}
\label{FQED}
F^{\rm QED} = -\frac{d \langle \Psi^{\langle r^2 \rangle} | \sum_i H^{\rm QED}(\mathbf{r}_i, \langle r^2 \rangle) | \Psi^{\langle r^2 \rangle} \rangle}{d \langle r^2 \rangle}.
\end{equation}
where the minus sign is opposite to that used in Ref.~\cite{skripnikov2024isotopeQED} and corresponds to the definition~(\ref{Fdef1}) of the FS constant used in the present paper.

To calculate the VP contribution to $F^{\rm QED}$ one can use the Uehling potential~\cite{Uehl} and directly apply Eq.~(\ref{FQED}).
The main difficulty is to calculate the SE contribution. In Ref.~\cite{skripnikov2024isotopeQED}, it was started with the model SE approach introduced in the seminal paper~\cite{Shabaev:13} to calculate the SE contribution to transition energies in atoms. The idea of this approach is to scale the SE contribution of interest to the Lamb shift result for the Coulomb potential. The scaling is possible due to the fact that the dominant part of the QED effects comes from the vicinity of the nucleus and the proportionality property of the radial parts of atomic spinors having different principal quantum numbers $n$ and the same relativistic quantum number $\kappa=(-1)^{j+l+1/2}(j+1/2)$. The model SE operator can be written as follows~\cite{Shabaev:13,Skripnikov:2021a}
\begin{equation}
    H^{\rm SE} =\sum_{k,k',ljm} |h_{kljm}\rangle X_{kljm,k'ljm} \langle h_{k'ljm}|,
\label{Xmod}    
\end{equation}
where $h_{kljm}(\mathbf{r})$ is an orthonormalized set of functions~\cite{Skripnikov:2021a}, which are linear combinations of functions of the type
\begin{equation}
  \widetilde{h}_{nljm}(\mathbf{r})=\eta_{nljm}(\mathbf{r}) \theta(R_{\rm cut}-|\mathbf{r}|).
\label{hfuns}  
\end{equation}
Here $\eta_{nljm}(\mathbf{r})$ are H-like functions (with a principal quantum number $n \le 5$),  $\theta(R_{\rm cut}-|\mathbf{r}|)$ is the Heaviside step function and $R_{\rm cut}$ is a small radius, which can be varied slightly to study the stability of the results obtained. $X_{kljm,k'ljm}$ in Eq.~(\ref{Xmod}) are matrix elements of the SE operator over the $h_{kljm}$ functions and to a good accuracy they are the linear combinations of the corresponding matrix elements over the H-like functions. The SE operator is diagonal in $|ljm \rangle$. Both diagonal and off-diagonal matrix elements of this operator were calculated and tabulated in Ref.~\cite{Shabaev:13}. 

The operator, given by Eq.~(\ref{hfuns}), can be used to calculate the SE contribution to energies in atoms, but not the contribution to the FS. This is because it is not enough to know $X_{kljm,k'ljm}$ to calculate $F^{\rm SE}$. To see this, let us formally introduce a one-particle density matrix $D_{p,q}^{\langle r^2 \rangle}$ calculated using a many-electron wave function for an atom of interest and expressed in the basis of functions $h_{p}(\mathbf{r})$, where for brevity the set of indices $kljm$ are replaced by only one index $p$ ($q$). The index $\langle r^2 \rangle$ for the density matrix $D_{p,q}^{\langle r^2 \rangle}$ indicates its dependence on the RMS charge radius. Now it follows from Eq.~(\ref{FQED}) that
\begin{eqnarray}\label{Fse}
    F^{\rm SE} = -\sum_{p,q} 
\frac{d X_{p,q}^{\langle r^2 \rangle}}
    {d \langle r^2 \rangle} D_{p,q}^{\langle r_0^2 \rangle} 
    -\sum_{p,q} 
    X_{p,q}^{\langle r_0^2 \rangle}
 \frac{d D_{p,q}^{\langle r^2 \rangle}}{d \langle r^2 \rangle},    
\end{eqnarray}
where $\langle r_0^2 \rangle$ is the mean-squared nuclear charge radius corresponding to the isotope used as a reference in the IS measurement,  all derivatives are taken at this point,
and the minus sign corresponds to the definition of the FS constant~(\ref{Fdef1}).
In contrast to the calculation of the SE contribution to the transition energy, one should know not only the $X_{kljm,k'ljm}$ matrix elements, but also their derivatives with respect to the mean-squared nuclear charge radius change.
This can be implemented~\cite{skripnikov2024isotopeQED} by using the nuclear size correction to the SE matrix elements calculated and tabulated in Ref.~\cite{Shabaev:13} and analytic properties of such matrix elements explored in Refs.~\cite{Milstein:2003,Milstein:2004,Yerokhin:2011b,Ulrich:2003,Pachucki:1993,Milstein:2002,Eides:1997}.

To probe the approach outlined above, it was applied to calculate QED contribution to the FS factors in a series of Li-like HCIs~\cite{skripnikov2024isotopeQED}. The method was able to reproduce the results of a rigorous QED treatment~\cite{Zubova:2016} with a deviation of roughly 15\%. Such an accuracy for the QED contribution, which is in itself small, is enough for neutral and singly-charged many-electron systems. For neutral atoms, the method has been recently applied to calculate QED contributions to FS factors in Al~\cite{skripnikov2024isotopeQED} (see below).

\subsubsection{Model QED approach for the mass shift.}\label{QED-MS}
Eq.~(\ref{nmsexp}) gives terms of order $(\alpha Z)^4m^2/M$ for the MS. Terms of order $(\alpha Z)^5m^2/M$ and higher orders in $\alpha Z$ can be calculated within the rigorous QED approach (see, e.g.,~\cite{Shabaev:02a} and references therein). In Ref.~\cite{Anisimova:2022}, an approach was formulated to account for these terms within the model operator technique. This operator has the same structure as those introduced in Ref.~\cite{Shabaev:13} for the SE effect calculation. Both diagonal and off-diagonal matrix elements over the H-like functions were calculated and tabulated and pilot test applications have been made~\cite{Anisimova:2022}.
The approach was also adapted~\cite{skripnikov2024isotopeQED} 
to the operator form of Eq.~(\ref{Xmod}) that can be used (interfaced) with relativistic program packages allowing to treat electron correlation effects at the relativistic CC level.
Note that an additional QED contribution to the nuclear recoil effect should be considered, which is induced by the perturbation of the electronic wave function by the SE and VP interactions~\cite{King:2022:43,skripnikov2024isotopeQED}. This effect can be calculated as the difference between the values of the mass shift factors computed with and without inclusion of the model VP and SE operators~(\ref{Xmod}) into the electron correlation calculation.

It was shown~\cite{skripnikov2024isotopeQED} using the example of the neutral Al atom (Z=13) that the QED corrections to both the FS and MS can be significant compared to the achievable uncertainty of the electronic calculation within the DCB approximation. It was also demonstrated that there is a strong interference between QED and correlation effects in such a neutral many-electron system.

\subsubsection{QED effects for HCIs.}\label{subsec:QED_HCIs}

Rigorous calculations of QED effects for HCIs are possible to all orders in $Z\alpha$, that is, including the nuclear Coulomb interaction in a non-perturbative manner.
The subject was recently reviewed by Indelicato~\cite{2019-Paul} and Shabaev~\cite{Shabaev:2024:94:inbook}, so here we will make a few general comments. While these methods can be very accurate for high-$Z$ HCIs, they tend to lose accuracy when applied to neutral atoms or low-charge ions, partly because of near cancellations between different terms in the formalism, and partly because of the difficulty of treating correlation corrections to the QED effects adequately in a neutral system. Therefore the most successful applications of these approaches have been reported for high-$Z$ few-electron ions (see, for example, Refs.~\cite{blundell-93-is,kozhedub-10-qed,sapirstein-15-is,2020-QEDNMS,2014-LiLike,2019-QEDSMS,2020-QEDNMS}).

Bound-state QED perturbation theory can be used to evaluate the Feynman diagrams for the VP and SE effects directly to all-orders in $Z\alpha$. Such calculations were performed for a single electron in the bare nuclear potential~\cite{mohr-74-qed}, of direct applicability to H-like ions, but it is also possible to include a mean-field potential for core electrons in the electron propagators to improve treatments for many-electron ions~\cite{blundell-93-is,blundell-09-qed,kozhedub-10-qed,sapirstein-15-is}. The Feynman diagrams for the SE and VP contain the well-known ultraviolet divergences of QED and, to extract the finite physically observable effect, they must be renormalized using methods analogous to those used for scattering diagrams in QED. For this renormalization program to be consistent, the mean-field potential (if included) must be a local potential. Therefore, it is not possible to include a Dirac-Hartree-Fock (DHF) mean-field potential in the electron propagators of the SE or VP, because the exchange terms of this potential are nonlocal. Instead, one chooses a suitable local mean-field potential and evaluates the exchange effects separately using extra diagrams; see Refs.~\cite{blundell-93-is,kozhedub-10-qed,sapirstein-15-is} for examples.

Another field-theoretic effect is retardation, arising from the finiteness of the speed of light, for which one must consider the full frequency-dependence of the photon propagator rather than approximating it by frequency-independent or instantaneous forms (e.g., the DCB Hamiltonian). Retardation in this sense is necessarily included in the SE and VP mentioned above, but other diagrams of bound-state QED perturbation theory can be regarded as field-theoretical generalizations of the familiar correlation-type diagrams of RMBPT. The rigorous evaluation of these correlation diagrams within bound-state QED allows one to compute the retardation corrections to correlation effects (see, for example, Ref.~\cite{blundell-93-qed}). Also, negative-energy electron states can be included in the electron propagators of the correlation diagrams, but the diagram must then be evaluated correctly using the rules of QED to avoid spurious divergences.

Another important types of QED effects for IS work are the higher-order relativistic corrections to nuclear recoil. The NMS [given by Eq.~(\ref{nmsexp})] and SMS [given by Eq.~(\ref{smsexp})] operators include only the leading relativistic correction of order $(Z\alpha)^{2}$ to the nonrelativistic mass shift. Higher-order in $Z\alpha$ QED corrections to the MS have been evaluated in Refs.~\cite{1988-Shab,shabaev1985mass,1995-Shab,ShabaevRecoil:98,2007-Adkins,kozhedub-10-qed,zubova-14-is,2020-QEDNMS}.

\section{Calculating IS factors with the FF approach}\label{secFF}

The most straightforward way to calculate the IS factors of Eq.~(\ref{eq:IS}) is to perform two sequential many-body calculations, the first one with, and the second one without, the effect of interest, and then take the difference of the two calculations. For the FS, this means performing two calculations for the different nuclear charge radii (of isotopes $A$ and $A'$) and taking the difference between them. Similarly, for the MS one can perform one calculation with the NMS and SMS terms omitted from the Hamiltonian (i.e., assuming an infinite nuclear mass) and one calculation with the NMS and SMS operators present. The total mass shift effect is then the difference of the two calculations. The difficulty with this direct-subtraction approach is that it involves finding a small difference of two large numbers and can easily be overwhelmed by numerical errors in the calculations. Nevertheless, it can be used successfully in some cases, particularly in RMBPT methods (see Sec.~\ref{subsec:HCI_d<r^2>} for an example), where the truncation errors can be controlled more easily in every order of the perturbation expansion.
In principle, this approach can be used for neutral atoms, but certain caution should be exercised.

To have a better control of the numerical stability in the (R)CC methods, it is customary to apply the FF approach, which is a commonly-used generalization of the above simple procedure. In this approach, the IS factor of an atomic state due to the corresponding IS operator $O$ can be determined by using an effective Hamiltonian $H_{\text{eff}}=H + \lambda O$, where $\lambda$ is a small parameter, which can still be larger than the ``natural'' choice for the operator; e.g. $1/M$ for mass effects.
The energy correction due to $O$ can be expressed as
\begin{equation}
 E(\lambda) = E^{(0)} + \lambda \langle O\rangle + {\cal O}(\lambda)^{2} ,
\label{eqff}   
\end{equation}
where the superscripts (0), (1), etc. denote the order of perturbation. The first-order energy correction can be calculated as
\begin{equation}
\langle O\rangle \equiv \left. \frac{\partial E(\lambda)}{\partial \lambda} \right|_{\lambda=0}  \approx \frac{ E(+\lambda)-E(-\lambda)}{2 \lambda}.
\label{eqff1}
\end{equation}

The application of the FF approach for the case of NMS and SMS operators is straightforward: the operator $O$ in the equations above is substituted with one of the operators given by (\ref{nmsexp}) and~(\ref{smsexp}). For the case of the FS factor, there are two options for applying the FF approach. The first one implies a substitution of the operator $O$ with the FS operator given by~(\ref{Fdef1}). Within the second option, the $\lambda$ parameter in Eq.~(\ref{eqff1}) corresponds to a change in the RMS charge radius in the nuclear potential operator $V_n(r)$ in Eq.~(\ref{eq:DC}). Thus, the FF approach reduces to the calculation of the following derivative:
%\LS{This formula leads to opposite sign with respect to Eq.~(\ref{Fdef1}). I would correct sign in Eq.~(\ref{Fdef1}).}
% Ben: Thanks, corrected
\begin{equation}
\label{FScalc}
F=-\frac{E(\langle r_0^2\rangle+h) - E(\langle r_0^2\rangle-h)}{2h}. 
\end{equation}
Here, $h$ describes a small change in the mean-squared charge radius.

From the aforementioned discussions it is evident that the advantage of the FF method is that it can be easily combined with almost any method of calculating electronic energy with high accuracy. On the other hand, depending on the (i) concrete electronic structure computational method, (ii) the structure of the IS operator and (iii) the electronic state of interest, certain numerical instabilities in calculating the energy differences with $+\lambda$ and $-\lambda$ can arise. The appropriate choice of the $\lambda$ parameter for the respective IS factor requires knowledge of their strengths {\it a priori}. Uncertainties arising due to such an approximation can be overcome by choosing different $\lambda$ values for the FS, NMS and SMS factors. A possible remedy to this issue is to use more numerically stable differentiation formulae to estimate first-order energy derivatives~\cite{prasana}. In the latter part of this review, we discuss approaches which do not require to choose a perturbative parameter $\lambda$ explicitly for the estimation of IS factors.

\section{Solving the many-electron problem}\label{sec:many}
\subsection{Many-body perturbation theory}\label{subsec:mbpt}

The patterns of electron correlations in HCIs are relatively easier to understand than in a neutral atom or a lightly-charged ion. This happens because each order of RMBPT brings in an additional Coulomb matrix element ($\sim Z)$ and energy denominator ($\sim1/Z^{2}$), and thus scales approximately as $1/Z$ compared to the previous order. RMBPT therefore becomes increasingly convergent as $Z$ increases, with $1/Z$ acting as the ``small parameter'' of the expansion. The many-body problem is generally harder in neutral atoms, molecules, or low-charge ions, where for high-accuracy results it is usually necessary to use methods that sum subsets of many-body diagrams to all-orders of perturbation theory, such as the CC method discussed in the next subsections. In simple HCIs (notably single-valence systems, but also closed-shell and two-valence systems), one can use RMBPT directly; in Sec.~\ref{subsec:HCI_d<r^2>} we will discuss the use of RMBPT up to the third order for Na-like ions. 
Computations using RMBPT, to a low order in the expansion, are generally much faster than for the case of all-order methods such as RCC or RCI, and in particular one can use atomic basis sets of different size for different orders of perturbation theory. On the other hand, the great advantage of a method like RCI for HCIs is that it can be applied to more complex systems such as trivalent ions~\cite{2020-CAR,2022-HCIclock}, within the limits of the basis-set size and configuration space possible, which have to be assessed on a case-by-case basis.
Against this, the treatment of relativistic effects is more difficult in HCIs. In addition to the Breit interaction, it may also be necessary to include field-theoretic effects as discussed in Sec.~\ref{subsec:QED_HCIs}.

An order-by-order treatment of electron correlation for the IS in Na-like ion is given in Ref.~\cite{silwal-18-is}. This employed RMBPT up to third-order in the DCB interaction Hamiltonian, following the general scheme of Ref.~\cite{johnson-88-mbpt}. For the MS, the leading relativistic corrections to nuclear recoil, of order $(Z\alpha)^{2}$ relative to the nonrelativistic MS, were included using the one-body NMS operator and two-body SMS operator introduced earlier. The DHF method with the DCB Hamiltonian was extended to include both the NMS and SMS terms, with the SMS term treated analogously to the two-body Coulomb and Breit interactions. In this way, the basis states and their energies are perturbed by the nuclear-recoil operator. In the RMBPT terms through the third-order, the residual SMS interaction also enters explicitly along with the residual Coulomb and Breit interactions. Two calculations were performed: one including nuclear recoil and the other one without it. Their differences, in each order of RMBPT, were determined and tested carefully for numerical significance. Another direct-subtraction procedure was employed for the FS, where calculations were performed for different nuclear charge distributions. In this way, it was possible to construct a perturbation series through the third-order for the FS coefficient. See Sec.~\ref{subsec:HCI_d<r^2>} for further discussion.

\subsection{All-order methods}

In neutral atoms or weakly charged many-electron ions, it is necessary to employ methods which account for relativistic and high-order electron correlation effects simultaneously to accurately calculate IS factors. The CI method and its variants are typically employed to evaluate the IS factors of atomic systems~\cite{1997-Per, Kozlov:15, NAZE20132187}. However, a truncated CI method is prone to size-consistency issues~\cite{2007-Bartlett}. The CC theory is considered to be a more potent many-body method for calculating 
energies and electronic properties of multi-electron systems~\cite{Cizek,bartlett,crawford}. The RCC theory is thus a natural choice to apply for accurate determination of the IS factors. There are many variants of CC methods proposed in the literature~\cite{bartlett, bishopbook, arponen}. These methods have not yet been explored extensively for the determination of the IS factors. Furthermore, in most cases, many-body methods are applied in the FF framework for the estimation of the IS factors~\cite{1997-Per, Kozlov:15, NAZE20132187}.

\subsection{Single-reference RCC theory}\label{SRRCC}

The ground state wave function, $\ket{\Psi}$, of an atomic system in the (R)CC theory is expressed as (see e.g. Refs.~\cite{Cizek,Visscher:96a,2007-Bartlett} and references therein for a comprehensive review)
\begin{equation} 
\label{expAnsatz}
    \ket{\Psi} = e^{T} \ket{\Phi_0} ,
\end{equation}
where $\ket{\Phi_0}$ is the reference determinant obtained using the DHF method, $T = {T}_1 + {T}_2 + {T}_3 + \dots$ is the cluster operator consisting of $n$-body terms representing single (${T}_1$), double (${T}_2$), triple (${T}_3$), etc., up to $n$-tuple excitations for an $n$-electron system, with respect to the reference determinant $\ket{\Phi_0}$. The cluster operator amplitudes-determining equations are given by
\begin{equation}
\label{eq:srcc-ampl-eq}
\langle \Phi_0^* | \bar{H}| \Phi_0 \rangle = 0 ,
\end{equation}
where $\ket{\Phi_0^*}$ denotes all possible determinants excited with respect to $\ket{\Phi_0}$, and $\bar{H} = \left( H e^T \right)_c$, with the subscript $c$ indicating that only the connected terms are retained in amplitude equations~\cite{2007-Bartlett}.

Commonly, the approximation within the single reference theory, accounting for single and double excitations (the SR-RCCSD method), ${T} \approx {T}_1 + {T}_2$, is used in routine electronic structure calculations. However, for an accurate determination of the IS factors, it is necessary to consider the singles, doubles, and triples excitations (the SR-RCCSDT method) by expressing ${T} \approx {T}_1 + {T}_2 + {T}_3$~\cite{Noga:87}. Such an electronic-structure model is computationally very costly, scaling with $O(N^8)$, where $N$ stands for the number of atomic or molecular spinors considered in the calculations. In some cases, different kinds of CC schemes are applied to include triple and higher-level excitations in an approximate manner. To account for the most important triple excitation effects at the 
%L.S. the computational cost of CCSD is O(N^6), while the computational cost of CCSD(T) is O(N^7)
%computational cost of the SR-RCCSD method level, 
relatively low computational cost
they are sometimes included through a perturbative approach along with the SR-RCCSD method approximation (SR-RCCSD(T) method). This involves a solution of the SR-RCCSD amplitude equations followed by a non-iterative second-order estimate of the $T_3$ cluster amplitudes. These amplitudes are then used to calculate the energy corrections; accurate up to the fifth-order of many-body perturbation theory~\cite{Raghavachari:89}. In a similar approach, contributions from the quadruple excitation operator from the ${T}_4$ operator can also be accounted for using a perturbative approach along with the SR-RCCSDT method (the SR-RCCSDT(Q) correction)~\cite{Bomble:05,Kallay:6}. It should be noted that the computational cost grows rapidly while proceeding from double to triple and from triple to quadruple excitations.

Single-reference methods for energy calculations are implemented for example in the {\sc exp-t}~\cite{EXPT_website,Oleynichenko_EXPT}, {\sc dirac}~\cite{DIRAC19,Saue:2020}, and {\sc mrcc}~\cite{MRCC2020,Kallay:1,Kallay:2} program packages. As it was noted in Sec.~\ref{secFF}, if the electronic-structure method provides a value of electronic energy for a given system, one can implement the FF approach to compute IS factors. For the MS calculation, the matrix elements of one- and two-electron integrals of the MS operators (Eqs.~(\ref{nmsexp}) and~(\ref{smsexp})) over atomic bispinors are necessary. The code for computing them using Gaussian-type basis functions employed in the~\textsc{dirac}~\cite{DIRAC19, Saue:2020} package was implemented in Ref.~\cite{Penyazkov:2023}. The use of Gaussian-type basis functions enables fast and analytical computation of all required integrals. Within this approach at the first stage all integrals are evaluated within the set of primitive Gaussian basis functions. Subsequently, a transformation to the basis of atomic bispinors derived from the DHF calculation using the \textsc{dirac} code, is performed.

In addition to the FF technique, one can also calculate the FS factor as an expectation value of the FS operator of Eq.~(\ref{Fdef1}). For this purpose, one should calculate the effective one-particle density matrix $D$ using the coupled cluster method supplied by the $\Lambda$-equations technique~\cite{Kallay:3,2007-Bartlett}, and then contract this matrix with
\begin{equation}\label{FSmethodB}
F=\sum_{p,q}D_{p,q} F_{p,q},
\end{equation}
where $F_{p,q}$ are the matrix elements of the FS operator (\ref{Fdef1}) expressed in the basis of atomic bispinors. Such a calculation can be used to control the numerical stability of the FF procedure, as was recently done in Al (See Ref.~\cite{skripnikov2024isotopeQED}, and section~\ref{SRapplication}). Similarly, one can employ this approach to compute the expectation value of the NMS operator of Eq,~(\ref{nmsexp}). 

The {\it ans\"atz} given by Eq.~(\ref{expAnsatz}) works well for single-reference problems in which the weight of the reference determinant is dominant in the expansion of Eq.~(\ref{expAnsatz}). Note that this {\it ans\"atz} also covers multi-valence high-spin problems.

\subsection{FS-RCC method for multi-valence systems}\label{sec:fscc}

Except for the simplest cases, atomic and molecular states possess a multi-reference character, i.e. several reference Slater determinants are required just to provide a qualitative description of an electronic wave function of the target state. In such cases single-reference approaches can give inadequate results. Thus, an inherently multi-reference approach assuming several leading configurations in an electronic wave function could to be used. Here we will review only the multi-reference Fock-space relativistic coupled cluster (FS-RCC) method~\cite{Kaldor1991,Eliav1998,Visscher:01,Eliav:Review:22}.Its version accounting for single and double excitations has been used for some time to calculate atomic excitation energies~\cite{Kaldor:98,Eliav:15}, as well as different properties. However, it was successfully applied to calculate IS factors just recently, which was facilitated by fully accounting for triple excitations~\cite{Oleynichenko:20} needed for calculations of states with complicated electronic structure.
Moreover, the treatment of triple excitations is also of high importance in order to assess the convergence of properties with increasing excitation level and so it assists with estimation the calculation errors.

The Fock-space \textit{ans\"atz} implies the use of the exponentially parameterized wave operator normal-ordered with respect to the common Fermi-vacuum state to generate an exact target wavefunction $\Psi_n$ by acting on a model vector $\tilde{\Psi}_n$:
\begin{equation}
\ket{\Psi_n} = \left\{ e^{\hat T}\right\} \ket{\tilde{\Psi}_n},
\quad\quad
\ket{\tilde{\Psi}_n} = \sum_\mu C_{\mu n} \ket{\Phi_\mu},
\label{woper}
\end{equation}
where curly braces denote normal-ordering and $\tilde{\Psi}_n$ is represented by a linear combination of model determinants $\Phi_\mu$ spanning a model space.
The latter model space is typically chosen to be complete in order to ensure the connectedness of working equations~\cite{mukherjee1989use} and is obtained by distribution of $N_h$ holes and $N_p$ particles among active one-electron spinors in all possible ways. Such a model space is referred to as the $(N_h h,N_p p)$ Fock-space sector.

The Fermi-vacuum state is typically chosen to be a closed-shell state, usually obtained within the DHF method. The cluster operator $T$ in Eq.~(\ref{woper}) is partitioned according to the number of valence holes ($n_{h}$) and valence particles ($n_{p}$), with $n_h \le N_h$ and $n_p \le N_p$, to be attached (detached) with respect to the Fermi-vacuum state:
\begin{equation}
{\hat T}=\sum_{n_{h} = 0}^{N_{h}} \sum_{n_{p}=0}^{N_{p}} \, {\hat T}^{(n_{h} h, n_{p} p)} . \label{sectors}%
\end{equation}
Here the ${\hat T}^{(n_{h} h, n_{p} p)}$ operator is related to the ($n_h h$,$n_p p$) sector of the Fock-space.

To describe electronic states in the target ($N_{h} h$,$N_{p} p$) Fock-space sector, one has to determine ${\hat T}^{(n_{h} h, n_{p} p)}$ operators with $n_{h}\le N_{h}$ and $n_{p}\le N_{p}$ \cite{mukherjee1989use}, thus the FS-RCC equations are solved sector-by-sector~\cite{Kaldor1991,Eliav1998}. As in the single-reference case considered in Section~\ref{SRRCC}, the partitioning of cluster operators with respect to their excitation rank is further introduced. For example, in the Fock-space RCCSDT approximation (FS-RCCSDT method), the ${\hat T}^{(n_{h} h, n_{p} p)}$ operator includes up to triple excitations:
\begin{equation}
{\hat T}^{(n_{h} h, n_{p} p)} = {\hat T}^{(n_{h} h, n_{p} p)}_{1} + {\hat T}^{(n_{h} h,n_{p} p)}_{2} + {\hat T}^{(n_{h} h,
n_{p} p)}_{3}.
\end{equation}
Note that the Fock-space (R)CC theory for the $(0h,0p)$ sector is essentially equivalent to the single-reference (R)CC approach (\ref{expAnsatz}). Moreover, the (R)CC approach for single-valence systems, described further in the next section, can be regarded as the FS-RCC method formulated for the $(0h,1p)$ sector with additional restriction imposed on model vectors, which must be represented by a single determinant (this requirement actually holds for almost all single-valence atomic systems).

Within the FS-RCC methodology, the relativistic counterpart of the Schr\"{o}dinger equation is reformulated in the model space resulting in the so-called Bloch equations defining cluster amplitudes~\cite{Lindgren:78,Lindgren:87,Kaldor1991,Mani:11},
\begin{equation}
\label{eq:fscc-ampl-eq}
Q[{\hat T}^{(n_{h} h, n_{p} p)},H_0]P =
Q\left( V \Omega - \Omega (V\Omega)_{cl} \right)_{conn} P,
\end{equation}
where the $H = H_0 + V$ partitioning is introduced for the electronic Hamiltonian, $\Omega$ denotes the wave operator $\{e^T\}$, $P$ and $Q$ stands for projectors onto the model space and its orthogonal complement, respectively, \textit{cl} indicates the operator is closed, i.e. is acting only within the model space, and \textit{conn} means that only the terms represented by connected Brandow diagrams are retained. Energy levels together with model vectors $\tilde{\Psi}_n$ are obtained by diagonalization of the effective Hamiltonian operator $\tilde{H}$ defined as
\begin{equation}
\label{eq:fscc-heff}
\tilde{H} = (H\Omega)_{cl,conn}.
\end{equation}
Amplitude equations Eq.~(\ref{eq:fscc-ampl-eq}) are very similar to that for the vacuum sector~(\ref{eq:srcc-ampl-eq}) except for the additional renormalization term.

The manifold of electronic states accessible by FS-RCC is completely determined by the structure of the model spaces used. In radii determinations with IS studies, one typically considers transitions between different electronic shells, such as those driven by the $s\rightarrow p$ or $p \rightarrow d$ type transitions. The shells under consideration, possibly supplemented by other shells close to them in energy, comprise an active space which is further used to construct model space determinants. The need to use a complete model-space to ensure exact size-consistency (and to keep FS-RCC equations simple and computationally tractable) typically leads to a situation where some model space determinants are not well-separated in energy from determinants belonging to the orthogonal complement space $Q$. As a result, energy denominators that are positive or close to zero appear in the FS-RCC amplitude equations, leading to deteriorated or even completely destroyed convergence, making it impossible to obtain solutions. There are two basic ways to overcome such an intruder state problem (for details see~\cite{Eliav:Review:22} and references therein). The first one consists in shifting energy denominators to the negative energy domain \cite{Forsberg:97,Eliav:IH:05,Zaitsevskii:RbCs:17,Oleynichenko:HFS:20} with possible further extrapolation to the zero shift limit~\cite{Zaitsevskii:Pade:18}. This approach distorts FS-RCC equations and slightly breaks exact size-extensivity while still preserving core separability, but it allows one to obtain a solution in the most difficult cases. It can be used to obtain corrections for some effects not very sensitive to the shifting parameters like contributions of triple excitations. The second approach, known as the intermediate Hamiltonian (IH) technique (see Ref.~\cite{Malrieu:IH:85}), is more subtle and is based on the assumption that only a small number of low-lying (main) electronic states are of interest, while other (intermediate) electronic states generated by the model space are considered as buffer states which are used to prevent intruder states from interfering with convergence. There are several formulations of the IH-FS-RCC method~\cite{Meissner:98,Landau:IH:01,Eliav:IH:05,Musial:IH:08,Dutta:IH:14,Zaitsevskii:QED:22}. The recently developed IH for incomplete main model-space technique~\cite{Zaitsevskii:QED:22} seems to be one of the most computationally affordable and at the same time stable with respect to the IH parameters; it is already used in ongoing applications on IS and the results are published elsewhere~\cite{Skripnikov:24}.

Until recently, there were two well-established codes for FS-RCC calculations in multivalence sectors, the Tel-Aviv relativistic atomic code {\sc trafs-3c}~\cite{Eliav:94}, and {\sc dirac}~\cite{DIRAC19,Saue:2020}. Both programs implement the FS-RCCSD model with no more than two quasiparticles over the Fermi-vacuum. The latest implementation of the FS-RCC theory was presented recently in the {\sc exp-t} program~\cite{EXPT_website,Oleynichenko_EXPT}.
{\sc exp-t} also includes the first implementations of FS-RCCSDT method and its simplified versions FS-RCCSDT-n ($n=1,2,3$).
These features are available not only for the lowest Fock-space sectors~\cite{Oleynichenko:20}, but also allowing model spaces with three quasiparticles~\cite{SkripnikovBiCCSDT:2021,Eliav:Review:22}, thus extending the scope of applicability of FS-RCC.
The intermediate Hamiltonian formulation for incomplete main model spaces as well as the above-mentioned technique of property matrix element calculations were also recently implemented in {\sc exp-t}.

\subsubsection{IS factors calculations in multi-valence systems.}

In its practical implementation, the FS-RCC method uses the same integrals as the single-reference methods described in Sec.~\ref{SRRCC}. Therefore, there is a close similarity in approaches to calculate IS factors, in particular, one can apply the FF method. The specific feature of the FS-RCC approach is that this method can be regarded as a multi-state approach and allows one to obtain IS factors for all the electronic states under consideration simultaneously. 
The first application of the FS-RCCSDT for the two-valence system, for the Si atom, is given below.

A fully analytic approach to calculate FS-RCC density matrices was proposed~\cite{Szalay:95,Shamasundar:04}, but it has never been used in any large-scale relativistic applications due to the lack of a program implementation. More importantly, within this approach, the $\Lambda$-equations explicitly depend on the coefficients of a particular model vector (Eq.~(\ref{woper})) and thus have to be solved separately for each electronic state of interest. This circumstance makes the computational cost of the analytic method very high compared to the FF approach.

Recently, an approximate but computationally inexpensive 
method to obtain expectation values and even transition matrix elements of one-particle operators within the FS-RCC approach was proposed~\cite{Zaitsevskii:ThO:23,Oleynichenko:Optics:23}. It assumes that a matrix element for a pair of electronic states
%$\Psi_n$ and $\Psi_m$ ($n = m$ 
$\ket{\Psi_n}$ and $\ket{\Psi_m}$ ($n = m$ 
for expectation value calculations) can be expressed  via a matrix element of the corresponding effective property operator as
\begin{eqnarray}\label{eq:eff-oper-general}
    \braket{O}_{nm} &=&N_n N_m^{-1} \braket{\tilde{\Psi}_n^{\perp\perp}|\tilde{O}|\tilde{\Psi}_m}, \nonumber \\
    \quad N_n &=& \braket{\Psi_n|\Psi_n}^{1/2},
\end{eqnarray}
where $\tilde{\Psi}_n^{\perp\perp}$ stands for the left eigenvector of an effective Hamiltonian and $\braket{\tilde{\Psi}_n^{\perp\perp}|\tilde{\Psi}_m} = \delta_{nm}$. Effective operator $\tilde{O}$ in FS-RCC is represented by an actually infinite series and one has to retain only terms of some maximum power of the cluster operator $T$. The $\tilde{O}$ operator exact to the second order in $T$ is given by the expression~\cite{Zaitsevskii:ThO:23}:
\begin{eqnarray}
\tilde{O}(2) &\approx & \left( O + T^\dagger O + OT + \frac{\{ (T^\dagger)^2 \}}{2}O + T^\dagger O T \right. \nonumber \\ && \left. + O \frac{\{ T^2 \}}{2} - (T^\dagger T)_{cl} O \right)_{cl,conn},
\label{eq:eff-oper-2nd-order}
\end{eqnarray}
%
%[{\bf in normal-order formulation $(T+T)_{cl}$ should not appear}]
%AO: such terms indeed arise due to the (e^T)^\dagger operator from the bra-vector
%
where \textit{cl} denotes the closed operator, i.e. restricted to the model space, and \textit{conn} means that the expression contains only connected Brandow diagrams, thus ensuring size-consistency of calculated property operator matrix elements. Note that the presented formulation again lacks for the orbital relaxation effects. Eqs.~(\ref{eq:eff-oper-general}) and~(\ref{eq:eff-oper-2nd-order}) can be regarded as a generalization of Eq.~(\ref{evexp}) to the multi-valence systems (see also Ref.~\cite{Gopakumar:02}). The described approach has not been yet applied to calculate IS factors, but due to its very modest computational cost, it suits well for obtaining fast estimates of matrix elements of interest. 

\subsection{FS-RCC method for one-valence systems}\label{sec:single-valence-cc}

Here we discuss the FS-RCC method tailored to one-valence atomic systems.
The ground state wave function of a one-valence atomic system can be obtained by using the V$^n$ potential in the RCC theory. However, it would require to apply the equation-of-motion based CC theory (the EOMCC method) to determine the excited states of a system. Within the EOMCC method, one diagonalizes the effective Hamiltonian, $\tilde{H}$, in a similar fashion to the CI method to obtain solutions for the desired excited states. However, one could obtain simultaneously solutions for both the ground and excited states having a valence orbital with a common closed-shell vacuum easily by using the single reference FS-RCC method. The single reference FS-RCC method for one-valence system is equivalent to the EOMCC method.

Within the FS-RCC method for one-valence system, we first evaluate the exact wave function ($\ket{\Psi_0}$) of the ground state of the closed-shell core using the (R)CC theory {\it ans\"atz} (see e.g. Refs.~\cite{Cizek,Visscher:96a,2007-Bartlett} and references therein for a comprehensive review) as
\begin{equation} 
\label{expAnsatzFS}
    \ket{\Psi_0} = e^{T^{(0,0)}} \ket{\Phi_0},
\end{equation}
where $\ket{\Phi_0}$ is the DHF determinant approximation of the wave function of a closed-shell core. For brevity, we denote the RCC excitation operator (wave operator) as $T^{(0,0)} \equiv T$ which includes single, double, etc. excitations from $\ket{\Phi_0}$ that are denoted by $T_1$, $T_2$, etc. respectively. 
The amplitude-determining equations for the $T$ operators were discussed in Sec.~\ref{SRRCC}.

To determine wave function of the actual state of a single-valence system in the FS-RCC method, an electron can be attached to a valence orbital spinor of the closed-shell core and correlation of this valence electron should be included. For this purpose a new reference determinant is defined as $|\Phi_v \rangle = a_v^{\dagger} |\Phi_0 \rangle$ with $a_v^{\dagger}$ denoting attachment of the valence orbital $v$; this construction is equivalent to the V$^{n-1}$ potential DHF wave function. Note that for one-valence systems, the multi-reference FS-RCC theory boils down to its single-reference counterpart naturally. Following the definition of Eq.~(\ref{sectors}), the exact wave function of the new state can be expressed sing the definition of the RCC operator~\cite{lindgren,Lindgren:78,Lindgren:87,mukherjee}
\begin{equation}
 |\Psi_v \rangle = e^{T^{(0,0)}+T^{(0,1v)}} |\Phi_v \rangle = e^T \{ 1+ S_v \} |\Phi_v \rangle ,
 \label{eqcc}
\end{equation}
where we have again replaced the notation from $T^{(0,1v)}$ to $S_v$ in the above expression which is responsible for accounting the electron correlation effects involving the valence electron. It can be noticed that we have written $e^{S_v}= 1 + S_v$ due to the fact that we deal with only one valence electron in this case.

Again following Eq.~(\ref{eq:fscc-ampl-eq}), amplitudes of the $S_v$ operator are obtained by solving the set of equations
\begin{equation}
 \langle \Phi_v^* | \{ (\tilde{H}-E_v) S_v \} + \bar{H} | \Phi_v \rangle = 0,
 \label{eqamp}
\end{equation}
where $|\Phi_v^* \rangle$ denote all possible excited determinants with respect to $|\Phi_v \rangle$ and $E_v$ is the energy of the one-valence atomic state which can be obtained using Eq.~(\ref{eq:fscc-heff}) as
\begin{equation}
E_v = \langle \Phi_v | \tilde{H}\{ 1+ S_v \} | \Phi_v \rangle. \label{eqeng}
\end{equation}
Both Eqs.~(\ref{eqamp}) and (\ref{eqeng}) are solved self-consistently. When a normal-ordered form of $H$ with respect to the Fermi-vacuum state $|\Phi_0 \rangle$ is used, the above energy expression gives an electron affinity (EA).

In the RCCSD method approximation, we define 
\begin{equation}
 S_v = S_{1v} + S_{2v}  ,
\end{equation}
while in the RCCSDT method the $S_v$ cluster operator is defined as
\begin{equation}
 S_v = S_{1v} + S_{2v} + S_{3v} .
\end{equation}
Here subscripts 1, 2 and 3 denote single, double and triple excitations, respectively. In open-shell atomic systems, correlation due to valence electron(s) are usually stronger than the correlation effects due to core electrons. Thus, it can be appropriate to make an approximation like $T = T_1 + T_2$ and $S_v = S_{1v} + S_{2v} + S_{3v}$ to account for the dominant triple excitation contributions involving the valence electron, $v$. This single, double and dominant triple excitation involving the valence electron approximation in the RCC method is referred to as the RCCSDTv method. 

As it was mentioned in the Sec.~\ref{secFF}, it is possible to estimate IS factors using the FS-RCC method using the FF approach. However, the numerical accuracy of the results can depend on the choice of a perturbation parameter $\lambda$. Below we discuss two approaches, namely EVE and AR, that are applied to estimate the FS factors of one-valence atomic systems in the FS-RCC theory framework. These approaches were discussed elaborately in Refs.~\cite{SahooLi,Sahoo_2020}, and are outlined here for the sake of completeness.

\subsubsection{The EVE approach.}

As discussed in the previous two sections, the IS factors are basically the expectation values of the IS operators. In the EVE approach of the FS-RCC method for single-valence systems, the IS factor of the corresponding operator $O$ is given by 
\begin{eqnarray}
\langle O \rangle &=&  \frac{\langle \Psi_v | O | \Psi_v \rangle}{\langle \Psi_v |\Psi_v \rangle} \nonumber \\
&=& \frac{\langle \Phi_v | \{ 1+ S_v^{\dagger} \} e^{T^{\dagger}} O e^T \{1+S_v \} | \Phi_v \rangle} {\langle \Phi_v | \{ 1+ S_v^{\dagger} \} e^{T^{\dagger}} e^T \{1+S_v \} | \Phi_v \rangle}  .
\label{evexp}
\end{eqnarray}
As can be seen, the above expression contains two non-terminating series, namely, $e^{T^{\dagger}} O e^T$ and $e^{T^{\dagger}} e^T$ in the numerator and denominator, respectively. Moreover, the above expression does not satisfy the Hellmann-Feynman theorem~\cite{bishopbook}. These are the main disadvantages of adopting the EVE approach for the determination of IS factors in the FS-RCC theory framework. 
%\textcolor{red}{However, the EVE expression in low orders can be implemented as a quite efficient computer code and thus allows for rather fast (but not extremely accurate) estimates of expectation values of property operators.} [{\bf This is not my statement. I am not sure what is being conveyed here.}]

\subsubsection{The AR approach.}

The problems of the FF and EVE approaches in the determination of the IS factors can be circumvented by applying the AR approach formulated for the FS-RCC method~\cite{Monkhorst}. The basic notion of this approach lies in the fact that it evaluates the IS factors as the first-order energy correction like in the FF approach, {\it albeit} its starting point is the same as in the EVE approach. However, unlike the FF approach, the AR approach does not depend on the perturbative parameter $\lambda$ and higher-order perturbative corrections do not appear in the expression. Also, all series are naturally terminated.

To derive the IS expression in the AR approach, we consider the total Hamiltonian of the FF approach. i.e. in the AR approach,  the wave function and energy obtained for the total Hamiltonian $H_{\text{eff}}=H + \lambda O$ are expanded as
\begin{equation}
|\Psi_v \rangle = |\Psi_v^{(0)} \rangle + \lambda  |\Psi_v^{(1)}\rangle + {\cal O}(\lambda)^{2}
\end{equation}
and
\begin{equation}
E_v = E_v^{(0)} + \lambda E_v^{(1)} + {\cal O}(\lambda)^{2} ,
\end{equation}
where superscripts denote orders of perturbation. Different orders of a wave function can be obtained by expanding the cluster operators as~\cite{Sahoo_2020,SahooLi}
\begin{equation}
T= T^{(0)} + \lambda T^{(1)}+ {\cal O}(\lambda)^{2} 
\end{equation}
and
\begin{equation}
S_{v}= S_{v}^{(0)} + \lambda S_v^{(1)}+ {\cal O}(\lambda)^{2} .
\end{equation}
The resulting expression for the first-order energy is
\begin{equation}
E_v^{(1)} = \langle \Phi_v | \bar{H} S_v^{^{(1)}}  + (\bar{H} T^{(1)} + \bar{O}) \{ 1 + S_v^{^{(0)}} \} | \Phi_v \rangle .  \label{eqeng1}
\end{equation}

It is evident from these equations that the above procedure of evaluating the IS factors does not depend on the choice of $\lambda$, and that the lowest-order contributions are the same as the values obtained in the EVE approach. The amplitudes of the $T^{(1)}$ and $S_v^{(1)}$ operators are obtained by solving the following additional set of equations
\begin{equation}
\langle \Phi_0^* | \overline{H}_0T^{(1)} + \overline{O}  | \Phi_0 \rangle = 0 \label{eqt1}
\end{equation}
and
\begin{eqnarray}
\langle \Phi_v^* | \{ (\overline{H}_0 &-& E_v^{(0)}) S_v^{(1)}\} +  \nonumber \\
&+& \left(\overline{H}_0 T^{(1)} + \overline{O} \right) \{1+ S_v^{(0)}\}| \Phi_v \rangle = 0. \label{eqamp1}
\end{eqnarray}
The AR approach contains few terms if the RCCSD approximation is considered. Therefore, the contributions from the higher-level excitations are necessary to achieve more accurate results. Moreover, it requires to compute amplitudes of cluster operators twice (for the unperturbed and perturbed excitation operators). 
However, this would be less expensive compared to the FF approach where it is necessary to compute the total energies considering different values of $\lambda$. 

It is worth mentioning here that the DHF results from EVE and AR methods are same, while they are different in the FF approach. It is due to the orbital relaxation effects that are included at the DHF level in the FF approach, but they are missing in the DHF method of other two approaches. However, these effects are compensated through the core-polarization effects, equivalent to the random phase approximation, in the EVE and AR approaches (for details see Refs.~\cite{SahooCa,SahooCd}).

\section{Selected Results and Discussion}\label{sec:res}

\subsection{Applying RMBPT to ISs in highly-charged ions}

\label{subsec:HCI_d<r^2>}Examples of ISs are much rarer in the literature for HCIs than for neutral atoms or low-charge ions. Soria Orts \textit{et al}.~\cite{orts-06-is} studied the IS in Be- and B-like ions of $^{36,40}$Ar using magnetic-dipole transitions in the visible range.
The IS in this case is dominated by the MS, which is larger than the FS by two orders of magnitude, and a detailed comparison with theory confirmed the importance of including the relativistic corrections to nuclear recoil, even for an intermediate nuclear charge $Z=18$.
The experimental precision was however not sufficient to probe the much smaller FS. 
Recently, the IS for B-like Ar was remeasured in Ref.~\cite{King:2022:43} with many orders of magnitude higher accuracy. It was observed that the QED contribution to the mass shift is comparable to the field shift, and an excellent agreement between the experimental results and the theoretical predictions was found~\cite{King:2022:43}.
Another approach is the dielectric recombination, which has been successfully used to probe the IS and the FS in a variety of few-electron HCIs~\cite{schuch-05-is,brandau-08-is}, and in Ref.~\cite{brandau-08-is} an improved value of $\delta\langle r^{2}\rangle$ for the isotopes $^{142,150}$Nd was inferred from the measurements. Elliott \textit{et al}.~\cite{elliott-98-is} have used precision x-ray spectroscopy of 3- to 8-electron ions of $^{235,238}$U to infer an improved value of $\delta\langle r^{2}\rangle^{235,238}$.

In this section we will consider in some detail a recent measurement of $\delta\langle r^{2}\rangle$ for $^{124,136}$Xe using precision EUV spectroscopy of highly charged Na-like ions~\cite{silwal-18-is}. This example will allow us to illustrate how the atomic theory in HCIs is simplified compared to that for neutral atoms or low-charge ions, and to discuss the issues involved in the atomic theory and the extraction of $\delta\langle r^{2}\rangle$ from the measurements.

First, we note that the MS in a HCI with nuclear charge $Z$ scales approximately as $p^{2}/M\sim Z$, while the FS of an $s$- or $p_{1/2}$-electron scales as $|\psi(0)|^{2}V_{\text{nuc}}(0)\sim Z^{4}$, so that as $Z$ increases one expects the FS to become increasingly dominant. The details depend on the transition and the isotopes involved, but as examples, in Ref.~\cite{orts-06-is} (B- and Be-like ions with $Z=18$), the MS was two orders of magnitude larger than the FS, in Ref.~\cite{silwal-18-is} (Na-like ions with $Z=54$), the MS was about half the FS, and in Ref.~\cite{elliott-98-is} (3--8-electron ions with $Z=92$), the MS was estimated to be at most about 3\% of the FS. An immediate consequence of this is that the error incurred in subtracting the MS from the IS to isolate the FS (and hence infer $\delta\langle r^{2}\rangle$) is naturally reduced as $Z$ increases. Often, it is possible to eliminate the MS using a precise calculation of nuclear recoil effects.

The FS and MS are comparable in the IS of the Na-like $3s$--$3p_{1/2}$ $D1$ line of $^{124,136}$Xe~\cite{silwal-18-is}. In this reference, order-by-order RMBPT was employed to calculate the MS (see Sec.~\ref{subsec:mbpt}), including the leading relativistic corrections to nuclear recoil of order $(Z\alpha)^{2}$ relative to the nonrelativistic MS. In this way, one obtains the order-by-order RMBPT expansion of the MS (in units of cm$^{-1}$): $E_{\text{MS}}^{(0)}=4.452$, $E_{\text{MS}}^{(1)}\equiv0$, $E_{\text{MS}}^{(2)}=-0.088$, $E_{\text{MS}}^{(3)}=-0.001$, giving a total MS of $E_{\text{MS}}^{\text{tot}}=4.363$~cm$^{-1}$. The third-order MS term is $<0.1$\% of the total MS, implying that the treatment of correlation is very complete. Another calculation using the RCI module of \textsc{grasp}2\textsc{k}~\cite{silwal-18-is} found $E_{\text{MS}}^{\text{tot}}=4.37$~cm$^{-1}$, and an earlier RCI calculation by Tupitsyn \textit{et al}.~\cite{2003-RelIS} found
$E_{\text{MS}}^{\text{tot}}=4.40$~cm$^{-1}$, both in good agreement with the RMBPT result. The leading uncertainty in this MS value is the omitted higher-order relativistic corrections, of which the leading one is of order $(Z\alpha)^{5}$~\cite{erickson-65-is}. In Ref.~\cite{silwal-18-is}, an error of 5\% was assigned to the MS to account for these omitted terms. Such higher-order relativistic corrections have been evaluated explicitly in some cases (see, e.g., Refs.~\cite{orts-06-is} and
\cite{zubova-14-is}, and references therein).

The FS in Ref.~\cite{silwal-18-is} was handled in an analogous way.
RMBPT calculations of the transition energy were performed for two
different nuclear charge distributions (for example, Fermi distributions with different skin thicknesses)
with $\langle r^{2}\rangle$ and $\delta\langle r^{2}\rangle$ close
to the physical values. The difference between the two calculations
was then found in each order of RMBPT and the numerical significance was carefully tested. 
For the Na-like $D1$ transition in Ref.~\cite{silwal-18-is}
this yielded (in units of cm$^{-1}$fm$^{-2}$): $F^{(0)}=-32.12$,
$F^{(1)}\equiv0$, $F^{(2)}=-0.03$, and $F^{(3)}=5\times10^{-4}$,
giving a total FS factor $F=-32.16$~cm$^{-1}$fm$^{-2}$. Once
again, the treatment of correlation can be seen to be very complete.
Using \textsc{grasp}2\textsc{k}, it was determined that QED effects
(self-energy and vacuum polarization) modify $F$ at about the 0.1\%
level in this case~\cite{silwal-18-is}.

The leading uncertainty in $F$ here is due to the nuclear-model dependence.
This arises because the FS is not strictly proportional to $\delta\langle r^{2}\rangle^{A,A'}$, 
but to the Seltzer moment $\lambda^{A,A'}$~(\ref{eq:Seltzer}),
which depends on the higher-order nuclear moments $\delta\langle r^{4}\rangle^{A,A'}$
and $\delta\langle r^{6}\rangle^{A,A'}$. The fractional contribution of these higher-order moments grows as $Z$ increases and can be quite important for heavy elements. For example, for the $D1$ transition in Na-like Xe, $\lambda^{124,136}$ is about 4\% smaller than $\delta\langle r^{2}\rangle^{124,136}$~\cite{silwal-18-is}.
For the method used above to calculate $F$, the leading-order effect of the higher-order moments is automatically subsumed into the predicted value of $F$. The problem is that the ratio between the Seltzer moment of Eq.~(\ref{eq:Seltzer}) to the RMS charge radius varies according to the model assumed for the change in nuclear charge distribution $\delta\rho_{\text{nuc}}(r)$, which may be concentrated at the nuclear surface, in the nuclear volume, or some combination of the two. The physical form for $\delta\rho_{\text{nuc}}(r)$ is highly sensitive to whether the nuclei are deformed. To investigate this effect, in Ref.~\cite{silwal-18-is} the entire calculation procedure for $F$ was repeated for many pairs of nuclear charge distribution
and the fluctuation in values of $F$ obtained was estimated at about
2\%. This represents a fundamental limit of precision with which $\delta\langle r^{2}\rangle^{A,A'}$
can be extracted in a model-independent way from the spectroscopic
data. Note that the 2\% fluctuation observed here is on the order
of the effect of the higher-order terms ($-4$\%). In Ref.~\cite{silwal-18-is},
the final value obtained was $\delta\langle r^{2}\rangle^{124,136}=0.269(42)$~fm$^{2}$,
including an experimental uncertainty that was about five times the combined theoretical uncertainties discussed above.

\subsection{IS applications for single-valence systems using FS-RCC method.}

Here we discuss applications of the FS-RCC method to determine the IS factors of the atomic states which can be considered as composed of a closed-shell core and a valence electron. Obviously, this method can be employed to all the alkali-like systems. In addition, other atomic systems with a closed-core configuration (total angular momentum $J=0$) and a valence electron in the ground state, such as Cd$^+$ and Tl,  can also be studied (see Fig.~\ref{Fig:overview}). 

\begin{table*}[h]
\small
\caption{Comparison of the calculated electron affinities (in cm$^{-1}$) of the first three low-lying states of the Li-like systems from different methods with the experimental values. Corrections for the Breit and QED interactions are given as $+$Breit and $+$QED, respectively.}
\begin{tabular}{l  rrr rrr rrr}
\hline \\
Method &  \multicolumn{3}{c}{Li atom} & \multicolumn{3}{c}{ Be$^+$ ion} & \multicolumn{3}{c}{Ar$^{15+}$ ion} \\
\hline \\
       & $2S_{1/2}$ & $2P_{1/2}$ & $2P_{3/2}$ & $2S_{1/2}$ & $2P_{1/2}$ & $2P_{3/2}$ & $2S_{1/2}$ & $2P_{1/2}$ & $2P_{3/2}$\\
\hline \\
DHF      &  43087.33 & 28232.86 & 28232.30 & 146210.22 & 114005.30 & 113996.38 & 7408273 & 7150039 & 7123169 \\
RMBPT(2) &  43444.25 & 28530.50 & 28529.85 & 146836.83 & 114856.95 & 114847.32 & 7409478 & 7152426 & 7125492  \\
RCCSD    &  43483.22 & 28577.94 & 28577.27 & 146884.35 & 114943.68 & 114933.96 & 7409503 & 7152481 & 7125545  \\
RCCSD(T) &  43479.68 & 28575.42 & 28574.72 & 146880.76  & 114939.21  &  114929.28 & 7409502 & 7152481 & 7125545 \\
RCCSDTv  &  43488.68 & 28582.18 & 28581.44 & 146888.97 & 114948.86 & 114938.87 & 7409503 & 7152482 & 7125546 \\
$+$Breit &  $-0.79$  &  $-0.40$ & $-0.15$  &  $-4.21$ & $-4.72$  &  $-1.97$ & $-1329$ &  $-2566$ & $-1155$\\
$+$QED   &  $-0.32$  &   0.07 &    0.06    &  $-1.50$  &  0.32  & 0.29 & $-742.73$ & 0.81 & 16.13 \\
%SMS & $-0.22$ & $0.77$ & $0.77$ & $-0.43$ & $3.39$ & $3.39$\\
\\
Final    &  43487.57 & 28581.85 & 28581.35 & 146883.26 & 114944.46 & 114937.19 & 7407432 & 7149917 & 7124408 \\

Experiment &  43487.11 & 28583.45 & 28583.11 & 146882.86 & 114954.12 & 114947.54 & 7407190 & 7150164 & 7124587 \\
\hline
\label{LiEA}
\end{tabular}
\end{table*}

\begin{table*}[h]
%\small
\caption{IS factors of the Li-like systems calculated at different levels of approximations using the FS-RCC method for single-valence problems and the expectation value evaluation (EVE) approach.}
\begin{tabular}{l  rrr rrr rrr}
\hline \\
Method &  \multicolumn{3}{c}{Li atom} & \multicolumn{3}{c}{ Be$^+$ ion} & \multicolumn{3}{c}{Ar$^{15+}$ ion} \\
\hline \\
       & $2S_{1/2}$ & $2P_{1/2}$ & $2P_{3/2}$ & $2S_{1/2}$ & $2P_{1/2}$ & $2P_{3/2}$ & $2S_{1/2}$ & $2P_{1/2}$ & $2P_{3/2}$\\
\hline \\
\multicolumn{10}{c}{\underline{$F$ values in MHz/fm$^2$}} \\
DHF      & $-2.43$  & $\sim 0.0$ & $\sim 0.0$ & $-15.71$  &  $\sim 0.0$  & $\sim 0.0$  & $-19203$  & $-52.65$ & $-0.001$ \\
RCCSD    & $-2.04$  & 0.41 & 0.41 & $-14.03$  & 2.98 & 2.98 & $-18821$ & 960.28 & 988.75 \\
%RCCSD(T) &  $-2.04$  & 0.41 & 0.41 & $-14.03$ & 2.97 & 2.96 & $-18821$ & 960.14 & 988.50 \\
RCCSDTv  & $-2.03$  & 0.42 & 0.42 & $-13.99$  & 2.99 & 2.98 & $-18820$ & 960.59 & 989.03 \\
&  &  &  & \\
\multicolumn{10}{c}{\underline{$K^{\rm{NMS}}$ values in GHz amu}} \\
DHF      & 747.09 & 508.60 & 508.61 & 2506.87 & 2080.43 & 2080.39  & 122796 & 120795 & 120368 \\
RCCSD    & 713.78 & 472.03 & 469.76 & 2413.85 & 1898.66 & 1890.05  & 121656 & 117426 & 117039 \\
%RCCSD(T) & 711.33 & 471.01 & 467.84 &  2410.92 & 1898.93 & 1886.14 & 161817 & 117403 & 117038 \\
RCCSDTv  & 711.89  & 471.29 & 469.05 & 2408.64 & 1897.18 & 1886.14 & 121655 & 117425 & 117038 \\
&  &  &  & \\
\multicolumn{10}{c}{\underline{$K^{\rm{SMS}}$ values in GHz amu}} \\
DHF      &  0.0   & $-150.22$ & $-150.22$ & 0.0 & $-950.73$  & $-950.67$ & 0.0        & $-73703$ & $-73431$ \\
RCCSD    & 42.79 & $-158.70$ & $-158.86$  &  112.03 & $-936.77$ & $-937.33$ & 1177 & $-72236$ & $-72009$ \\
%RCCSD(T) & 42.73 & $-160.15$  & $-160.15$ & 112.33 & $-940.37$ & $-940.80$ & 2065.87  & $-72242$ & $-72014$ \\
RCCSDTv  & 43.40  & $-158.86$ & $-158.68$ & 112.97 & $-936.25$ & $-936.74$ & 1178  & $-72242$ & $-72013$ \\
\hline
\label{LiEVE}
\end{tabular}
\end{table*}

\begin{table*}[h]
%\small
\caption{IS factors of the Li-like systems calculated using the finite-field (FF) approach.
Note that in FF approach the DHF results include orbital relaxation effects.
}
\begin{tabular}{l  rrr rrr rrr}
\hline \\
Method &  \multicolumn{3}{c}{Li atom} & \multicolumn{3}{c}{ Be$^+$ ion} & \multicolumn{3}{c}{Ar$^{15+}$ ion} \\
\hline \\
       & $2S_{1/2}$ & $2P_{1/2}$ & $2P_{3/2}$ & $2S_{1/2}$ & $2P_{1/2}$ & $2P_{3/2}$ & $2S_{1/2}$ & $2P_{1/2}$ & $2P_{3/2}$\\
\hline \\
\multicolumn{10}{c}{\underline{$F$ values in 
MHz/fm$^{2}$}} \\
DHF      & $-5.88$ & $0.80$ & $0.79$ & $-15.56$ & $2.69$ & $2.69$ & $-18819$ & $936$ & $967$ \\
RMBPT(2) & $-5.74$ & $1.09$ & 1.08 & $-15.36$ & $3.21$ & $3.21$ & $-18811$ & $958$ & $988$ \\
RCCSD    & $-5.71$ & $1.14$ & 1.14 & $-15.37$ & $3.26$ & $3.26$ & $-18811$ & $958$ & $989$\\
RCCSD(T) & $-5.72$ & $1.13$ & $1.13$ & $-15.38$ & 3.28 & 3.29 & $-18811$ & 958 & 988 \\
RCCSDTv  & $-5.71$ & $1.14$ & $1.14$ & $-15.37$ & 3.30  & 3.30 &  $-18811$ & 958 & 988 \\
 &    & \\
\multicolumn{10}{c}{\underline{$K^{\rm{NMS}}$ values in GHz amu}} \\
DHF      &  708.43 & 464.20 & 464.25 & 2404.41 & 1874.33 & 1874.44 & 121634 & 117356 & 117006 \\
RMBPT(2) &  714.25 & 469.06 & 469.11 & 2414.54 & 1888.09 & 1888.18 & 121649 & 117392 & 117040 \\
RCCSD    &  714.90 & 469.85 & 469.90 & 2415.36 & 1889.58 & 1889.67 & 121649 & 117393 & 117041 \\
RCCSD(T) &  714.84 & 469.81 & 469.86 & 2415.30 & 1889.51 & 1889.60 & 121649 & 117394 & 117041 \\
RCCSDTv  &  714.98 & 469.92 & 469.97 & 2415.43 & 1889.67 & 1889.75 & 121649 & 117394 & 117041 \\
 & & \\
\multicolumn{10}{c}{\underline{$K^{\rm{SMS}}$ values in GHz amu}} \\
DHF      & 0.0  & $-150.16$ & $-150.24$ & 0.0 & $-951.02$ & $-950.86$ & 0.0 & $-73703$  & $-73431$ \\
RMBPT(2) & 39.02 & $-149.66$ & $-149.74$ & 105.00 & $-913.44$ & $-913.35$ & 1169 & $-72177$ & $-71949$ \\
RCCSD    & 45.44 & $-150.52$ & $-150.61$ & 115.45 & $-913.07$ & $-912.98$ & 1193 & $-72155$ & $-71927$ \\
RCCSD(T) & 45.13 & $-152.48$ & $-152.56$ & 115.33 & $-918.48$ & $-918.44$ & 1194 & $-72169$ & $-71941$ \\
RCCSDTv  & 46.21 & $-152.37$ & $-152.44$ & 116.80 & $-917.69$ & $-917.60$ & 1195 & $-72168$ & $-71940$ \\
\hline
\end{tabular}
\label{LiFF}
\end{table*}

\begin{table*}[h]
%\small
\caption{
Dependencies of the FS factor values (in MHz/fm$^{-2}$) in the Li atom on the choices of the perturbative parameter $\lambda$ in the FF approach. It is noted that the DHF calculations do not converge for $\lambda\ge 5\cdot10^{-5}$. Calculations are carried out using the Fermi-charge distribution approximation.
Note that in FF approach the DHF results include orbital relaxation effects.
}
\begin{tabular}{l  rrr rrr rrr}
\hline \\
State &  \multicolumn{3}{c}{$2S_{1/2}$ } & \multicolumn{3}{c}{$2P_{1/2}$ } & \multicolumn{3}{c}{$2P_{3/2}$ } \\
\hline \\
& DHF & RMBPT(2) & RCCSD & DHF & RMBPT(2) & RCCSD &  DHF & RMBPT(2) & RCCSD \\
\hline \\
$\lambda=10^{-5}$    & $-5.88$ & $-5.74$ & $-5.71$ & $0.80$ & $1.09$ & $1.14$ & $0.79$ & 1.08  & 1.14\\
$\lambda=10^{-6}$   & $-2.08$ & $-2.03$ & $-2.02$ & 0.33 & 0.44 & 0.45 & 0.25 & 0.35 & 0.37 \\
$\lambda=10^{-7}$  & $-1.07$ & $-1.05$ & $-1.05$ & $-0.27$ & $-0.21$ & $-0.19$ & $-0.05$ & 0.01 & 0.02\\
\hline
\end{tabular}
\label{LiFF1}
\end{table*}

\begin{table*}[h]
%\small
\caption{Extraction of the FS factor values (in MHz/fm$^{-2}$) directly from the energy differences of two different isotopes via $F=-(E^A - E^{A'}) / (\langle r^2 \rangle^A - \langle r^2 \rangle^{A'})$,
 using the Fermi-charge distribution in the FF framework.
Note that in FF approach the DHF results include orbital relaxation effects.
}
\begin{tabular}{l  rrr rrr rrr}
\hline \\
Isotopes &  \multicolumn{3}{c}{$2S_{1/2}$ } & \multicolumn{3}{c}{$2P_{1/2}$ } & \multicolumn{3}{c}{$2P_{3/2}$ } \\
\hline \\
$A,A'$ & DHF & RMBPT(2) & RCCSD & DHF & RMBPT(2) & RCCSD &  DHF & RMBPT(2) & RCCSD \\
\hline \\
Li$^{6,7}$ &  $-2.09$  & $-34.55$ & $-5.38$ & $-25.35$ & $-25.24$ & $-90.83$ & $-41.15$ & $-41.17$  & $-99.72$ \vspace{1 mm} \\ 
Be$^{9,10}$ &  $-1.95$  & $-2.26$ & $-368.52$ & 17.77 & 17.41 &  310.64 & 37.06 & 37.76 &  309.51 \vspace{1 mm} \\
Ar$^{38,40}$ &  $-18815$  & $-18807$  &  $-18807$ & 932 & 953 & 954 & 976 & 997 & 998 \\
\hline
\end{tabular}
\label{LiFF2}
\end{table*}

\begin{table*}[h]
%\small
\caption{IS factors from the analytical response (AR) approach using the FS-RCC method for single-valence systems.}
\begin{tabular}{l  rrr rrr rrr}
\hline \\
Method &  \multicolumn{3}{c}{Li atom} & \multicolumn{3}{c}{ Be$^+$ ion} & \multicolumn{3}{c}{Ar$^{15+}$ ion} \\
\hline \\
       & $2S_{1/2}$ & $2P_{1/2}$ & $2P_{3/2}$ & $2S_{1/2}$ & $2P_{1/2}$ & $2P_{3/2}$ & $2S_{1/2}$ & $2P_{1/2}$ & $2P_{3/2}$\\
\hline \\
\multicolumn{10}{c}{\underline{$F$ values in MHz/fm$^2$}} \\
DHF      & $-2.43$  & $\sim 0.0$ & $\sim 0.0$ & $-15.71$ & $\sim 0.0$ & $\sim 0.0$ & $-19203$ & $-53$ & $-0.001$\\
RCCSD    & $-2.04$  &  0.42  &  0.42 & $-14.01$  &  3.02 & 3.02 & $-18820$ & 961 & 989 \\
RCCSD(T) & $-2.04$  & 0.42   &  0.42 & $-14.02$ & 3.01 & 3.01 & $-18820$ & 961 & 989 \\
RCCSDTv  & $-2.03$  &  0.42  &  0.42 & $-14.00$ & 3.01 & 3.01 & $-18820$ & 961 & 989 \\
$+$Breit & $0.01$   & $\sim 0.0$ & $\sim 0.0$ & $\sim 0.0$ & $\sim 0.0$ & $\sim 0.0$ & 21.54 & $-2.17$ & $-3.06$ \\
$+$QED   & $0.01$ & $\sim 0.0$ & $\sim 0.0$  &  0.03 & $-0.01$ & $-0.01$ & 165.18 & $-8.22$ & $-8.30$ \\
 & & \\
Final  &  $-2.01$  & 0.42 & 0.42 &  $-13.97$  & 3.0  & 3.01  & $-18655$ & 953 & 981 \\
 & & \\
%\hline \\
\multicolumn{10}{c}{\underline{$K^{\rm{NMS}}$ values in GHz amu}} \\
DHF      & 747.09  & 508.60 & 508.61 & 2506.87 & 2080.43 & 2080.39 & 122796 & 120795 & 120368 \\
RCCSD    & 712.66  & 468.41 & 468.43 & 2412.50 & 1887.25 & 1887.33 & 121655 & 117392 & 117038 \\
RCCSD(T) & 713.05  & 468.41 & 468.43 & 2412.91 & 1887.07 & 1887.15 & 121655 & 117391 & 117038 \\
RCCSDTv  & 712.36  & 468.29 & 468.33 & 2412.09 & 1886.76 & 1886.88 & 121655 & 117390 & 117038 \\
$+$Breit & $-0.03$ & $-0.03$ & $-0.01$ & $-0.21$ & $-0.30$ & $-0.11$ & $-64.55$ & $-123.91$ & $-47.37$ \\
$+$QED   & $-0.02$ & $\sim 0.0$ & 0.01 & $-0.07$ & 0.01 & 0.01 & $-41.49$ & 0.39 & 1.20 \\
 & & \\
Final    &  712.31 &  468.26 &  468.33 & 2411.81 & 1886.47 & 1886.78 & 121549 & 117267 & 116992 \\
Semiempirical & 715.19 & 470.05 & 470.05 & 2415.67  & 1890.37  &  1890.25 & 121824 & 117588 & 117168 \\
& &  \\
\multicolumn{10}{c}{\underline{$K^{\rm{SMS}}$ values in GHz amu}} \\
DHF      & 0.0   & $-150.22$ & $-150.22$  & 0.0 & $-950.73$ & $-950.67$ & 0. & $-73703$ & $-73431$ \\
RCCSD    & 44.48 & $-150.71$ & $-150.71$  & 113.81 & $-912.94$ & $-912.93$ & 1190 & $-72155$ & $-71927$ \\
RCCSD(T) & 45.29 & $-151.90$ & $-151.91$ & 115.82 & $-916.61$ & $-916.61$ & 1196 & $-72166$ & $-71938$ \\
RCCSDTv  & 46.38 & $-151.60$ & $-151.55$ & 117.11 & $-915.75$ & $-915.60$ & 1196 & $-72166$ & $-71937$ \\
$+$Breit & 0.04  &  $\sim 0.0$ & $\sim 0.0$ & 0.25 & 0.02 & 0.05 & 68.52 & 26.75 & 35.21 \\
$+$QED   &  $\sim 0.0$ & $\sim 0.0$  & $-0.01$ & $\sim 0.0$ & $-0.03$ & $-0.03$ & $-6.06$ & $-6.20$ & $-6.57$ \\
 & & \\
Final    &  46.42 & $-151.60$ & $-151.55$ & 117.35 & $-915.73$ & $-915.58$ & 1259 & $-72145$ & $-71909$ \\
\hline
\end{tabular}
\label{LiAR}
\end{table*}
As mentioned earlier, it is possible to apply the FS-RCC method to calculate IS factors of the one-valence systems by adopting FF, EVE and AR approaches. 
Here, we consider, for example, applications of these three approaches to the Li atom and Li-like ions. These systems have three electrons in total, two of them form a closed-shell configuration, i.e. $1s^2$, and an electron on the valence spinor. Depending on the valence spinor of the configuration, we can have the ground and excited states. So these systems are suitable for applying the one-valence FS-RCC method.
Moreover, the RCCSDTv method is an exact many-body method for Li-like systems.
In an earlier work~\cite{SahooLi}, we had considered the neutral Li atom, and Li-like Be (Be$^+$) and Ar (Ar$^{15+}$) ions to show the differences in the results from the EVE, FF and AR approaches in a number of atomic states. In the meantime, we have improved the code and here we use a larger basis to analyze the IS factors of the ground $2s~^2S_{1/2}$ state and the first two excited states ($2p~^2P_{1/2}$ and $2p~^2P_{3/2}$) of these systems.
We demonstrate results from the EVE, FF and AR approaches using the FS-RCC theory. By taking differences in the results between the ground and the first excited state, one can study the IS in the $D1$ transition. Similarly using the differences of the IS factors between the ground and $2p~^2P_{3/2}$ states, IS of the $D2$ line can be investigated. We initially chose $\lambda=10^{-5}$ in the FF analysis to evaluate the IS factors, later the dependency of the the FS factors on $\lambda$ is demonstrated. The values obtained using the EVE and AR approaches are free from any external parameter. The DHF calculations are performed using the same basis functions in all the three approaches. 

\begin{table*}[h]
%\small
\caption{The FS factors (in MHz/fm$^2$) of the Li-like systems using the Fermi and uniform charge distributions from the AR-RCCSD method with DC Hamiltonian. We have given FS factors using Eq.~(\ref{Fdef4}) as Fermi-1 and Eq.~(\ref{Fdef5}) as Fermi-2 in the Fermi-charge distribution.}
\begin{tabular}{l  ccc ccc ccc}
\hline \\
Method &  \multicolumn{3}{c}{Li atom} & \multicolumn{3}{c}{ Be$^+$ ion} & \multicolumn{3}{c}{Ar$^{13+}$ ion} \\
\hline \\
Type & $2S_{1/2}$ & $2P_{1/2}$ & $2P_{3/2}$ & $2S_{1/2}$ & $2P_{1/2}$ & $2P_{3/2}$ & $2S_{1/2}$ & $2P_{1/2}$ & $2P_{3/2}$\\
\hline \\
Fermi-1  &  $-2.035$ & 0.424 & 0.424 & $-14.012$ & 3.018 & 3.017 & $-18820$ & 961 & 989 \\
Fermi-2 & $-2.036$ & 0.424 & 0.424 & $-14.017$ & 3.019 & 3.018 & $-18906$ & 965 & 994 \\
Uniform & $-2.037$ & 0.425 & 0.424 & $-14.064$ & 3.030 & 3.028 & $-18827$  & 961  &  990  \\
\hline
\end{tabular}
\label{LiAR1}
\end{table*}

Before presenting the IS factors for the Li-isoelectronic systems, we summarize the electron affinities (EAs) calculated at different levels of approximations in Table~\ref{LiEA}. As can be seen from the table, the results improve gradually from the DHF values with the inclusion of the electron correlation effects. The corrections from the Breit and QED interactions are large in the ground state and increase with the size of the system. The final values are in close agreement with the experimental values~\cite{2007-LiE,Kramida_2005,2010-Saloman,2019-Paul}. In fact, the final theoretical EA values of Li and Be$^+$ agree with the experimental results within 0.1\% indicating the calculated wave functions are reliable enough to estimate the IS factors of the first three states. 

\begin{table*}[htb]
%\small
\caption{IS factors of the $3s ~ ^2S_{1/2}$, $3p ~ ^2P_{1/2}$ and $3p ~ ^2P_{3/2}$ states in Na and Mg$^+$ calculated using the DHF, AR-RCCSD and AR-RCCSDTv methods. Previously calculated values using other methods are also given.}
\begin{tabular}{l rrr rrr}
\hline \hline \\
 Method &  \multicolumn{3}{c}{Na atom} & \multicolumn{3}{c}{Mg$^+$ ion} \\
    & $3s ~ ^2S_{1/2}$  & $3p ~ ^2P_{1/2}$ & 	$3p ~ ^2P_{3/2}$  & $3s ~ ^2S_{1/2}$  & $3p ~ ^2P_{1/2}$ & 	$3p ~ ^2P_{3/2}$ \\
 \hline  \\
\multicolumn{7}{c}{\underline{$F$ values in MHz/fm$^2$}} \\
 DHF    &  $-29.717$ & $-0.008$ & $-0.000$ & $-104.568$  & $-0.059$  & $-0.000$ \\
 AR-RCCSD  & $-37.317$  &  $1.591$ & $1.603$ & $-116.388$  & $9.836$    & $9.860$ \\
 AR-RCCSDTv & $-37.762$  & 1.495  & 1.480 & $-116.686$  & $9.724$  &  9.746  \\
 $+\Delta$Breit & 0.022 & 0.003  & 0.001 & $0.096$  & $-0.006$ & $-0.009$ \\
 $+\Delta$QED  & 0.261 & $-0.010$  & $-0.010$ & $0.778$  & $-0.061$ &  $-0.062$ \\
 & & \\
 Final & $-37.48$ & 1.49 & 1.47  & $-115.81$   & $9.66$  & $9.68$ \\
  Ref. \cite{2001-Saf} & $-36.825$ & $1.597$  & $1.603$ & $-116.01$   & $9.800$  & $9.811$  \\
  Ref. \cite{Sahoo_2010} & $-37.035$ & $1.725$ & $1.765$ & $-116.102$ & $10.119$   & $10.222$ \\
% \hline  \\
& & \\
\multicolumn{7}{c}{\underline{$K^{\rm{NMS}}$ values in GHz amu}} \\
  DHF & 955.68 &  507.54 & 507.28 & 2649.86 & 1854.17 & 1851.87  \\
 AR-RCCSD & 665.44 & 396.56 & 396.75 & 1975.57 & 1396.42 & 1396.37 \\
 AR-RCCSDTv & 660.93 & 393.44 & 393.26 & 1960.13 & 1385.49 & 1386.29 \\
 $+\Delta$Breit & $-0.23$ & $\sim 0.0$ & $-0.01$ & $-0.50$ & $-0.56$ & $-0.19$ \\
 $+\Delta$QED & $-0.18$  & 0.01 & 0.01 & $-0.43$ & 0.04 & $\sim 0.0$\\
 & & \\
 Final &  660.52 & 393.45 & 393.26 & 1952.20 & 1384.97 & 1386.10 \\
 Semi-empirical & 681.68 & 402.82 & 402.53 & 1994.37 & 1407.75 & 1406.25 \\
 %\hline \\
 & & \\
\multicolumn{7}{c}{\underline{$K^{\rm{SMS}}$ values in GHz amu}} \\
  DHF & $-221.967$ &  $-115.735$ & $-115.537$ & $-563.278$  & $-600.518$ &  $-598.181$ \\
 AR-RCCSD & $101.658$  &  $-29.765$ &  $-29.810$ &  125.451    & $-278.034$ & $-277.721$ \\
 AR-RCCSDTv & 72.121   &  $-37.074$ &  $-36.960$ & 69.965   & $-303.543$ & $-303.232$ \\
 $+\Delta$Breit & 0.235 &  0.141 & 0.028 & 1.051 & 0.471 &  0.026 \\
 $+\Delta$QED & 0.027   & 0.029 & $-0.017$ & 0.137  & $-0.128$ ~ &  $-0.119$\\
  & & \\
 Final & $72.38$ &  $-36.90$ &  $-36.95$ & 71.15  & $-303.20$ &  $-303.33$ \\
  Ref. \cite{2001-Saf} & 53.94 & $-43.36$  & $-43.39$  & 38.0 & $-324.0$  & $-323.0$ \\
    Ref. \cite{berengut2003} & 69 & $-40$  & $-39$ & 83.0 & $-296.0$  & $-290.0$ \\
  Ref. \cite{Sahoo_2010} & 73.2 & $-41.2$ & $-39.1$ & $78.9$ & $-319.9$ & $-311.0$ \\
   Ref. \cite{2015-Roy}  &  &   &  & $-206.5$  & $-571.6$  & $-572.1$  \\
\hline \hline
\end{tabular}
\label{namg+}
\end{table*}

\begin{table*}[htb]
%\small
\caption{Differential IS factors of the D1 and D2 lines of Zn$^+$, Cd$^+$ and In calculated using the DHF, AR-RCCSD and AR-RCCSDTv methods.}
\begin{tabular}{l rr rr rr}
\hline \hline \\
 Method &  \multicolumn{2}{c}{Zn$^+$ ion} & \multicolumn{2}{c}{Cd$^+$ ion} & \multicolumn{2}{c}{In atom} \\
    & D1 & D2 & D1 & D2 & D1 & D2 \\
 \hline  \\
\multicolumn{7}{c}{\underline{$F$ values in MHz/fm$^2$}} \\
 DHF    &  $-1216.22$ & $-1221.23$ & $-4718.63$ & $-4777.65$ & 350.85 & 413.78 \\
 AR-RCCSD  & $-1547.13$ & $-1552.47$ & $-5962.37$ & $-6039.34$ & 1851.59 & 1816.49 \\
 AR-RCCSDTv & $-1520.39$ & $-1525.26$ & $-5994.44$ & $-6072.90$  & 1924.41 & 1901.85 \\
 $+\Delta$Breit & 3.43 & 3.42 & 20.80 & 21.02 & $-0.19$ & $-2.39$ \\
 $+\Delta$QED  & 18.48 & 18.52 & 143.02 & 144.27 &  $-11.35$ & $-11.01$ \\
 & & \\
 Final & $-1498.48$ & $-1503.32$ & $-5830.62$ & $-5907.61$ & 1912.87 & 1888.45 \\
 %\hline  \\
  & & \\
 \multicolumn{7}{c}{\underline{$K^{\rm{NMS}}$ values in GHz amu}} \\
  DHF &   2031.27 &  2082.64 & 2577.58 & 2788.12 & 2550.76 & 2286.40 \\
 AR-RCCSD & 758.68 &  775.41 & 732.38 & 773.93 &  465.59 & 439.56 \\
 AR-RCCSDTv & 737.64 & 754.76 & 733.90 & 777.80 &  403.01 & 386.32 \\
 $+\Delta$Breit & 0.57  & $-0.43$ & 0.32 & $-1.60$ & 2.39 & $-0.88$ \\
 $+\Delta$QED & $-1.69$  & $-1.36$ & $-6.50$ & $-6.45$ &  $1.58$ & $-0.16$\\
 %\hline \\
  & & \\
 Final &  736.52 & 752.97 & 727.72 & 769.75 7 & 406.98 & 385.28  \\
 Semi-empirical & 797 & 812 & 726 & 767 & 401 & 365 \\
 %\hline \\
 & & \\
\multicolumn{7}{c}{\underline{$K^{\rm{SMS}}$ values in GHz amu}} \\
  DHF &  $-400.05$  & $-455.50$  & $-1022.31$ & $-1286.55$ & $-1861.18$ & $-1595.75$ \\
 AR-RCCSD & 1296.89 & 1275.20 & 1165.07  & 1032.27 &  $-700.02$  & $-593.43$  \\
 AR-RCCSDTv & 1300.47 & 1279.83 & 1214.39 & 1083.44 & $-561.55$ & $-482.57$ \\
 $+\Delta$Breit & 2.87 & 2.76 & 6.72 & 7.62 & $-4.19$ & 4.26 \\
 $+\Delta$QED & 0.82 & 0.64 & 5.67 & 5.34 & 0.74 & 1.56 \\
 %\hline \\
 & & \\
 Final &  1304.16 & 1283.23  & 1226.78 & 1096.40 & 565.00 & $-476.75$ \\
\hline \hline
\end{tabular}
\label{ZnCdIn}
\end{table*}

Since the employed RCCSDTv method is exact for the Li-like systems, the differences between the IS values that came from the FF, EVE and AR approaches can be attributed to numerical instabilities in the respective method.
In Table~\ref{LiEVE}, we present the IS factors calculated within the DHF method and different variants of the RCC method using the EVE approach. The FS factors were determined using the Fermi-charge distribution model (eq.~\ref{eq:fermi}). As can be seen, the correlation effects are non-negligible in these few-electron systems. Since Li-like Ar, Ar$^{15+}$, is a HCI, the electron correlation effects can be expected to be small. However, the table shows that roles of electron correlation effects are quite significant in the determination of IS factors in this HCI. Nevertheless, the differences in the results from the RCCSD and RCCSDTv methods are insignificant. 

We now present IS factors of the Li-like systems calculated using the FF approach (see Table~\ref{LiFF}) and compare them with the EVE results from Table~\ref{LiEVE}. The values from the second-order RMBPT (RMBPT(2)) method from the FF approach are also given. It can be seen that the DHF values of the FF approach differ from those obtained within the EVE approach. 
%\LS{Above it is written ``The DHF calculations are performed using the same basis functions in all the three approaches''. So, if I correctly understood in the FF method the $\lamba F(r)$ operator was added after DHF stage and before CC stage. I do not fully understand the statement below. It sounds like one calls the RPA result the DHF one? If it is not a ``clean'' DHF results, maybe one can use some other designation (RPA)?}
%%%% Ben: I added a setnece to the captions of the relavant tables to emphasize.
These differences are attributed to the orbital relaxation effects, which are absent in the DHF method of EVE approach but are included through the RPA contributions in the RCC methods within the latter approach. The correlation effect trends for the FF approach are found to be similar to the EVE approach, but the final results differ noticeably. In fact, the FS factor of all the three states of Li differ by more than 200\% between both approaches whereas results for the NMS and SMS factors from these approaches almost agree with each other. The large differences in the FS factors are mainly coming from the DHF values, so we analysed the FS results obtained by the FF technique further by varying the $\lambda$ values. The $\lambda$-dependent values are given in Table~\ref{LiFF1}. As can be seen from this table, the FS factors of Li vary significantly with $\lambda$ and the EVE and FF values agree nost with the results for $\lambda=0.10^{-6}$ a.~u. It appears from this analysis that for large values of $\lambda$, the numerical differentiation gives large errors while for very small $\lambda$ there are significant round-off errors. It means that optimized values of $\lambda$ to be used in the FF approach could depend on the state and system under consideration for the study. 

To support this explanation further, we use the approximate Eq.~(\ref{eq:rad}) to estimate the FS factors by calculating energies of two stable isotopes of each considered Li-like systems. 
The results are given in Table~\ref{LiFF2}. 
The energy differences between the $^6$Li and $^7$Li isotopes  results in unreliable values of the FS factors. It stems from the tiny energy differences between these two isotopes as the FS is tiny. A similar trend has been observed for the pair of the singly charged $^{9}$Be and $^{10}$Be isotopes. However, the inferred FS factors from the energy differences between the Li-like $^{38}$Ar and $^{40}$Ar isotopes are found to be quite reliable. 
Note that we do not claim that the IS factors obtained using the FF approach could be wrong, but controlling the numerical instability could be a tedious procedure. The AR approach allows an easier determination of the IS factors and without dependencies on any external parameter, provided that higher level excitations are included for attaining accurate results.

In Table \ref{LiAR}, the IS factors of the Li-like systems calculated using the AR approach are given. As it was mentioned earlier, the DHF values of the AR approach are the same as returned by the EVE approach. However, the correlation trends arising through the RCC methods are found to follow the pattern of the FF approach rather than the EVE approach. The reason for it is as follows. Though the RCCSDTv method is exact for the Li-like systems, the EVE approach contains two terminating series. Moreover, program implementation of the atomic FS-RCC methods allows violation of Pauli's exclusion principle for easy dealing with the angular momentum couplings. This would introduce unphysical contributions to the calculated values if all the direct and exchange integrals are not included as in the case of the EVE approach. From this point of view, both the FF and AR approaches are better than the EVE approach for the evaluation of atomic properties. In the above table, we also give the final values from our RCC method by adding corrections from the Breit and QED interactions. 

We also analyze FS factors from the AR approach using the Fermi and uniform charge distributions in the Li-like systems in Table~\ref{LiAR1}. Further, we use Eqs.~(\ref{Fdef4}) and~(\ref{Fdef5}) for the Fermi-charge distribution and the results are denoted as `Fermi-1' and `Fermi-2', respectively, in the table. The nuclear-model dependency is found to be negligible for Li, but not so for Ar$^{13+}$.

After comparing various approaches and expressions for accurate determination of the IS factors in the Li-like systems, we discuss results for the heavier atomic systems using the AR-RCC methods and compare them with the previously reported values using other many-body methods. For better understanding of the trends of correlation effects from light to heavy atomic systems. We discuss results for medium size atomic systems Na and Mg$^+$. Then, present results in the heavier systems: Zn$^+$, Cd$^+$, and In. It is worth mentioning here that trends in the electron correlation effects are found to be unique to each atomic system, thus emphasizing that calculations must be performed for each system.

In Table \ref{namg+}, we present the IS factors calculated at different levels of approximation~\cite{2022-Na}. Our results were compared with the earlier calculations from Refs.~\cite{2001-Saf,berengut2003,Sahoo_2010,2015-Roy}. Detailed discussions on the trends of results from the RCC methods and that reported earlier can be found in Ref.~\cite{2022-Na}. However, one can notice that the differences between the RCCSD and RCCSDTv methods in Na and Mg$^+$ are very strong compared to the Li-like systems that were presented above. 
The results from the AR-RCC method agree reasonably well with the previously reported values except for the data of Ref. \cite{2015-Roy}. Also, the calculated NMS factors differ significantly from the semi-empirical values.

We now turn to the analysis of IS factors for another class of atomic systems, Zn$^+$~\cite{2023-Zn}, Cd$^+$~\cite{2022-Cd}, and In~\cite{Sahoo_2020, kar24}, with ground state configurations $(n-1)d^{10}ns$ and $ns^2np_{1/2}$. The trends for these atomic systems are expected to be different from those for the iso-electronic systems of alkali atoms. In Table~\ref{ZnCdIn}, the differential values of the IS factors of the D1 and D2 lines of Zn$^+$, Cd$^+$ and In, calculated at different levels of approximations, are given. The contributions of electron correlation effects to the calculated IS factors in these atomic systems significantly differ from the Li-like systems, Na, and Mg$^+$ as discussed above. In Na and Mg$^+$, the differences between the SMS factors from the RCCSD and RCCSDTv methods were quite large. Though Zn$^+$, Cd$^+$ and In are much heavier than Na and Mg$^+$, contributions from the triple excitations in the latter systems are found to be stronger. A detailed analysis, however, reveals that the contributions from individual diagrams arising from triple excitation in Zn$^+$, Cd$^+$, and In are very strong but they all cancel each other resulting in smaller final values of contributions. Contributions from the Breit and QED effects are also found to be small compared to the total values of IS factors. Furthermore, electron correlation effects in Cd$^+$ are larger than in Zn$^+$. It can also be noticed that the differences in the IS factors between that obtained within the DHF and AR-RCCSD methods for In are quite significant. We also observe large differences between the NMS factors from the semi-empirical estimates and that obtained using the AR-RCCSD/T methods. This suggests that it may not be appropriate to combine semi-empirical values of the NMS factors and {\it ab initio} results for the SMS factors to estimate the total MS factors in the IS studies.

\subsection{IS applications of Single-reference RCC}\label{SRapplication}

\begin{table*}[htb]
\caption{Values of the IS atomic factors for Al calculated within the single-reference RCC method~\cite{skripnikov2024isotopeQED}.
%We reversed the sign of $F$ relative to Ref.~\cite{skripnikov2024isotopeQED} due to the opposite sign definition (Eq.~\ref{Fdef1}) of the FS factor used in this study.
}
\begin{tabular}{lrr}
\hline
\hline
\multicolumn{3}{c}{$K^{\rm NMS}$ values in GHz u}                                                          \\
                 & {$3s^23p~^2P^o_{1/2} \to 3s^24s~^2S_{1/2}$} & {$3s^23p~^2P^o_{3/2} \to 3s^24s~^2S_{1/2}$}    \\
\hline                 
SR-RCCSDT              & $-415.4$ & $-414.4$  \\
SR-RCCSDT(Q) $-$ SR-RCCSDT & 0.0    & 0.0     \\
basis set corr.    & $-0.7$   & $-0.7$    \\
high virt.         & 0.1    & 0.1     \\
Breit              & 0.4    & 0.1     \\
QED                & 1.0    & 1.0     \\
Total              & $-414.7(0.5)$ & $-413.9(0.3)$ \\
                   &             &                \\
\multicolumn{3}{c}{$K^{\rm SMS}$ values in GHz u}                                                          \\
\hline
%                 & P1/2 – 4S            & P3/2 – 4S  & P1/2 – 5S    & P3/2 – 5S  \\
SR-RCCSDT              & 653.4 & 655.1 \\
SR-RCCSDT(Q) $-$ SR-RCCSDT & $-0.1$  & $-0.6$  \\
basis set corr.    & 0.8   & 0.8   \\
high virt.         & 0.5   & 0.5   \\
Breit              & $-0.5$  & 0.0   \\                
QED                & 0.2    & 0.2   \\
Total              & 654.3$(0.9)$ & 656.0$(1.0)$  \\
                   &                      &                  \\
\multicolumn{3}{c}{$K^{\rm NMS} + K^{\rm SMS}$, GHz u   }   \\
\hline
Total              & 239.6$(1.1)$   &  242.1$(1.0)$   \\
%MCDF(CV+VV)~\cite{Filippin:2016,Heylen:2021}  & 240.0$(5.0)$ $^a$ &  243.0$(4.0)$    &              &               \\
\hline                   
                   &                      &                   \\
\multicolumn{3}{c}{ $F$ values in MHz/fm$^2$}                                                          \\
\hline
SR-RCCSDT                  & $76.95$ & $76.85$  \\
SR-RCCSDT(Q) $-$ SR-RCCSDT & $-$0.02   & $-$0.01  \\
basis set corr.            & $0.04$  & $0.05$   \\
high virt.                 & $0.01$  & $0.01$   \\
Breit                      & $-$0.07 & 0.00   \\
QED-VP                     & $0.07$  & $0.06$   \\
QED-SE                     & $-$0.17 & $-$0.17  \\
Total                      & $76.81(12)$ & $76.80(6)$  \\
%Refs.~\cite{Filippin:2016,Heylen:2021}    &   76.45$(1.95)$$^a$        & 76.20$(2.20)$           \\
\hline
\hline
\end{tabular}
%\begin{flushleft}
%$^a$ Derived from the data given in Table III of Ref.~\cite{Filippin:2016} (columns RIS3/Sep. and RATIP/Sep.) following the approach described in Ref.~\cite{Heylen:2021}, see Refs.~\cite{Filippin:2016,Heylen:2021} for the description and abbreviations of these methods.
%\end{flushleft}
\label{AlISfactors}
\end{table*}

As an example of the application of the single-reference relativistic CC theory outlined in section~\ref{SRRCC}, let us consider a recent study of the IS factors of Al~\cite{skripnikov2024isotopeQED}. Four transitions were considered: $3s^23p~^2P^o_{1/2} \to 3s^24s~^2S_{1/2}$, $3s^23p~^2P^o_{3/2} \to 3s^24s~^2S_{1/2}$, $3s^23p~^2P^o_{1/2} \to 3s^25s~^2S_{1/2}$, and $3s^23p~^2P^o_{3/2} \to 3s^25s~^2S_{1/2}$. Each of the considered states formally has one valence electron. In the zero-order approximation, the electronic wave function can be written as a single Slater determinant $\ket{\Phi_0}$ with 13 occupied atomic orbitals. The correlated electronic wave function is then written in the form of Eq.~(\ref{expAnsatz}). The illustrative results of the calculations of the IS factors for two of these transitions are given in Table~\ref{AlISfactors}. 

The leading contributions to the IS factors were obtained within the full iterative SR-RCCSDT model, where all triple excitation were considered for all electrons. To account for higher-order excitations, the correction for quadruple excitation amplitudes was calculated as the difference between the results obtained within the SR-RCCSDT(Q)~ \cite{Bomble:05,Kallay:6} and SR-RCCSDT models. To have a better estimation of the uncertainty, the correction for the increase of the basis set was calculated at the SR-RCCSD model. In the SR-RCCSDT and SR-RCCSDT(Q) correlation calculations, all virtual orbitals with orbital energies below 500 Hartree were included. The correlation contribution of higher-lying orbitals was calculated at the SR-RCCSD(T) and SR-RCCSD levels for FS, NMS, and SMS factors, respectively. The contribution of the Breit electron-electron interaction was obtained at the SR-RCCSDT level.

Finally, the contributions of the QED effects to the FS factor were calculated using the approach outlined in section~\ref{QED-FS}. 
It was found that the dominant QED contribution arises from the first term in Eq.~(\ref{Fse}).
To calculate the QED contribution to the nuclear recoil effect, the approach described in section~\ref{QED-MS} was applied. 
It was shown that, in this case, the dominant nuclear recoil QED effect is described by the model  approach~\cite{Anisimova:2022}, while the QED contribution arising from the perturbation of the electronic wave function by the SE and VP interactions is negligible.
One can see that for FS and NMS factors, the QED contribution is larger than the overall uncertainty. Thus, at the achieved level of accuracy for the neutral Al atom, these contributions become important.

Calculations outlined in Table~\ref{AlISfactors} were performed within the FF approach. In particular, the FS constant was evaluated according to Eq.~(\ref{FScalc}). However, as it was shown in Ref.~\cite{skripnikov2024isotopeQED}, in the considered case, the uncertainty due to the numerical differentiation in Eq.~(\ref{FScalc}) is negligible. To demonstrate this, several test calculations were performed, where results obtained using the FF approach were compared with those obtained by taking the trace of the product of the density matrix calculated within the $\Lambda$-equations~\cite{Kallay:3,2007-Bartlett} coupled cluster approach, along with the matrix of the operator given by Eq.~(\ref{Fgauss}). This approach does not involve numerical differentiation as per Eq.~(\ref{FSmethodB}).
The calculated values of the IS factors in Al are in agreement with MCDF calculations~\cite{2016-AlI,2021-Al}, but have a reduced theoretical uncertainty.
A similar computational scheme, though without accounting for QED effects, was recently applied in the calculation of IS factors for neutral Tl~\cite{Penyazkov:2023} and Au~\cite{Cubiss:Au:23} atoms.

\subsection{IS applications of Multi-reference Fock-Space RCC}\label{MRapplication}

While the single-reference RCC approach demonstrated very high accuracy for the low-lying states of Al ($Z=13$), this approach is not suitable to describe electronic states of Si ($Z=14$), that are of experimental interest, since they possess multireference character. For this case, the multi-reference FS-RCC method, outlined in section~\ref{sec:fscc}, was applied in a recent calculation of the IS factors for the $3s^2 3p^2\;^1\mathrm{S}_0 \to 3s^2 3p 4s\;^1\mathrm{P}_1^o$ transition in the silicon atom~\cite{kon24}. The results are given in Table~\ref{SiResults}.As in the Al case described above, the FF approach in the formulation of Eq.~(\ref{eqff1}) was used to calculate the IS factors.
\begin{table}[]
\caption{Mass shift factors $K^{\text{NMS}}$ and $K^{\text{SMS}}$ (in GHz u) for the $3s^2 3p^2\;^1\mathrm{S}_0 \to 3s^2 3p 4s\;^1\mathrm{P}_1^o$ transition in the Si atom.}
\begin{tabular}{lrr}
\hline 
\hline 
\vspace{1 mm}
                  & $K^{\text{NMS}}$ & $K^{\text{SMS}}$ \\
\hline                
FS-RCCSD                &   $-407$  & 792 \\
FS-RCCSDT $-$ FS-RCCSD	&   $-12$   & $-1$  \\
Total                   &   $-418$  & 791 \\
\hline
\end{tabular}
\label{SiResults}
\end{table}

The calculation was performed using the Dirac-Coulomb Hamiltonian. Both target states $3s^2 3p^2\;^1\mathrm{S}_0$ and $3s^2 3p 4s\;^1\mathrm{P}_1^o$ belong to sector $(0h,2p)$ of the Fock-space and possess strong multi-reference character. The calculation was completed in two steps. First, the FS-RCCSD calculation was performed using the large atomic basis set. Then, the correlation correction was calculated as the difference between FS-RCCSDT and FS-RCCSD results obtained using the the reduced-size basis set. The ``Total'' line of Table \ref{SiResults} gives the final values of the MS factors. In all calculations, the virtual cut-off was set to 500 Hartree, and all electrons were correlated. Such a choice of a cut-off is required to properly take into account correlation effects of the inner-core electrons~\cite{Skripnikov:17a}. In the FS-RCCSDT model, all triple excitation cluster amplitudes were considered, i.e., including triple excitations in all sectors of the FS and for all electrons.

%L.S.: added to illustrate the importance of FS-RCCSDT
As noted above, the FF approach was used to calculate MS constants in Si. This can be explored in more detail. There are two strategies for applying the FF method~(\ref{eqff1}). In the first strategy, the perturbing operator $O$ is added to the electronic Hamiltonian at the DHF stage of the calculation. In the second strategy, the operator $O$ is added only at the RCC stage. In the former case, one allows for the orbital relaxation due to the perturbation $O$ already at the DHF level. For the exact wave function, the results obtained using these two methods should be close to each other, up to the contribution of the negative energy Dirac spectrum. The contribution of the first term in Eq.~(\ref{nmsexp}) to the NMS atomic factor $K^{\text{NMS}}$ was calculated using the first strategy, while all other contributions were obtained using the second strategy.
To illustrate the importance of triple excitation amplitudes, Table \ref{TCompare} compares the contributions of the first term in Eq.~(\ref{nmsexp}) to the NMS atomic factor $K^{\text{NMS}}$ calculated using the two described options. These calculations were performed in the same basis set. One can see that strategy one allows for faster convergence with respect to high-order correlation effects for the considered operator. However, at the FS-RCCSDT level, the results are nearly invariant to the FF method employed.

\begin{table}[t]
\caption{The contribution of the first term in Eq.~(\ref{nmsexp}) to the NMS atomic factor $K^{\text{NMS}}$ (in GHz u) for the $3s^2 3p^2\;^1\mathrm{S}_0 \to 3s^2 3p 4s\;^1\mathrm{P}_1^o$ transition in the Si atom using different methods.}
\begin{tabular}{lrr}
\hline
\hline
    & Strategy 1 & Strategy 2 \\
\hline    
FS-RCCSD   & $-381$  & $-253$  \\
FS-RCCSDT  & $-396$  & $-393$  \\
\hline
\hline
\end{tabular}
\label{TCompare}
\end{table}

\section{Theory input to upcoming experiments}

With the ever-increasing capabilities of radioactive beam facilities around the world, our access to precision measurements for isotopes at the limits of existence has greatly expanded. Laser spectroscopy measurements of ISs remain as a unique tool for measuring the nuclear charge radii of these short-lived isotopes, for which key questions remain open in our understanding of nuclear size at large proton-to-neutron ratios. However, the precision of extracting nuclear charge radii is, in many cases, limited by the atomic theory input. This is certainly the case for all odd-Z elements, which do not have enough stable isotopes to perform independent IS measurements, as well as for the light ($Z<20$) and heavy elements ($Z>82$). 

\subsection{Light and Medium Mass Nuclei}

\begin{figure*}
    \centering
    \includegraphics[width=0.98\textwidth]{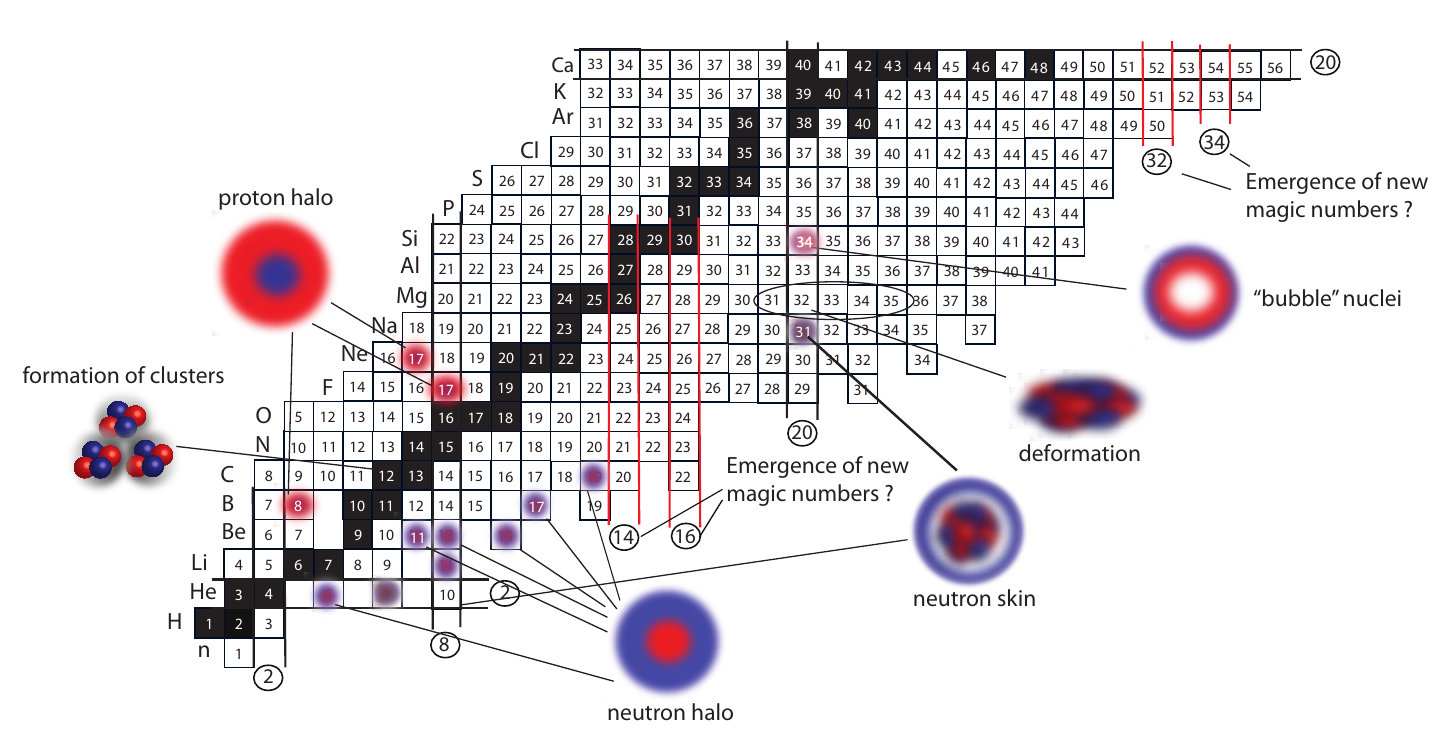}
    \caption{Lightest isotopes of the nuclear chart. Distinct physical phenomena are highlighted for particular isotopes.}
    \label{fig:light}
\end{figure*}

Light and medium mass nuclear systems are critical in understanding fundamental nuclear phenomena and their connection to the strong force described by Quantum Chromodynamics (QCD). Figure \ref{fig:light} shows the lightest isotopes of the nuclear chart ($Z$ vs $N$). Distinct physical phenomena are highlighted for particular isotopes. 
Some recent experimental efforts are ongoing for boron ($Z=5$, \cite{maa19}), carbon ($Z=6$, \cite{Img23}), fluorine ($Z=9$, \cite{Gar16a}), magnesium ($Z=12$,~\cite{2020-MIRACLES}), aluminium ($Z=13$, \cite{2021-Al,Pla23}), silicon ($Z=14$, \cite{kon24}), potassium ($Z=19$, \cite{Kre14,2021-K}), cobalt ($Z=27$ \cite{Kos23}), nickel ($Z=28$ \cite{Mal22}), and germanium ($Z=32$ \cite{2024-wang}). 

Calculating IS factors in atomic boron with five electrons is the state-of-the-art in the few-electron methods which can be solved to high precision~\cite{2015-BIS}. Above $Z=5$, IS calculations for elements with $Z<30$ pose two challenges, the first is that the MS can be of the order of or much larger than the FS~\cite{Hey16}. Therefore, a high accuracy (sub percent) is needed in its calculation in order to extract radii differences. 
The second is that it is challenging to reach high accuracy calculations, as electron correlations play an important role in elements with low $Z$~\cite{2022-Na}. 
%\LS{??? At least, for Al (Z=13), already 2016-MCDF calculations were more or less enough to extract radii differences ??? Ben: In my opinion, the MCDF result had a bit of cherry picking :)}

\subsection{Mirror nuclei and nuclear matter}

A main open question in nuclear physics research is understanding how nuclear matter behaves at different densities and asymmetries between protons and neutrons. The thermodynamic properties of nuclear matter can be described by the nuclear equation of state. However, at present, the parameters of this equation are poorly constrained. In particular, there is a significant uncertainty regarding how the symmetry energy changes with nuclear density, commonly referred to as the slope of the symmetry energy, $L$ \cite{Roca-Maza.2018}. Hence, there is a need to measure observables that could provide information related to the densities of protons and neutrons in nuclei. Parity-violating electron scattering experiments offer powerful probes to measure the radius difference between neutrons and protons, known as the neutron skin, which has been shown to be proportional to $L$~\cite{Adh21}. More recently, a complementary approach has been proposed, using the charge radii differences of mirror nuclei, whose differences can be correlated to $L$~\cite{Wang.2013, Brown.2017, Yang.2018, Reinhard.2022b}. 
Atomic theory is of critical importance, as the uncertainty of recent measurements is limited by the accuracy of the theoretical input~\cite{Brown.2017,kon24}.
Future highly sensitive prospects for these studies include the mirror pairs of $^{16}$S-$^{16}$Mg, $^{22}$O-$^{22}$Si, $^{40}$Ti-$^{40}$Ar, $^{44}$Cr-$^{44}$Ca, $^{46}$Fe-$^{46}$Ca, $^{48}$Ni-$^{48}$Ca, $^{50}$Ni-$^{50}$Ti,  $^{52}$Ni-$^{52}$Cr. Input from the atomic theory is needed to compare the differences in charge radii across all of these cases.
Moreover, it has recently been recognized that the absolute radii of mirror nuclei play a key role in extraction of the $ft$ values from superallowed nuclear beta-decays~\cite{2023-weak,2023-EW,2024-ft}, crucial for a precise determination of the largest CKM matrix element, $V_{ud}$, and low-energy tests of the electroweak sector of the standard model~\cite{2020-HT,2024-Super}.

\subsection{Heavy and super-heavy nuclei}
There is ongoing interest in the study of isotopes near the proton closed shell $ Z = 50 $. New isotope shift measurements are expected for In (\( Z = 49 \))~\cite{wil22} and Sb (\( Z = 51 \))~\cite{lyc23}, and the extraction of their charge radii values will depend on the accuracy of atomic physics calculations. 
Future experimental programs focused on isotope shift measurements of proton emitter nuclei, such as Tm ($Z=69$)~\cite{che22} and Lu ($Z=71$)~\cite{lyc24} , are also expected in the near future.

For heavier nuclei, the region of heavy actinide  ($Z > 82$) is especially interesting, as it is expected to exhibit pear-like density distributions, known as static octupole deformation~\cite{Gaffney2013,Lyn18}.Octupole nuclear deformation is expected to result in a large enhancement of symmetry-violating nuclear properties, such as the Schiff and the nuclear magnetic quadrupole moments (See e.g. Ref.~\cite{behr2022nuclei}, and references therein). Measurements of these parity- and time-reversal-violating nuclear properties would be a major scientific discovery. Such measurements would provide constraints to unknown sources of $\mathcal{CP}$-violation, which have been proposed as necessary ingredients in our understanding of the observed matter-antimatter asymmetry in the universe~\cite{Arr24}. Moreover, dark matter particles can interact with the atomic nucleus inducing an oscillating nuclear Schiff moment~\cite{Dal23}. Thus, the theoretical and experimental study of nuclei with large octupole deformation is of major interest in several fields of modern physics~\cite{Arr24,Wil24}. 

Until now, direct experimental evidence of static octupole deformation has been reported for only a couple of isotopes, e.g. $^{222,224}$Ra~\cite{Gaffney2013,Butler2020}. 
The search for a non-zero electric octupole moment is of high interest as it violates time-reversal symmetry, which has not been measured yet in atomic spectroscopy.
However, the odd-even staggering of charge radii, defined as 
\begin{equation}
\gamma_N = \frac{\delta\langle r^2 \rangle^{N-1,N}}{\delta\langle r^2 \rangle^{N-1,N+1}},   
\end{equation}
has been suggested as an indicator for the emergence of octupole deformation in nuclei (see e.g.~\cite{AHMAD1988244,1989-Otten,2014-Fr,Lyn18,2019-Ac,bar19}). For isotopes that do not possess octupole deformation, this parameter takes typical values between 0 and 1, referred to as ``normal'' staggering. On the other hand, isotopes approaching the region of octupole deformation exhibit ``inverted'' odd-even staggering, with $\gamma_N$ values larger than 1. However, our lack of experimental knowledge in the region does not allow firm conclusions to be established. 
Moreover, in lower-Z nuclei, other explanations than octupole phenomena are considered~\cite{2021-DFT}.
Future IS measurements of Ac, Th, Pa and U around the neutron number $N=136$, combined with atomic calculations, will provide essential information to understand the evolution of the nuclear size in actinide isotopes and clarify the proposed correlation between odd-even staggering of ISs and octupole deformation.
Isotope shifts have been measured in super-heavy nuclei such as Cf~\cite{Web23}, Es~\cite{Not22}. However, the lack of atomic theory calculations prevents the direct extraction of their nuclear charge radii. See reference~\cite{Blo21} for a recent review focused on laser spectroscopy of the actinides.

\subsection{High order moments and new physics searches}\label{sec:high}

High-precision IS measurements currently search for potential new fundamental particles and interactions manifesting as non-linearities in King plots~\cite{2017-HiggsLike,2017-Mik,2017-Yotam,Ber18,Sta18, 2020-KP}.
Recently observed non-linearities with Yb isotopes~\cite{2020-YbKP,2020-KP,2020-YbBud,Hur22,Ono22,2024-YbKP,2024-YbKP2} have been suggested to arise from higher-order effects of the nuclear charge distribution, such as $\langle r^4 \rangle$ and nuclear dipole polarizability~\cite{Kau20,All21,2022-NPOL}. Atomic theory calculations are essential for extracting these higher-order moments of the nuclear density distribution from precise isotope shift measurements. These results are needed to decouple new physics from the Standard Model physics. Moreover, in the absence of the new physics, these measurements will allow access to elusive observables, significantly enhancing our understanding of the nuclear density distribution and properties of the nuclear surface~\cite{Reinhard2020}. An empirical knowledge of nuclear dipole polarizability provides crucial complementary information on the nuclear response, which has been demonstrated to correlate with the equation of state for nuclear matter~\cite{Ros13,Roc18b}.
Isotopic chains of several elements, in various charge states, are already utilised or are proposed to be used in these studies, namely: Li~\cite{2021-LiKP}, Be~\cite{2024-Be}, Ar~\cite{2020-CAR, 2022-HCIclock}, Ca~\cite{Geb15,Ber18,Flam18,2020-KP,2020-g,2020-CaKP,2021-CaHCI,2022-NPOL,2023-CaNL, 2024-simon, 2024-free}, Ni~\cite{2020-g}, Sr~\cite{Ber18,Flam18,Man19,2022-NPOL,2024-SrM}, Cd~\cite{2022-Wang,2022-Cd,2023-Cd2}, Sn~\cite{2024-Sn}, Xe~\cite{2023-XeKP}, Ba~\cite{Flam18}, Yb~\cite{Ber18, Flam18, 2020-YbKP,2020-KP,2020-YbBud,Hur22,Ono22,2022-NPOL,2024-YbKP,2024-YbKP2}, Hg~\cite{Flam18, 2022-Hg}, and Ra~\cite{Hur22,Flam18,2019-Ra,2022-Ra,2023-Ra}.

\section{Outlook}

\subsection{Benchmarking calculations}

\subsubsection{With muonic atoms:}

The energy levels of Muonic atoms, which are simple and compact hydrogen-like  systems, are highly sensitive to nuclear effects. As such, muonic atoms are used to measure both absolute radii, and their difference between (mostly stable) isotopes~\cite{1969-muonic,1979-Friar,2022-muonic}.
For most naturally abundant nuclei, except in the range $Z=3-10$~\cite{2024-QUARTET}, the radii (and isotopic differences) extracted from muonic atoms are limited by nuclear theory calculations, mainly of the nuclear polarizability.
In even-Z elements where several stable isotopes exist, determining two (or more) differential radii pairs enables to extract the IS factors of Eq.~\ref{eq:IS} via a calibrated King Plot~\cite{king}. Odd-Z elements do not possess enough stable isotopes to perform this procedure. To address this by extending muonic atom x-ray spectroscopy to long-lived radioactive targets, techniques for measurements with microgram target material are being developed~\cite{2018-MuX,2019-MuX,2020-Abs,2023-MUX,2023-MUX2,2024-Ra}.

The new generation of atomic many body methods reviewed here enables accurate determinations of $\delta r^2$ from optical isotope shifts. In some cases the claimed accuracy of these \textit{all-optical} radii is comparable or even higher than $\delta r^2$ measured with muonic atoms~\cite{silwal-18-is,2021-K,2021-Yb,2022-Cd,2022-CdHG,2023-Zn,Penyazkov:2023,2024-Ag, skripnikov2024isotopeQED}.
Thus, the two techniques of extracting $\delta r^2$ are complementary; advancements in nuclear theory enable better tests of atomic many-electron theory, and vice-versa.

\subsubsection{With g-factors of HCIs:}

A new experimental technique has recently come into fruition; measuring isotopic differences in g-factors of single electrons bound to bare nuclei. 
In combination with well-known QED theory of ISs in one-these systems~\cite{1997-bound, 2002-bound, 2019-bound, 2020-bound}, the difference in radii between even-even isotopes can be extracted with record precision. This is in part due to the fact that the nuclear polarization contribution fraction of the finite-size effect in these systems is much smaller than in muonic atoms~\cite{1979-Friar}.
Currently, only the pair $^{20,22}$Ne~\cite{2022-g} has been subjected to this measurement, returning $\delta r^2$ with an order of magnitude higher accuracy the value from muonic atoms~\cite{1992-Ne,2019-Ne}.
If and when such measurements are applied to more cases, then the many-body techniques discussed in this review could be tested with high-precision, allowing for more realistic uncertainty estimations as well as the guiding of new calculation methods, such as the MR-CC approach, discussed next.

\subsection{Other multireference coupled cluster methods for calculations of IS factors}
\label{sec:other-mrcc}

Despite the enormous progress in relativistic coupled cluster methods made in recent years, electronic states with complicated configurations involving more than three unpaired electrons on open shells remain generally inaccessible within highly accurate CC methods. Nevertheless, for systems with several valence electrons, but with single-reference character of electronic wave functions, one can still successfully apply single-reference CC methods. See e.g. the calculation of the neutral polonium atom~\cite{Skripnikov2023_Po}.

To reveal a new page in very accurate IS calculations of atoms and other small systems, further developments in multireference CC theories other than the Fock-space CC are highly desirable (see, for example, Refs.~\cite{Lyakh:11,Kohn:12} for a review of other MR-CC approaches). There are three basic requirements to such a theory able to treat IS factors in systems with complicated electronic configuration. First of all, its formulation and implementation must be compatible with the use of relativistic Hamiltonians. To the authors' best knowledge, currently only the complete active space coupled cluster (CAS-CC) method implemented in the {\sc mrcc} package~\cite{MRCC2020,Kallay:1,Kallay:2} fits this requirement (but only for highly symmetric systems where the real arithmetics can be used~\cite{Saue:99}). Secondly, it is desirable that the MR-CC theory would not be biased towards one of reference determinants; except for FS-CC such a property is guaranteed for the formulations based on the Jeziorski-Monkhorst (JM) ans\"atz~\cite{Jeziorski:81,Kucharski:91} and for the internally-contracted (ic) MR-CC~\cite{Evangelista:11,Kohn:12}. However, the JM-based formulations of MR-CC do not meet the last requirement, i.e. acceptable computational cost comparable to that of the single-reference CC, due to the dependence on the model space dimension (which scales as a factorial of the number of unpaired electrons). These arguments allow us to conclude that the internally contracted formulation seems to be the most promising for the relativistic adaption and further use in highly accurate computational studies of isotope shifts and other atomic properties needed for state-of-the-art studies of nuclei.

%\subsubsection{Other all-order methods}
%Fock-space coupled cluster theory outlined here is not the only tool successfully applied to two-valence atomic problems. Another methodology to be mentioned here is the so-called ``CI+all order'' approach developed in the series of paper first for the case of single-valence atomic systems
%~\cite{Blundell:91,Safronova:99,Pal:07,Safronova:08} and then for multivalence ones~\cite{Kozlov:04,Safronova:09}.
%It is very close to FS-RCC in its construction and possibilities and was also very successful in predicting isotope shifts in atoms and atomic ions (\cite{2001-Saf,Safronova:08,Safronova:Th:18} and references therein).
%The modern computer implementation of the CI+all order method is very high performance~\cite{Cheung:21}, and is oriented at systems with purely atomic symmetry. [{\bf I am not sure the reason why CI+all method is highlighted here. It is not equivalent to FS-RCC method. The philosophy of this method is different than the FS-RCC method. Again, if CI+all-order method has to be discussed here then it is necessary to discuss the CC+CV+VV approaches of the MCDF method. The MCDF method is better than the FS-RCC and CI+all-order methods for small atomic systems.}]

\subsection{Mixed field and mass shifts}

Beyond the nonrelativistic approximation, the calculation of the nuclear-recoil effect with a finite-size nucleus may only be achieved within the framework of QED~\cite{Anisimova:2022,2023-QEDRec}, as the simple approach of calculating the MS with the DC Hamiltonian including an extended nucleus potential leads to a spurious contribution linear in the RMS charge radius~\cite{1998-Spurious}.
A rigorous QED approach to treat this problem has been developed recently, for hydrogenic systems, using a generalized photon propagator~\cite{2023-QEDRec,2023-recQED,pachucki2024heavy}. This formulation has not been extended yet to many-electron systems, and so for all IS factor calculations, the relativistic mixed MS and FS contribution should be considered unknown. 
On top of the nuclear polarization and deformation mentioned in section~\ref{sec:high}, the mixed contribution leads to nonlinearity in King Plots.
Thus, extending relativistic-recoil calculations to multiple-electron systems with a finite-size nucleus is crucial in order to disentangle known and new physics effects in these searches.

\section{Summary}

In this topical review, we described some recent advancements in atomic many-body methods for high-precision studies of isotope shifts, with an emphasis on extractions of mean-squared radii differences. We focused on two frontiers: The first is single-valence highly charged ions. Here, electron-correlation effects can be treated perturbatively while QED corrections must be calculated more rigorously. As an example, we presented previously-published results of third-order many-body perturbation theory applied to Na-like xenon.
The second frontier is related to neutral and singly charged systems. Here electron correlations are needed to be treated to all-orders while QED effect are weaker, making them amenable to approximate methods. We discussed recent advancements in relativistic coupled-cluster theory and calculation methods, for both single- and multi-valence systems.
We have shown that, in single and two-valence systems, the new generation of relativistic coupled-cluster calculations is able to provide accurate isotope shift factors. These enable RMS radii differences to be reliably extracted from optical data independently of experiments with muonic atoms.
Nevertheless, the latter are still responsible for providing absolute reference radii.
With the evolving optical isotope shift measurements and the corresponding theory, the uncertainties in the reference radii are becoming limiting.
Measuring these with electronic atoms, as well as performing modern rigorous extractions from existing muonic atom data, is highly desirable. 

\section*{Acknowledgement}

The authors thank Prof. E.~Krishnakumar for inviting us to write this review article and Dr. Dillon Eastoe for all his cooperation in the review process. B.~K.~S. acknowledges DST of India for the financial support under the Grant No. CRG/2023/002558 and Department of Space, Government of India, and the use of ParamVikram-1000 HPC facility at Physical Research Laboratory (PRL), Ahmedabad for carrying out the atomic calculations.
B.~O. acknowledges support from the Helen-Diller quantum center, as well as helpful discussions with J.~Behr, K.~Pachucki, V.~Yerokhin, T.~Cocolios, and Y.~Soreq.
R.~F.~G.~R. acknowledges support from the Office of Nuclear Physics, U.S. Department of Energy, under grants DE-SC0021176 and DE-SC0021179, and the National Science Foundation under the grant 2207996.
L.~V.~S. and A.~V.~O. are grateful to E.~Eliav, D.~E.~Maison, G.~Penyazkov, S.~D.~Prosnyak, A.~V.~Titov, A.~V.~Zaitsevskii, A.~V. Malyshev and V.~M.~Shabaev for useful discussions and joint work.
The work of A.~V.~O. on preparing Sections~\ref{sec:fscc},~\ref{MRapplication} and~\ref{sec:other-mrcc} at NRC ``Kurchatov Institute'' -- PNPI was supported by the Russian Science Foundation under grant no.~20-13-00225, \url{https://rscf.ru/project/23-13-45028/}.
The work of L.V.S. at NRC ``Kurchatov Institute'' -- PNPI on Sections~\ref{QED-FS}, \ref{QED-MS},~\ref{SRRCC} and ~\ref{SRapplication} were supported by the Russian Science Foundation under grant no.~24-12-00092, \url{https://rscf.ru/project/24-12-00092/}.
Electronic structure calculations of L.~V.~S. and A.~V.~O. have been carried out using computing resources of the federal collective usage center Complex for Simulation and Data Processing for Mega-science Facilities at National Research Centre ``Kurchatov Institute'', \url{http://ckp.nrcki.ru/}.

\bibliography{Bibliography}

\providecommand{\newblock}{}
\begin{thebibliography}{100}
\expandafter\ifx\csname url\endcsname\relax
  \def\url#1{{\tt #1}}\fi
\expandafter\ifx\csname urlprefix\endcsname\relax\def\urlprefix{URL }\fi
\providecommand{\eprint}[2][]{\url{#2}}

\bibitem{Geb15}
Gebert F, Wan Y, Wolf F, Angstmann C~N, Berengut J~C and Schmidt P~O 2015 {\em Physical Review Letters\/} {\bf 115} 053003 ISSN 0031-9007 \urlprefix\url{https://link.aps.org/doi/10.1103/PhysRevLett.115.053003}

\bibitem{2017-HiggsLike}
Delaunay C, Ozeri R, Perez G and Soreq Y 2017 {\em Phys. Rev. D\/} {\bf 96}(9) 093001 \urlprefix\url{https://link.aps.org/doi/10.1103/PhysRevD.96.093001}

\bibitem{2021-Ag}
Reponen M, de~Groote R, Al~Ayoubi L, Beliuskina O, Bissell M, Campbell P, Ca{\~n}ete L, Cheal B, Chrysalidis K, Delafosse C {\em et~al.\/} 2021 {\em Nature Communications\/} {\bf 12} 4596

\bibitem{Ber18}
Berengut J~C, Budker D, Delaunay C, Flambaum V~V, Frugiuele C, Fuchs E, Grojean C, Harnik R, Ozeri R, Perez G and Soreq Y 2018 {\em Physical Review Letters\/} {\bf 120} 91801 \urlprefix\url{https://link.aps.org/doi/10.1103/PhysRevLett.120.091801}

\bibitem{Bar21}
Barzakh A, Andreyev A~N, Raison C, Cubiss J~G, Van~Duppen P, P\'eru S, Hilaire S, Goriely S, Andel B, Antalic S, Al~Monthery M, Berengut J~C, Biero\ifmmode~\acute{n}\else \'{n}\fi{} J, Bissell M~L, Borschevsky A, Chrysalidis K, Cocolios T~E, Day~Goodacre T, Dognon J~P, Elantkowska M, Eliav E, Farooq-Smith G~J, Fedorov D~V, Fedosseev V~N, Gaffney L~P, Garcia~Ruiz R~F, Godefroid M, Granados C, Harding R~D, Heinke R, Huyse M, Karls J, Larmonier P, Li J~G, Lynch K~M, Maison D~E, Marsh B~A, Molkanov P, Mosat P, Oleynichenko A~V, Panteleev V, Pyykk\"o P, Reitsma M~L, Rezynkina K, Rossel R~E, Rothe S, Ruczkowski J, Schiffmann S, Seiffert C, Seliverstov M~D, Sels S, Skripnikov L~V, Stryjczyk M, Studer D, Verlinde M, Wilman S and Zaitsevskii A~V 2021 {\em Phys. Rev. Lett.\/} {\bf 127}(19) 192501 \urlprefix\url{https://link.aps.org/doi/10.1103/PhysRevLett.127.192501}

\bibitem{2023-Review}
Yang X~F, Wang S~J, Wilkins S~G and {Garcia Ruiz} R~F 2023 {\em Progress in Particle and Nuclear Physics\/} {\bf 129} 104005 ISSN 0146-6410 \urlprefix\url{https://www.sciencedirect.com/science/article/pii/S0146641022000631}

\bibitem{2024-Struct}
Koszor{\'u}s {\'A}, de~Groote R, Cheal B, Campbell P and Moore I 2024 {\em The European Physical Journal A\/} {\bf 60} 20

\bibitem{king}
King W 2013 {\em {Isotope Shifts in Atomic Spectra}\/} vol~11 (Springer Science {\&} Business Media) ISBN 1489917861 \urlprefix\url{https://books.google.com/books?id=eEgGCAAAQBAJ{\&}pgis=1}

\bibitem{Mil19}
{Miller} A~J, {Minamisono} K, {Klose} A, {Garand} D, {Kujawa} C, {Lantis} J~D, {Liu} Y, {Maa{\ss}} B, {Mantica} P~F, {Nazarewicz} W, {N{\"o}rtersh{\"a}user} W, {Pineda} S~V, {Reinhard} P~G, {Rossi} D~M, {Sommer} F, {Sumithrarachchi} C, {Teigelh{\"o}fer} A and {Watkins} J 2019 {\em Nature Physics\/} {\bf 15} 432--436

\bibitem{Gar16}
{Garcia Ruiz} R~F, {Bissell} M~L, {Blaum} K, {Ekstr{\"o}m} A, {Fr{\"o}mmgen} N, {Hagen} G, {Hammen} M, {Hebeler} K, {Holt} J~D, {Jansen} G~R, {Kowalska} M, {Kreim} K, {Nazarewicz} W, {Neugart} R, {Neyens} G, {N{\"o}rtersh{\"a}user} W, {Papenbrock} T, {Papuga} J, {Schwenk} A, {Simonis} J, {Wendt} K~A and {Yordanov} D~T 2016 {\em Nature Physics\/} {\bf 12} 594--598 (\textit{Preprint} \eprint{1602.07906})

\bibitem{Reinhard2020}
Reinhard P~G, Nazarewicz W and Garcia~Ruiz R~F 2020 {\em Phys. Rev. C\/} {\bf 101}(2) 021301(R) \urlprefix\url{https://link.aps.org/doi/10.1103/PhysRevC.101.021301}

\bibitem{Allehabi:20}
Allehabi S~O, Dzuba V~A, Flambaum V~V, Afanasjev A~V and Agbemava S~E 2020 {\em Phys. Rev. C\/} {\bf 102}(2) 024326

\bibitem{Blaum_2011}
Blaum K, Block M, Cakirli R~B, Eliseev S, Kowalska M, Kreim S, Litvinov Y~A, Nagy S, Nörtershäuser W and Yordanov D~T 2011 {\em Journal of Physics: Conference Series\/} {\bf 312} 092001 \urlprefix\url{https://dx.doi.org/10.1088/1742-6596/312/9/092001}

\bibitem{2013-Blaum}
Blaum K, Dilling J and Nörtershäuser W 2013 {\em Physica Scripta\/} {\bf 2013} 014017 \urlprefix\url{https://dx.doi.org/10.1088/0031-8949/2013/T152/014017}

\bibitem{2018-Mass}
Dilling J, Blaum K, Brodeur M and Eliseev S 2018 {\em Annual Review of Nuclear and Particle Science\/} {\bf 68} 45--74 ISSN 1545-4134 \urlprefix\url{https://www.annualreviews.org/content/journals/10.1146/annurev-nucl-102711-094939}

\bibitem{2022-Mass}
Blaum K, Eliseev S and Goriely S 2022 Masses of exotic nuclei {\em Handbook of Nuclear Physics\/} (Springer) pp 1--38

\bibitem{Gor19}
Gorges C, Rodr\'{\i}guez L~V, Balabanski D~L, Bissell M~L, Blaum K, Cheal B, Garcia~Ruiz R~F, Georgiev G, Gins W, Heylen H, Kanellakopoulos A, Kaufmann S, Kowalska M, Lagaki V, Lechner S, Maa\ss{} B, Malbrunot-Ettenauer S, Nazarewicz W, Neugart R, Neyens G, N\"ortersh\"auser W, Reinhard P~G, Sailer S, S\'anchez R, Schmidt S, Wehner L, Wraith C, Xie L, Xu Z~Y, Yang X~F and Yordanov D~T 2019 {\em Phys. Rev. Lett.\/} {\bf 122}(19) 192502 \urlprefix\url{https://link.aps.org/doi/10.1103/PhysRevLett.122.192502}

\bibitem{2020-Cu}
De~Groote R, Billowes J, Binnersley C~L, Bissell M~L, Cocolios T~E, Day~Goodacre T, Farooq-Smith G~J, Fedorov D, Flanagan K~T, Franchoo S {\em et~al.\/} 2020 {\em Nature Physics\/} {\bf 16} 620--624

\bibitem{kar24}
Karthein J {\em et~al.\/} 2024 {\em arXiv:2310.15093v2\/}

\bibitem{kon24}
K\"onig K, Berengut J~C, Borschevsky A, Brinson A, Brown B~A, Dockery A, Elhatisari S, Eliav E, Garcia~Ruiz R~F, Holt J~D, Hu B~S, Karthein J, Lee D, Ma Y~Z, Mei\ss{}ner U~G, Minamisono K, Oleynichenko A~V, Pineda S~V, Prosnyak S~D, Reitsma M~L, Skripnikov L~V, Vernon A and Zaitsevskii A 2024 {\em Phys. Rev. Lett.\/} {\bf 132}(16) 162502 \urlprefix\url{https://link.aps.org/doi/10.1103/PhysRevLett.132.162502}

\bibitem{2017-Mik}
Mikami K, Tanaka M and Yamamoto Y 2017 {\em The European Physical Journal C\/} {\bf 77} 1--11

\bibitem{2017-Yotam}
Delaunay C, Frugiuele C, Fuchs E and Soreq Y 2017 {\em Phys. Rev. D\/} {\bf 96}(11) 115002 \urlprefix\url{https://link.aps.org/doi/10.1103/PhysRevD.96.115002}

\bibitem{2017-Fuchs}
Frugiuele C, Fuchs E, Perez G and Schlaffer M 2017 {\em Phys. Rev. D\/} {\bf 96}(1) 015011 \urlprefix\url{https://link.aps.org/doi/10.1103/PhysRevD.96.015011}

\bibitem{Sta18}
Stadnik Y~V 2018 {\em Physical Review Letters\/} {\bf 120} 223202 \urlprefix\url{https://link.aps.org/doi/10.1103/PhysRevLett.120.223202}

\bibitem{2020-KP}
Berengut J~C, Delaunay C, Geddes A and Soreq Y 2020 {\em Phys. Rev. Res.\/} {\bf 2}(4) 043444 \urlprefix\url{https://link.aps.org/doi/10.1103/PhysRevResearch.2.043444}

\bibitem{Man19}
Manovitz T, Shaniv R, Shapira Y, Ozeri R and Akerman N 2019 {\em Phys. Rev. Lett.\/} {\bf 123}(20) 203001 \urlprefix\url{https://link.aps.org/doi/10.1103/PhysRevLett.123.203001}

\bibitem{2020-CaKP}
Solaro C, Meyer S, Fisher K, Berengut J~C, Fuchs E and Drewsen M 2020 {\em Phys. Rev. Lett.\/} {\bf 125}(12) 123003 \urlprefix\url{https://link.aps.org/doi/10.1103/PhysRevLett.125.123003}

\bibitem{2020-YbKP}
Counts I, Hur J, Aude~Craik D~P~L, Jeon H, Leung C, Berengut J~C, Geddes A, Kawasaki A, Jhe W and Vuleti\ifmmode~\acute{c}\else \'{c}\fi{} V 2020 {\em Phys. Rev. Lett.\/} {\bf 125}(12) 123002 \urlprefix\url{https://link.aps.org/doi/10.1103/PhysRevLett.125.123002}

\bibitem{2020-YbBud}
Figueroa N~L, Berengut J~C, Dzuba V~A, Flambaum V~V, Budker D and Antypas D 2022 {\em Phys. Rev. Lett.\/} {\bf 128}(7) 073001 \urlprefix\url{https://link.aps.org/doi/10.1103/PhysRevLett.128.073001}

\bibitem{Hur22}
Hur J, Aude~Craik D~P~L, Counts I, Knyazev E, Caldwell L, Leung C, Pandey S, Berengut J~C, Geddes A, Nazarewicz W, Reinhard P~G, Kawasaki A, Jeon H, Jhe W and Vuleti\ifmmode~\acute{c}\else \'{c}\fi{} V 2022 {\em Phys. Rev. Lett.\/} {\bf 128}(16) 163201 \urlprefix\url{https://link.aps.org/doi/10.1103/PhysRevLett.128.163201}

\bibitem{Ono22}
Ono K, Saito Y, Ishiyama T, Higomoto T, Takano T, Takasu Y, Yamamoto Y, Tanaka M and Takahashi Y 2022 {\em Phys. Rev. X\/} {\bf 12}(2) 021033 \urlprefix\url{https://link.aps.org/doi/10.1103/PhysRevX.12.021033}

\bibitem{2024-free}
Chang T~T, Awazi B~B, Berengut J~C, Fuchs E and Doret S~C 2024 {\em arXiv\/} (\textit{Preprint} \eprint{2311.17337}) \urlprefix\url{https://arxiv.org/abs/2311.17337}

\bibitem{2024-YbKP}
Door M, Yeh C~H, Heinz M, Kirk F, Lyu C, Miyagi T, Berengut J~C, Biero{\'n} J, Blaum K, Dreissen L~S {\em et~al.\/} 2024 {\em arXiv preprint arXiv:2403.07792\/}

\bibitem{Pap16}
Papoulia A, Carlsson B~G and Ekman J 2016 {\em Phys. Rev. A\/} {\bf 94}(4) 042502 \urlprefix\url{https://link.aps.org/doi/10.1103/PhysRevA.94.042502}

\bibitem{Flam18}
Flambaum V~V, Geddes A~J and Viatkina A~V 2018 {\em Phys. Rev. A\/} {\bf 97}(3) 032510 \urlprefix\url{https://link.aps.org/doi/10.1103/PhysRevA.97.032510}

\bibitem{Kau20}
Kaufmann S, Simonis J, Bacca S, Billowes J, Bissell M~L, Blaum K, Cheal B, Ruiz R~F~G, Gins W, Gorges C, Hagen G, Heylen H, Kanellakopoulos A, Malbrunot-Ettenauer S, Miorelli M, Neugart R, Neyens G, N\"ortersh\"auser W, S\'anchez R, Sailer S, Schwenk A, Ratajczyk T, Rodr\'{\i}guez L~V, Wehner L, Wraith C, Xie L, Xu Z~Y, Yang X~F and Yordanov D~T 2020 {\em Phys. Rev. Lett.\/} {\bf 124}(13) 132502 \urlprefix\url{https://link.aps.org/doi/10.1103/PhysRevLett.124.132502}

\bibitem{Sky21}
Pineda S~V, K\"onig K, Rossi D~M, Brown B~A, Incorvati A, Lantis J, Minamisono K, N\"ortersh\"auser W, Piekarewicz J, Powel R and Sommer F 2021 {\em Phys. Rev. Lett.\/} {\bf 127}(18) 182503 \urlprefix\url{https://link.aps.org/doi/10.1103/PhysRevLett.127.182503}

\bibitem{nortershauser2014nuclear}
N{\"o}rtersh{\"a}user W and Geppert C 2014 Nuclear charge radii of light elements and recent developments in collinear laser spectroscopy {\em The Euroschool on Exotic Beams, Vol. IV\/} (Springer) pp 233--292

\bibitem{2015-revExp}
Bradley~Cheal I~M and Nörtershäuser W 2015 {\em Nuclear Physics News\/} {\bf 25} 12--18 \urlprefix\url{https://doi.org/10.1080/10619127.2015.1104126}

\bibitem{2022-ReviewExp}
N{\"o}rtersh{\"a}user W and Moore I 2022 Nuclear charge radii {\em Handbook of Nuclear Physics\/} (Springer) pp 1--70

\bibitem{2012-Co}
Cheal B, Cocolios T~E and Fritzsche S 2012 {\em Phys. Rev. A\/} {\bf 86}(4) 042501 \urlprefix\url{https://link.aps.org/doi/10.1103/PhysRevA.86.042501}

\bibitem{1995-LiDrake}
Yan Z~C and Drake G~W~F 1995 {\em Phys. Rev. A\/} {\bf 52}(5) 3711--3717 \urlprefix\url{https://link.aps.org/doi/10.1103/PhysRevA.52.3711}

\bibitem{1998-LiDrake}
Yan Z~C, Tambasco M and Drake G~W~F 1998 {\em Phys. Rev. A\/} {\bf 57}(3) 1652--1661 \urlprefix\url{https://link.aps.org/doi/10.1103/PhysRevA.57.1652}

\bibitem{2002-LiDrake}
Yan Z~C and Drake G~W~F 2002 {\em Phys. Rev. A\/} {\bf 66}(4) 042504 \urlprefix\url{https://link.aps.org/doi/10.1103/PhysRevA.66.042504}

\bibitem{2008-LiBe}
Yan Z~C, N\"ortersh\"auser W and Drake G~W~F 2008 {\em Phys. Rev. Lett.\/} {\bf 100}(24) 243002 \urlprefix\url{https://link.aps.org/doi/10.1103/PhysRevLett.100.243002}

\bibitem{2009-Lit}
Bubin S, Komasa J, Stanke M and Adamowicz L 2009 {\em The Journal of Chemical Physics\/} {\bf 131} 234112 ISSN 0021-9606 (\textit{Preprint} \eprint{https://pubs.aip.org/aip/jcp/article-pdf/doi/10.1063/1.3275804/15701265/234112\_1\_online.pdf}) \urlprefix\url{https://doi.org/10.1063/1.3275804}

\bibitem{2010-C2}
Bubin S, Komasa J, Stanke M and Adamowicz L 2010 {\em Phys. Rev. A\/} {\bf 81}(5) 052504 \urlprefix\url{https://link.aps.org/doi/10.1103/PhysRevA.81.052504}

\bibitem{2011-Li}
Sharkey K~L, Bubin S and Adamowicz L 2011 {\em Phys. Rev. A\/} {\bf 83}(1) 012506 \urlprefix\url{https://link.aps.org/doi/10.1103/PhysRevA.83.012506}

\bibitem{2014-BeLikeQED}
Malyshev A~V, Volotka A~V, Glazov D~A, Tupitsyn I~I, Shabaev V~M and Plunien G 2014 {\em Phys. Rev. A\/} {\bf 90}(6) 062517 \urlprefix\url{https://link.aps.org/doi/10.1103/PhysRevA.90.062517}

\bibitem{2014-1sBe}
Sims J~S and Hagstrom S~A 2014 {\em The Journal of Chemical Physics\/} {\bf 140} 224312 ISSN 0021-9606 (\textit{Preprint} \eprint{https://pubs.aip.org/aip/jcp/article-pdf/doi/10.1063/1.4881639/13645369/224312\_1\_online.pdf}) \urlprefix\url{https://doi.org/10.1063/1.4881639}

\bibitem{2014-LiLike}
Zubova N~A, Kozhedub Y~S, Shabaev V~M, Tupitsyn I~I, Volotka A~V, Plunien G, Brandau C and St\"ohlker T 2014 {\em Phys. Rev. A\/} {\bf 90}(6) 062512 \urlprefix\url{https://link.aps.org/doi/10.1103/PhysRevA.90.062512}

\bibitem{2014-BeIS}
Puchalski M, Pachucki K and Komasa J 2014 {\em Phys. Rev. A\/} {\bf 89}(1) 012506 \urlprefix\url{https://link.aps.org/doi/10.1103/PhysRevA.89.012506}

\bibitem{2015-QEDLi}
Puchalski M and Pachucki K 2015 {\em Phys. Rev. A\/} {\bf 92}(1) 012513 \urlprefix\url{https://link.aps.org/doi/10.1103/PhysRevA.92.012513}

\bibitem{2015-Shab}
Yerokhin V~A and Shabaev V~M 2015 {\em Journal of Physical and Chemical Reference Data\/} {\bf 44} 033103 ISSN 0047-2689 (\textit{Preprint} \eprint{https://pubs.aip.org/aip/jpr/article-pdf/doi/10.1063/1.4927487/15981145/033103\_1\_online.pdf}) \urlprefix\url{https://doi.org/10.1063/1.4927487}

\bibitem{2015-HeIS}
Pachucki K and Yerokhin V~A 2015 {\em Journal of Physical and Chemical Reference Data\/} {\bf 44} 031206 ISSN 0047-2689 (\textit{Preprint} \eprint{https://pubs.aip.org/aip/jpr/article-pdf/doi/10.1063/1.4921428/15981576/031206\_1\_online.pdf}) \urlprefix\url{https://doi.org/10.1063/1.4921428}

\bibitem{2015-BIS}
Puchalski M, Komasa J and Pachucki K 2015 {\em Phys. Rev. A\/} {\bf 92}(6) 062501 \urlprefix\url{https://link.aps.org/doi/10.1103/PhysRevA.92.062501}

\bibitem{2019-C}
Strasburger K 2019 {\em Phys. Rev. A\/} {\bf 99}(5) 052512 \urlprefix\url{https://link.aps.org/doi/10.1103/PhysRevA.99.052512}

\bibitem{2019-QEDSMS}
Malyshev A~V, Anisimova I~S, Mironova D~V, Shabaev V~M and Plunien G 2019 {\em Phys. Rev. A\/} {\bf 100}(1) 012510 \urlprefix\url{https://link.aps.org/doi/10.1103/PhysRevA.100.012510}

\bibitem{2019-Be}
Horny\'ak I, Adamowicz L and Bubin S 2019 {\em Phys. Rev. A\/} {\bf 100}(3) 032504 \urlprefix\url{https://link.aps.org/doi/10.1103/PhysRevA.100.032504}

\bibitem{2020-CII}
Horny\'ak I, Adamowicz L and Bubin S 2020 {\em Phys. Rev. A\/} {\bf 102}(6) 062825 \urlprefix\url{https://link.aps.org/doi/10.1103/PhysRevA.102.062825}

\bibitem{2020-QEDNMS}
Malyshev A~V, Anisimova I~S, Glazov D~A, Kaygorodov M~Y, Mironova D~V, Plunien G and Shabaev V~M 2020 {\em Phys. Rev. A\/} {\bf 101}(5) 052506 \urlprefix\url{https://link.aps.org/doi/10.1103/PhysRevA.101.052506}

\bibitem{2021-BoronECG}
Horny\'ak I, Nasiri S, Bubin S and Adamowicz L 2021 {\em Phys. Rev. A\/} {\bf 104}(3) 032809 \urlprefix\url{https://link.aps.org/doi/10.1103/PhysRevA.104.032809}

\bibitem{2021-BenchBe}
Nasiri S, Adamowicz L and Bubin S 2021 {\em Journal of Physical and Chemical Reference Data\/} {\bf 50} 043107 ISSN 0047-2689 \urlprefix\url{https://doi.org/10.1063/5.0065282}

\bibitem{2023-C}
Shomenov T and Bubin S 2023 {\em Phys. Rev. E\/} {\bf 108}(6) 065308 \urlprefix\url{https://link.aps.org/doi/10.1103/PhysRevE.108.065308}

\bibitem{2023-CII}
Stanke M, Kedziorski A, Nasiri S, Adamowicz L and Bubin S 2023 {\em Phys. Rev. A\/} {\bf 108}(1) 012812 \urlprefix\url{https://link.aps.org/doi/10.1103/PhysRevA.108.012812}

\bibitem{2024-Be}
Qi X~Q, Zhang P~P, Yan Z~C, Drake G~W~F, Chen A~X, Zhong Z~X and Shi T~Y 2024 {\em Phys. Rev. A\/} {\bf 110}(1) 012810 \urlprefix\url{https://link.aps.org/doi/10.1103/PhysRevA.110.012810}

\bibitem{2024-CIS}
Saeed~Nasiri S~B and Adamowicz L 2024 {\em Molecular Physics\/} {\bf 0} e2325049 (\textit{Preprint} \eprint{https://doi.org/10.1080/00268976.2024.2325049}) \urlprefix\url{https://doi.org/10.1080/00268976.2024.2325049}

\bibitem{2019-QEDtest}
Indelicato P 2019 {\em Journal of Physics B: Atomic, Molecular and Optical Physics\/} {\bf 52} 232001 \urlprefix\url{https://dx.doi.org/10.1088/1361-6455/ab42c9}

\bibitem{1993-RCI}
Chen M~H, Cheng K~T and Johnson W~R 1993 {\em Phys. Rev. A\/} {\bf 47}(5) 3692--3703 \urlprefix\url{https://link.aps.org/doi/10.1103/PhysRevA.47.3692}

\bibitem{1994-RCI}
Cheng K~T, Chen M~H, Johnson W~R and Sapirstein J 1994 {\em Phys. Rev. A\/} {\bf 50}(1) 247--255 \urlprefix\url{https://link.aps.org/doi/10.1103/PhysRevA.50.247}

\bibitem{1995-RCI}
Chen M~H, Cheng K~T, Johnson W~R and Sapirstein J 1995 {\em Phys. Rev. A\/} {\bf 52}(1) 266--273 \urlprefix\url{https://link.aps.org/doi/10.1103/PhysRevA.52.266}

\bibitem{2003-RelIS}
Tupitsyn I~I, Shabaev V~M, Crespo L\'opez-Urrutia J~R, Dragani\ifmmode~\acute{c}\else \'{c}\fi{} I, Soria~Orts R and Ullrich J 2003 {\em Phys. Rev. A\/} {\bf 68}(2) 022511 \urlprefix\url{https://link.aps.org/doi/10.1103/PhysRevA.68.022511}

\bibitem{2007-Boronlike}
Artemyev A~N, Shabaev V~M, Tupitsyn I~I, Plunien G and Yerokhin V~A 2007 {\em Phys. Rev. Lett.\/} {\bf 98}(17) 173004 \urlprefix\url{https://link.aps.org/doi/10.1103/PhysRevLett.98.173004}

\bibitem{2008-RCI}
Yerokhin V~A 2008 {\em Phys. Rev. A\/} {\bf 78}(1) 012513 \urlprefix\url{https://link.aps.org/doi/10.1103/PhysRevA.78.012513}

\bibitem{2010-RCI}
Chen M~H and Cheng K~T 2010 {\em Journal of Physics B: Atomic, Molecular and Optical Physics\/} {\bf 43} 074019 \urlprefix\url{https://dx.doi.org/10.1088/0953-4075/43/7/074019}

\bibitem{2012-RCI}
Yerokhin V~A and Surzhykov A 2012 {\em Phys. Rev. A\/} {\bf 86}(4) 042507 \urlprefix\url{https://link.aps.org/doi/10.1103/PhysRevA.86.042507}

\bibitem{2014-RCI}
Yerokhin V~A, Surzhykov A and Fritzsche S 2014 {\em Phys. Rev. A\/} {\bf 90}(2) 022509 \urlprefix\url{https://link.aps.org/doi/10.1103/PhysRevA.90.022509}

\bibitem{2015-RCI}
Yerokhin V~A, Surzhykov A and Fritzsche S 2015 {\em Physica Scripta\/} {\bf 90} 054003 \urlprefix\url{https://dx.doi.org/10.1088/0031-8949/90/5/054003}

\bibitem{2017-RCI}
Yerokhin V~A, Surzhykov A and M\"uller A 2017 {\em Phys. Rev. A\/} {\bf 96}(4) 042505 \urlprefix\url{https://link.aps.org/doi/10.1103/PhysRevA.96.042505}

\bibitem{2020-CAR}
Yerokhin V~A, M\"uller R~A, Surzhykov A, Micke P and Schmidt P~O 2020 {\em Phys. Rev. A\/} {\bf 101}(1) 012502 \urlprefix\url{https://link.aps.org/doi/10.1103/PhysRevA.101.012502}

\bibitem{1996-CIMBPT}
Dzuba V~A, Flambaum V~V and Kozlov M~G 1996 {\em Phys. Rev. A\/} {\bf 54}(5) 3948--3959 \urlprefix\url{https://link.aps.org/doi/10.1103/PhysRevA.54.3948}

\bibitem{1998.CIMBPT}
Dzuba V~A and Johnson W~R 1998 {\em Phys. Rev. A\/} {\bf 57}(4) 2459--2465 \urlprefix\url{https://link.aps.org/doi/10.1103/PhysRevA.57.2459}

\bibitem{2003-CI}
Berengut J~C, Dzuba V~A and Flambaum V~V 2003 {\em Phys. Rev. A\/} {\bf 68}(2) 022502 \urlprefix\url{https://link.aps.org/doi/10.1103/PhysRevA.68.022502}

\bibitem{2006-CIMBPT}
Berengut J~C, Flambaum V~V and Kozlov M~G 2006 {\em Phys. Rev. A\/} {\bf 73}(1) 012504 \urlprefix\url{https://link.aps.org/doi/10.1103/PhysRevA.73.012504}

\bibitem{Kozlov:15}
Kozlov M~G, Porsev S~G, Safronova M~S and Tupitsyn I~I 2015 {\em Comput. Phys. Commun.\/} {\bf 195} 199–213

\bibitem{2017-CI}
Dzuba V~A, Berengut J~C, Harabati C and Flambaum V~V 2017 {\em Phys. Rev. A\/} {\bf 95}(1) 012503 \urlprefix\url{https://link.aps.org/doi/10.1103/PhysRevA.95.012503}

\bibitem{2019-CIMBPT}
Kaygorodov M~Y, Kozhedub Y~S, Tupitsyn I~I, Malyshev A~V, Glazov D~A, Plunien G and Shabaev V~M 2019 {\em Phys. Rev. A\/} {\bf 99}(3) 032505 \urlprefix\url{https://link.aps.org/doi/10.1103/PhysRevA.99.032505}

\bibitem{2019-ambit}
Kahl E and Berengut J 2019 {\em Computer Physics Communications\/} {\bf 238} 232--243 ISSN 0010-4655 \urlprefix\url{https://www.sciencedirect.com/science/article/pii/S0010465518304302}

\bibitem{2022-CIPT}
Kozlov M~G, Tupitsyn I~I, Bondarev A~I and Mironova D~V 2022 {\em Phys. Rev. A\/} {\bf 105}(5) 052805 \urlprefix\url{https://link.aps.org/doi/10.1103/PhysRevA.105.052805}

\bibitem{Ulrich:2003}
Jentschura U~D 2003 {\em J. Phys. A\/} {\bf 36} L229

\bibitem{2005-CIDFS}
Tupitsyn I~I, Volotka A~V, Glazov D~A, Shabaev V~M, Plunien G, Crespo L\'opez-Urrutia J~R, Lapierre A and Ullrich J 2005 {\em Phys. Rev. A\/} {\bf 72}(6) 062503 \urlprefix\url{https://link.aps.org/doi/10.1103/PhysRevA.72.062503}

\bibitem{kozhedub-10-qed}
Kozhedub Y~S, Volotka A~V, Artemyev A~N, Glazov D~A, Plunien G, Shabaev V~M, Tupitsyn I~I and St{\"o}hlker T 2010 {\em Phys.\ Rev.\ A\/} {\bf 81} 042513 \urlprefix\url{https://doi.org/10.1103/PhysRevA.81.042513}

\bibitem{Zubova:2016}
Zubova N~A, Malyshev A~V, Tupitsyn I~I, Shabaev V~M, Kozhedub Y~S, Plunien G, Brandau C and St\"ohlker T 2016 {\em Phys. Rev. A\/} {\bf 93}(5) 052502

\bibitem{2018-CIDFS}
Tupitsyn I~I, Zubova N~A, Shabaev V~M, Plunien G and St\"ohlker T 2018 {\em Phys. Rev. A\/} {\bf 98}(2) 022517 \urlprefix\url{https://link.aps.org/doi/10.1103/PhysRevA.98.022517}

\bibitem{Blundell:91}
Blundell S~A, Johnson W~R and Sapirstein J 1991 {\em Phys.~Rev.~A\/} {\bf 43}(7) 3407--3418

\bibitem{Safronova:99}
Safronova M~S, Johnson W~R and Derevianko A 1999 {\em Phys.~Rev.~A\/} {\bf 60} 4476–4487

\bibitem{Kozlov:04}
Kozlov M~G 2004 {\em Int. J. Quantum Chem.\/} {\bf 100} 336–342

\bibitem{Pal:07}
Pal R, Safronova M~S, Johnson W~R, Derevianko A and Porsev S~G 2007 {\em Phys.~Rev.~A\/} {\bf 75} 042515

\bibitem{Safronova:08}
Safronova M~S and Johnson W~R 2008 {\em All-Order Methods for Relativistic Atomic Structure Calculations\/} (Elsevier) p 191–233

\bibitem{Safronova:09}
Safronova M~S, Kozlov M~G, Johnson W~R and Jiang D 2009 {\em Phys.~Rev.~A\/} {\bf 80} 012516

\bibitem{2015-LCC}
Okhapkin M~V, Meier D~M, Peik E, Safronova M~S, Kozlov M~G and Porsev S~G 2015 {\em Phys. Rev. A\/} {\bf 92}(2) 020503 \urlprefix\url{https://link.aps.org/doi/10.1103/PhysRevA.92.020503}

\bibitem{Safronova:Th:18}
Safronova M~S, Porsev S~G, Kozlov M~G, Thielking J, Okhapkin M~V, G\l{}owacki P, Meier D~M and Peik E 2018 {\em Phys. Rev. Lett.\/} {\bf 121}(21) 213001

\bibitem{Raeder:18}
Raeder S, Ackermann D, Backe H, Beerwerth R, Berengut J~C, Block M, Borschevsky A, Cheal B, Chhetri P, D\"ullmann C~E, Dzuba V~A, Eliav E, Even J, Ferrer R, Flambaum V~V, Fritzsche S, Giacoppo F, G\"otz S, He\ss{}berger F~P, Huyse M, Kaldor U, Kaleja O, Khuyagbaatar J, Kunz P, Laatiaoui M, Lautenschl\"ager F, Lauth W, Mistry A~K, {Minaya Ramirez} E, Nazarewicz W, Porsev S~G, Safronova M~S, Safronova U~I, Schuetrumpf B, {Van Duppen} P, Walther T, Wraith C and Yakushev A 2018 {\em Phys. Rev. Lett.\/} {\bf 120}(23) 232503

\bibitem{Cheung:21}
Cheung C, Safronova M and Porsev S 2021 {\em Symmetry\/} {\bf 13} 621

\bibitem{1997-Per}
{J{\"o}nsson} P and {Froese Fischer} C 1997 {\em Computer Physics Communications\/} {\bf 100} 81--92

\bibitem{2007-grasp}
Jönsson P, He X, {Froese Fischer} C and Grant I 2007 {\em Computer Physics Communications\/} {\bf 177} 597--622 ISSN 0010-4655 \urlprefix\url{https://www.sciencedirect.com/science/article/pii/S0010465507003001}

\bibitem{2009-MCDHF}
Biero\ifmmode~\acute{n}\else \'{n}\fi{} J, Froese~Fischer C, Indelicato P, J\"onsson P and Pyykk\"o P 2009 {\em Phys. Rev. A\/} {\bf 79}(5) 052502 \urlprefix\url{https://link.aps.org/doi/10.1103/PhysRevA.79.052502}

\bibitem{2012-LiLike}
Li J, Naz\'e C, Godefroid M, Fritzsche S, Gaigalas G, Indelicato P and J\"onsson P 2012 {\em Phys. Rev. A\/} {\bf 86}(2) 022518 \urlprefix\url{https://link.aps.org/doi/10.1103/PhysRevA.86.022518}

\bibitem{jonsson-13-is}
J{\"o}nsson P, Gaigalas G, Bieron J, Fischer C~F and Grant I~P 2013 {\em Comput.\ Phys.\ Commun.\/} {\bf 184} 2197--2203 \urlprefix\url{https://doi.org/10.1016/j.cpc.2013.02.016}

\bibitem{NAZE20132187}
Nazé C, Gaidamauskas E, Gaigalas G, Godefroid M and Jönsson P 2013 {\em Computer Physics Communications\/} {\bf 184} 2187--2196 ISSN 0010-4655 \urlprefix\url{https://www.sciencedirect.com/science/article/pii/S0010465513000726}

\bibitem{2019-grasp}
{Froese Fischer} C, Gaigalas G, Jönsson P and Bieroń J 2019 {\em Computer Physics Communications\/} {\bf 237} 184--187 ISSN 0010-4655 \urlprefix\url{https://www.sciencedirect.com/science/article/pii/S0010465518303928}

\bibitem{2005-MCDFGME}
Indelicato P and Desclaux J 2005 {\em URL http://dirac. spectro. jussieu. fr/mcdf\/} {\bf 41}

\bibitem{2007-GME}
Indelicato P, Santos J, Boucard S and Desclaux J~P 2007 {\em The European Physical Journal D\/} {\bf 45} 155--170

\bibitem{2013-PaulQED}
Indelicato P 2013 {\em Phys. Rev. A\/} {\bf 87}(2) 022501 \urlprefix\url{https://link.aps.org/doi/10.1103/PhysRevA.87.022501}

\bibitem{bohr1922difference}
Bohr N 1922 {\em Nature\/} {\bf 109} 746--746

\bibitem{1922-Ehr}
Ehrenfest P 1922 {\em Nature\/} {\bf 109} 745--745

\bibitem{1930-HE}
Hughes D~S and Eckart C 1930 {\em Phys. Rev.\/} {\bf 36}(4) 694--698 \urlprefix\url{https://link.aps.org/doi/10.1103/PhysRev.36.694}

\bibitem{Martensson-Pendrill_1992}
Martensson-Pendrill A~M and Ynnerman A 1992 {\em Journal of Physics B: Atomic, Molecular and Optical Physics\/} {\bf 25} L551 \urlprefix\url{https://dx.doi.org/10.1088/0953-4075/25/22/001}

\bibitem{bartlett}
Shavitt I and Bartlett R~J 2009

\bibitem{lindgren}
I L and J M 1986

\bibitem{Dirac}
Dirac P~A~M 1982

\bibitem{breit}
Breit G 1929 {\em Phys. Rev.\/} {\bf 34} 553

\bibitem{2022-QEDbook}
Jentschura U~D and Adkins G~S 2022 {\em Quantum Electrodynamics: Atoms, Lasers and Gravity\/} (World Scientific)

\bibitem{virial}
Clausius R 1870 {\em The London, Edinburgh, and Dublin Philosophical Magazine and Journal of Science\/} {\bf 40} 122--127

\bibitem{2008-Li}
Puchalski M and Pachucki K 2008 {\em Phys. Rev. A\/} {\bf 78}(5) 052511 \urlprefix\url{https://link.aps.org/doi/10.1103/PhysRevA.78.052511}

\bibitem{1987-Palmer}
Palmer C~W~P 1987 {\em Journal of Physics B: Atomic and Molecular Physics\/} {\bf 20} 5987 \urlprefix\url{https://dx.doi.org/10.1088/0022-3700/20/22/011}

\bibitem{shabaev1985mass}
Shabaev V~M 1985 {\em Theor. Math. Phys.\/} {\bf 63} 588

\bibitem{1988-Shab}
Shabaev V~M 1988 {\em Sov. J. Nucl. Phys. (Engl. Transl.); (United States)\/} {\bf 47:1} \urlprefix\url{https://www.osti.gov/biblio/7021630}

\bibitem{1994-Shab}
Shabaev V~M and Artemyev A~N 1994 {\em Journal of Physics B: Atomic, Molecular and Optical Physics\/} {\bf 27} 1307 \urlprefix\url{https://dx.doi.org/10.1088/0953-4075/27/7/006}

\bibitem{johnson2007atomic}
Johnson W~R 2007 {\em Atomic structure theory\/} (Springer)

\bibitem{2021-Yb}
Schelfhout J~S and McFerran J~J 2021 {\em Phys. Rev. A\/} {\bf 104}(2) 022806 \urlprefix\url{https://link.aps.org/doi/10.1103/PhysRevA.104.022806}

\bibitem{2022-CdHG}
Schelfhout J~S and McFerran J~J 2022 {\em Phys. Rev. A\/} {\bf 105}(2) 022805 \urlprefix\url{https://link.aps.org/doi/10.1103/PhysRevA.105.022805}

\bibitem{2024-Ag}
Ohayon B, Padilla-Castillo J~E, Wright S~C, Meijer G and Sahoo B~K 2024 {\em Phys. Rev. Res.\/} {\bf 6}(3) 033040 \urlprefix\url{https://link.aps.org/doi/10.1103/PhysRevResearch.6.033040}

\bibitem{visscher:1997dirac}
Visscher L and Dyall K~G 1997 {\em At. Data Nucl. Data Tables\/} {\bf 67} 207--224

\bibitem{skripnikov2024isotopeQED}
Skripnikov L~V, Prosnyak S~D, Malyshev A~V, Athanasakis-Kaklamanakis M, Brinson A~J, Minamisono K, Cruz F~C~P, Reilly J~R, Rickey B~J and Ruiz R~F~G 2024 {\em Phys. Rev. A\/} {\bf 110}(1) 012807

\bibitem{1985-Fermi}
Estévez G and Bhuiyan L~B 1985 {\em American Journal of Physics\/} {\bf 53} 450--453 ISSN 0002-9505 \urlprefix\url{https://doi.org/10.1119/1.14198}

\bibitem{1985-R}
Johnson W and Soff G 1985 {\em Atomic Data and Nuclear Data Tables\/} {\bf 33} 405--446 ISSN 0092-640X \urlprefix\url{https://www.sciencedirect.com/science/article/pii/0092640X85900105}

\bibitem{1969-Seltzer}
SELTZER E~C 1969 {\em Phys. Rev.\/} {\bf 188}(4) 1916--1919 \urlprefix\url{https://link.aps.org/doi/10.1103/PhysRev.188.1916}

\bibitem{1987-Blund}
Blundell S~A, Baird P~E~G, Palmer C~W~P, Stacey D~N and Woodgate G~K 1987 {\em Journal of Physics B: Atomic and Molecular Physics\/} {\bf 20} 3663 \urlprefix\url{https://dx.doi.org/10.1088/0022-3700/20/15/015}

\bibitem{Pachucki-3P}
Pachucki K, Patk\'o\ifmmode~\check{s}\else \v{s}\fi{} V~c~v and Yerokhin V~A 2018 {\em Phys. Rev. A\/} {\bf 97}(6) 062511 \urlprefix\url{https://link.aps.org/doi/10.1103/PhysRevA.97.062511}

\bibitem{1979-Friar}
Friar J 1979 {\em Annals of Physics\/} {\bf 122} 151--196 ISSN 0003-4916 \urlprefix\url{https://www.sciencedirect.com/science/article/pii/0003491679903002}

\bibitem{1993-Shab}
Shabaev V~M 1993 {\em Journal of Physics B: Atomic, Molecular and Optical Physics\/} {\bf 26} 1103 \urlprefix\url{https://dx.doi.org/10.1088/0953-4075/26/6/011}

\bibitem{sapirstein-87-qed}
Sapirstein J 1987 {\em Phys.\ Scr.\/} {\bf 36} 801--808 \urlprefix\url{https://doi.org/10.1088/0031-8949/36/5/007}

\bibitem{Shabaev:2024:94:inbook}
Shabaev V~M 2024 Quantum electrodynamics effects in atoms and molecules {\em Comprehensive Computational Chemistry (First Edition)\/} ed Y{\'a}{\~n}ez M and Boyd R~J (Oxford: Elsevier) pp 94--128 ISBN 978-0-12-823256-9

\bibitem{Uehl}
Uehling E~A 1935 {\em Phys. Rev.\/} {\bf 48} 55

\bibitem{Shabaev:13}
Shabaev V~M, Tupitsyn I~I and Yerokhin V~A 2013 {\em Phys. Rev. A\/} {\bf 88}(1) 012513

\bibitem{Skripnikov:2021a}
Skripnikov L~V 2021 {\em J. Chem. Phys.\/} {\bf 154} 201101

\bibitem{Milstein:2003}
Milstein A~I, Sushkov O~P and Terekhov I~S 2003 {\em Phys. Rev. A\/} {\bf 67}(6) 062111

\bibitem{Milstein:2004}
Milstein A~I, Sushkov O~P and Terekhov I~S 2004 {\em Phys. Rev. A\/} {\bf 69}(2) 022114

\bibitem{Yerokhin:2011b}
Yerokhin V~A 2011 {\em Phys. Rev. A\/} {\bf 83}(1) 012507

\bibitem{Pachucki:1993}
Pachucki K 1993 {\em Phys. Rev. A\/} {\bf 48}(1) 120--128

\bibitem{Milstein:2002}
Milstein A~I, Sushkov O~P and Terekhov I~S 2002 {\em Phys. Rev. Lett.\/} {\bf 89}(28) 283003

\bibitem{Eides:1997}
Eides M~I and Grotch H 1997 {\em Phys. Rev. A\/} {\bf 56}(4) R2507--R2509

\bibitem{Shabaev:02a}
Shabaev V~M 2002 {\em Phys.\ Rep.\/} {\bf 356} 119--228

\bibitem{Anisimova:2022}
Anisimova I~S, Malyshev A~V, Glazov D~A, Kaygorodov M~Y, Kozhedub Y~S, Plunien G and Shabaev V~M 2022 {\em Phys. Rev. A\/} {\bf 106}(6) 062823

\bibitem{King:2022:43}
King S~A, Spie{\ss} L~J, Micke P, Wilzewski A, Leopold T, Benkler E, Lange R, Huntemann N, Surzhykov A, Yerokhin V~A, {Crespo L{\'o}pez-Urrutia} J~R and Schmidt P~O 2022 {\em Nature\/} {\bf 611} 43--47 ISSN 1476-4687

\bibitem{2019-Paul}
Indelicato P 2019 {\em Journal of Physics B: Atomic, Molecular and Optical Physics\/} {\bf 52} 232001 \urlprefix\url{https://doi.org/10.1088/1361-6455/ab42c9}

\bibitem{blundell-93-is}
Blundell S~A 1993 {\em Phys.\ Rev.\ A\/} {\bf 47} 1790--1803 \urlprefix\url{https://doi.org/10.1103/PhysRevA.47.1790}

\bibitem{sapirstein-15-is}
Sapirstein J and Cheng K~T 2015 {\em Phys.\ Rev.\ A\/} {\bf 91} 062508 \urlprefix\url{https://doi.org/10.1103/PhysRevA.91.062508}

\bibitem{mohr-74-qed}
Mohr P~J 1974 {\em Ann.\ Phys.\/} {\bf 88} 26--51 \urlprefix\url{http://dx.doi.org/10.1016/0003-4916(74)90398-4}

\bibitem{blundell-09-qed}
Blundell S~A 2009 {\em Can.\ J.\ Phys.\/} {\bf 87} 55--65 \urlprefix\url{http://dx.doi.org/10.1139/P08-065}

\bibitem{blundell-93-qed}
Blundell S~A, Mohr P~J, Johnson W~R and Sapirstein J 1993 {\em Phys.\ Rev.\ A\/} {\bf 48} 2615--2626 \urlprefix\url{http://dx.doi.org/10.1103/PhysRevA.48.2615}

\bibitem{1995-Shab}
Artemyev A~N, Shabaev V~M and Yerokhin V~A 1995 {\em Phys. Rev. A\/} {\bf 52}(3) 1884--1894 \urlprefix\url{https://link.aps.org/doi/10.1103/PhysRevA.52.1884}

\bibitem{ShabaevRecoil:98}
Shabaev V~M 1998 {\em Phys. Rev. A\/} {\bf 57}(1) 59--67

\bibitem{2007-Adkins}
Adkins G~S, Morrison S and Sapirstein J 2007 {\em Phys. Rev. A\/} {\bf 76}(4) 042508 \urlprefix\url{https://link.aps.org/doi/10.1103/PhysRevA.76.042508}

\bibitem{zubova-14-is}
Zubova N~A, Kozhedub Y~S, Shabaev V~M, Tupitsyn I~I, Volotka A~V, Plunien G, Brandau C and St{\"o}hlker T 2014 {\em Phys.\ Rev.\ A\/} {\bf 90} 062512 \urlprefix\url{https://doi.org/10.1103/PhysRevA.90.062512}

\bibitem{prasana}
Abe M, Prasannaa V~S and Das B~P 2018 {\em Phys. Rev. A\/} {\bf 97}(3) 032515 \urlprefix\url{https://link.aps.org/doi/10.1103/PhysRevA.97.032515}

\bibitem{2022-HCIclock}
King S~A, Spie{\ss} L~J, Micke P, Wilzewski A, Leopold T, Benkler E, Lange R, Huntemann N, Surzhykov A, Yerokhin V~A {\em et~al.\/} 2022 {\em Nature\/} {\bf 611} 43--47

\bibitem{silwal-18-is}
Silwal R, Lapierre A, Gillaspy J~D, Dreiling J~M, Blundell S~A, Dipti, Borovik A, Gwinner G, Villari A~C~C, Ralchenko Y and Takacs E 2018 {\em Phys.\ Rev.\ A\/} {\bf 98} 052502 \urlprefix\url{https://doi.org/10.1103/PhysRevA.98.052502}

\bibitem{johnson-88-mbpt}
Johnson W~R, Blundell S~A and Sapirstein J 1988 {\em Phys.\ Rev.\ A\/} {\bf 38} 2699--2706 \urlprefix\url{http://dx.doi.org/10.1103/PhysRevA.38.2699}

\bibitem{2007-Bartlett}
Bartlett R~J and Musia\l{} M 2007 {\em Rev. Mod. Phys.\/} {\bf 79}(1) 291--352 \urlprefix\url{https://link.aps.org/doi/10.1103/RevModPhys.79.291}

\bibitem{Cizek}
\v{C}\'{i}\v{z}ek J 1969 On the use of the cluster expansion and the technique of diagrams in calculations of correlation effects in atoms and molecules \urlprefix\url{http://dx.doi.org/10.1002/9780470143599.ch2}

\bibitem{crawford}
Crawford T 2000 {HF Schaefer III in Reviews in Computational Chemistry, Vol. 14, KB Lipkowitz, DB Boyd, Eds}

\bibitem{bishopbook}
Bishop R~F 1998   1

\bibitem{arponen}
Arponen J~S 1983 {\em Ann. Phys.\/} {\bf 151} 311

\bibitem{Visscher:96a}
Visscher L, Lee T~J and Dyall K~G 1996 {\em J. Chem. Phys.\/} {\bf 105} 8769--8776

\bibitem{Noga:87}
Noga J and Bartlett R~J 1987 {\em J. Chem. Phys.\/} {\bf 86} 7041–7050

\bibitem{Raghavachari:89}
Raghavachari K, Trucks G~W, Pople J~A and {Head-Gordon} M 1989 {\em Chem. Phys. Lett.\/} {\bf 157} 479--483

\bibitem{Bomble:05}
Bomble Y~J, Stanton J~F, K{\'{a}}llay M and Gauss J 2005 {\em J. Chem. Phys.\/} {\bf 123} 054101

\bibitem{Kallay:6}
K\'{a}llay M and Gauss J 2005 {\em J. Chem. Phys.\/} {\bf 123} 214105

\bibitem{EXPT_website}
Oleynichenko A, Zaitsevskii A and Eliav E 2023 {EXP-T}, an extensible code for {F}ock space relativistic coupled cluster calculations (see \url{http://www.qchem.pnpi.spb.ru/expt})

\bibitem{Oleynichenko_EXPT}
Oleynichenko A~V, Zaitsevskii A and Eliav E 2020 Towards high performance relativistic electronic structure modelling: the {EXP-T} program package {\em Supercomputing\/} vol 1331 ed Voevodin V and Sobolev S (Cham: Springer International Publishing) pp 375--386

\bibitem{DIRAC19}
{DIRAC}, a relativistic ab initio electronic structure program, Release {DIRAC19} (2019), written by A.~S.~P.~Gomes, T.~Saue, L.~Visscher, H.~J.~{\relax Aa}.~Jensen, and R.~Bast, with contributions from I.~A.~Aucar, V.~Bakken, K.~G.~Dyall, S.~Dubillard, U.~Ekstr{\"o}m, E.~Eliav, T.~Enevoldsen, E.~Fa{\ss}hauer, T.~Fleig, O.~Fossgaard, L.~Halbert, E.~D.~Hedeg{\aa}rd, B.~Heimlich--Paris, T.~Helgaker, J.~Henriksson, M.~Ilia{\v{s}}, Ch.~R.~Jacob, S.~Knecht, S.~Komorovsk{\'y}, O.~Kullie, J.~K.~L{\ae}rdahl, C.~V.~Larsen, Y.~S.~Lee, H.~S.~Nataraj, M.~K.~Nayak, P.~Norman, G.~Olejniczak, J.~Olsen, J.~M.~H.~Olsen, Y.~C.~Park, J.~K.~Pedersen, M.~Pernpointner, R.~di~Remigio, K.~Ruud, P.~Sa{\l}ek, B.~Schimmelpfennig, B.~Senjean, A.~Shee, J.~Sikkema, A.~J.~Thorvaldsen, J.~Thyssen, J.~van~Stralen, M.~L.~Vidal, S.~Villaume, O.~Visser, T.~Winther, and S.~Yamamoto (available at \url{http://dx.doi.org/10.5281/zenodo.3572669}, see also \url{http://www.diracprogram.org})

\bibitem{Saue:2020}
Saue T, Bast R, Gomes A~S~P, Jensen H~J~A, Visscher L, Aucar I~A, {Di Remigio} R, Dyall K~G, Eliav E, Fasshauer E, Fleig T, Halbert L, Hedegard E~D, {Helmich-Paris} B, Ilias M, Jacob C~R, Knecht S, Laerdahl J~K, Vidal M~L, Nayak M~K, Olejniczak M, Olsen J~M~H, Pernpointner M, Senjean B, Shee A, Sunaga A and {van Stralen} J~N~P 2020 {\em J. Chem. Phys.\/} {\bf 152} 204104

\bibitem{MRCC2020}
 {{\sc mrcc}} m. K\'{a}llay, P. R. Nagy, D. Mester, Z. Rolik, G. Samu, J. Csontos, J. Cs\'{o}ka, P. B. Szab\'{o}, L. Gyevi-Nagy, B. H\'{e}gely, I. Ladj\'{a}nszki, L. Szegedy, B. Lad\'{o}czki, K. Petrov, M. Farkas, P. D. Mezei, and \'{a}. Ganyecz: The {\sc mrcc} program system: Accurate quantum chemistry from water to proteins, J. Chem. Phys. 152, 074107 (2020); {\sc mrcc}, a quantum chemical program suite written by M. K\'{a}llay, P. R. Nagy, D. Mester, Z. Rolik, G. Samu, J. Csontos, J. Cs\'{o}ka, P. B. Szab\'{o}, L. Gyevi-Nagy, B. H\'{e}gely, I. Ladj\'{a}nszki, L. Szegedy, B. Lad\'{o}czki, K. Petrov, M. Farkas, P. D. Mezei, and \'{A}. Ganyecz. See \url{www.mrcc.hu}.

\bibitem{Kallay:1}
K\'{a}llay M and Surj\'{a}n P~R 2001 {\em J. Chem. Phys.\/} {\bf 115} 2945--2954

\bibitem{Kallay:2}
K\'{a}llay M, Szalay P~G and Surj\'{a}n P~R 2002 {\em J. Chem. Phys.\/} {\bf 117} 980--990

\bibitem{Penyazkov:2023}
Penyazkov G, Prosnyak S~D, Barzakh A~E and Skripnikov L~V 2023 {\em J. Chem. Phys.\/} {\bf 158} 114110 (\textit{Preprint} \eprint{https://doi.org/10.1063/5.0142202})

\bibitem{Kallay:3}
K\'{a}llay M, Gauss J and Szalay P~G 2003 {\em J. Chem. Phys.\/} {\bf 119} 2991--3004

\bibitem{Kaldor1991}
Kaldor U 1991 {\em Theor. Chim. Acta\/} {\bf 80} 427--439 ISSN 0040-5744

\bibitem{Eliav1998}
Eliav E, Kaldor U and Hess B~A 1998 {\em J. Chem. Phys\/} {\bf 108} 3409--3415 ISSN 0021-9606

\bibitem{Visscher:01}
Visscher L, Eliav E and Kaldor U 2001 {\em J. Chem. Phys.\/} {\bf 115} 9720–9726

\bibitem{Eliav:Review:22}
Eliav E, Borschevsky A, Zaitsevskii A, Oleynichenko A~V and Kaldor U 2022 Relativistic {F}ock-space coupled cluster method: theory and recent applications {\em Reference Module in Chemistry, Molecular Sciences and Chemical Engineering\/} (Elsevier) ISBN 978-0-12-409547-2

\bibitem{Kaldor:98}
Kaldor U and Eliav E 1998 {\em High-accuracy calculations for heavy and super-heavy elements\/} (Elsevier) p 313–336

\bibitem{Eliav:15}
Eliav E, Fritzsche S and Kaldor U 2015 {\em Nucl. Phys. A\/} {\bf 944} 518–550

\bibitem{Oleynichenko:20}
Oleynichenko A~V, Zaitsevskii A, Skripnikov L~V and Eliav E 2020 {\em {S}ymmetry\/} {\bf 12} 1101

\bibitem{mukherjee1989use}
Mukherjee D and Pal S 1989  {\bf 20} 291--373

\bibitem{Lindgren:78}
Lindgren I 1978 {\em Int. J. Quantum Chem.\/} {\bf 14} 33–58 \urlprefix\url{http://dx.doi.org/10.1002/qua.560140804}

\bibitem{Lindgren:87}
Lindgren I and Mukherjee D 1987 {\em Phys. Rep.\/} {\bf 151} 93–127 \urlprefix\url{http://dx.doi.org/10.1016/0370-1573(87)90073-1}

\bibitem{Mani:11}
Mani B~K and Angom D 2011 {\em Phys. Rev. A\/} {\bf 83}(1) 012501 \urlprefix\url{https://link.aps.org/doi/10.1103/PhysRevA.83.012501}

\bibitem{Forsberg:97}
Forsberg N and Malmqvist P~A 1997 {\em Chem. Phys. Lett.\/} {\bf 274} 196–204

\bibitem{Eliav:IH:05}
Eliav E, Vilkas M~J, Ishikawa Y and Kaldor U 2005 {\em J.~Chem.~Phys.\/} {\bf 122}

\bibitem{Zaitsevskii:RbCs:17}
Zaitsevskii A, Mosyagin N~S, Stolyarov A~V and Eliav E 2017 {\em Phys.~Rev.~A\/} {\bf 96}(2) 022516

\bibitem{Oleynichenko:HFS:20}
Oleynichenko A~V, Skripnikov L~V, Zaitsevskii A, Eliav E and Shabaev V~M 2020 {\em Chem. Phys. Lett.\/} {\bf 756} 137825

\bibitem{Zaitsevskii:Pade:18}
Zaitsevskii A and Eliav E 2018 {\em Int. J. Quantum Chem.\/} {\bf 118} e25772

\bibitem{Malrieu:IH:85}
Malrieu J~P, Durand P and Daudey J~P 1985 {\em J. Phys. A Math. Theor.\/} {\bf 18} 809–826

\bibitem{Meissner:98}
Meissner L 1998 {\em J.~Chem.~Phys.\/} {\bf 108} 9227–9235

\bibitem{Landau:IH:01}
Landau A, Eliav E and Kaldor U 2001 {\em Intermediate {H}amiltonian {F}ock-space coupled-cluster method\/} p 171–188

\bibitem{Musial:IH:08}
Musia\l{} M and Bartlett R~J 2008 {\em J.~Chem.~Phys.\/} {\bf 129} 044101

\bibitem{Dutta:IH:14}
Dutta A~K, Gupta J, Vaval N and Pal S 2014 {\em J. Chem. Theory Comput.\/} {\bf 10} 3656–3668

\bibitem{Zaitsevskii:QED:22}
Zaitsevskii A, Mosyagin N~S, Oleynichenko A~V and Eliav E 2022 {\em Int. J. Quantum Chem.\/} {\bf 123} e27077

\bibitem{Skripnikov:24}
Skripnikov L~V and Oleynichenko A~V 2024

\bibitem{Eliav:94}
Eliav E, Kaldor U and Ishikawa Y 1994 {\em Phys.~Rev.~A\/} {\bf 49} 1724–1729

\bibitem{SkripnikovBiCCSDT:2021}
Skripnikov L~V, Oleynichenko A~V, Zaitsevskii A~V, Maison D~E and Barzakh A~E 2021 {\em Phys.~Rev.~C\/} {\bf 104}(3) 034316

\bibitem{Szalay:95}
Szalay P~G 1995 {\em Int. J. Quantum Chem.\/} {\bf 55} 151–163

\bibitem{Shamasundar:04}
Shamasundar K~R, Asokan S and Pal S 2004 {\em J.~Chem.~Phys.\/} {\bf 120} 6381–6398

\bibitem{Zaitsevskii:ThO:23}
Zaitsevskii A, Oleynichenko A~V and Eliav E 2023 {\em Mol. Phys.\/}  e2236246

\bibitem{Oleynichenko:Optics:23}
Oleynichenko A~V, Zaitsevskii A~V, Kondratyev S~V and Eliav E 2023 {\em Optika i Spektroskopiya\/} {\bf 131}(11) 1549--1555

\bibitem{Gopakumar:02}
Gopakumar G, Merlitz H, Chaudhuri R~K, Das B~P, Mahapatra U~S and Mukherjee D 2002 {\em Phys.~Rev.~A\/} {\bf 66} 032505

\bibitem{mukherjee}
Haque M and Mukherjee D 1984 {\em J. Chem. Phys.\/} {\bf 80} 5058

\bibitem{SahooLi}
Sahoo B~K and Ohayon B 2021 {\em Phys. Rev. A\/} {\bf 103}(5) 052802 \urlprefix\url{https://link.aps.org/doi/10.1103/PhysRevA.103.052802}

\bibitem{Sahoo_2020}
Sahoo B~K, Vernon A~R, Ruiz R~F~G, Binnersley C~L, Billowes J, Bissell M~L, Cocolios T~E, Farooq-Smith G~J, Flanagan K~T, Gins W, de~Groote R~P, Koszor\'us A, Neyens G, Lynch K~M, Parnefjord-Gustafsson F, Ricketts C~M, Wendt K~D~A, Wilkins S~G and Yang X~F 2020 {\em New Journal of Physics\/} {\bf 22} 012001 \urlprefix\url{https://dx.doi.org/10.1088/1367-2630/ab66dd}

\bibitem{Monkhorst}
Monkhorst H~J 2009 {\em Int. J. Quantum Chem.\/} {\bf 12} 421–432

\bibitem{SahooCa}
Sahoo B~K 2019 {\em Phys. Rev. A\/} {\bf 99}(5) 050501 \urlprefix\url{https://link.aps.org/doi/10.1103/PhysRevA.99.050501}

\bibitem{SahooCd}
Sahoo B~K and Yu Y~m 2018 {\em Phys. Rev. A\/} {\bf 98}(1) 012513 \urlprefix\url{https://link.aps.org/doi/10.1103/PhysRevA.98.012513}

\bibitem{orts-06-is}
Soria~Orts R, Harman Z, Crespo L{\'o}pez-Urrutia J~R, Artemyev A~N, Bruhns H, Gonz{\'a}lez~Mart{\'i}nez A~J, Jentschura U~D, Keitel C~H, Lapierre A, Mironov V, Shabaev V~M, Tawara H, Tupitsyn I~I, Ullrich J and Volotka A~V 2006 {\em Phys.\ Rev.\ Lett.\/} {\bf 97} 103002 \urlprefix\url{https://doi.org/10.1103/PhysRevLett.97.103002}

\bibitem{schuch-05-is}
Schuch R, Lindroth E, Madzunkov S, Fogle M, Mohamed T and Indelicato P 2005 {\em Phys.\ Rev.\ Lett.\/} {\bf 95} 183003 \urlprefix\url{https://doi.org/10.1103/PhysRevLett.95.183003}

\bibitem{brandau-08-is}
Brandau C, Kozhuharov C, Harman Z, M{\"u}ller A, Schippers S, Kozhedub Y~S, Bernhardt D, Boehm S, Jacobi J, Schmidt E~W, Mokler P~H, Bosch F, Kluge H~J, St{\"o}hlker T, Beckert K, Beller P, Nolden F, Steck M, Gumberidze A, Reuschl R, Spillmann U, Currell F~J, Tupitsyn I~I, Shabaev V~M, Jentschura U~D, Keitel C~H, Wolf A and Stachura Z 2008 {\em Phys.\ Rev.\ Lett.\/} {\bf 100} 073201 \urlprefix\url{https://doi.org/10.1103/PhysRevLett.100.073201}

\bibitem{elliott-98-is}
Elliott S~R, Beiersdorfer P, Chen M~H, Decaux V and Knapp D~A 1998 {\em Phys.\ Rev.\ C\/} {\bf 57} 583--589 \urlprefix\url{https://doi.org/10.1103/PhysRevC.57.583}

\bibitem{erickson-65-is}
Erickson G~W and Yennie D~R 1965 {\em Ann. Phys.\/} {\bf 35} 271--313 \urlprefix\url{https://doi.org/10.1016/0003-4916(65)90081-3}

\bibitem{2007-LiE}
Bushaw B~A, N\"ortersh\"auser W, Drake G~W~F and Kluge H~J 2007 {\em Phys. Rev. A\/} {\bf 75}(5) 052503 \urlprefix\url{https://link.aps.org/doi/10.1103/PhysRevA.75.052503}

\bibitem{Kramida_2005}
Kramida A~E 2005 {\em Physica Scripta\/} {\bf 72} 309--319 \urlprefix\url{https://doi.org/10.1238/physica.regular.072a00309}

\bibitem{2010-Saloman}
Saloman E~B 2010 {\em Journal of Physical and Chemical Reference Data\/} {\bf 39} 033101 (\textit{Preprint} \eprint{https://doi.org/10.1063/1.3337661}) \urlprefix\url{https://doi.org/10.1063/1.3337661}

\bibitem{2001-Saf}
Safronova M~S and Johnson W~R 2001 {\em Phys. Rev. A\/} {\bf 64}(5) 052501

\bibitem{Sahoo_2010}
Sahoo B~K 2010 {\em Journal of Physics B: Atomic, Molecular and Optical Physics\/} {\bf 43} 231001 \urlprefix\url{https://dx.doi.org/10.1088/0953-4075/43/23/231001}

\bibitem{berengut2003}
Berengut J~C, Dzuba V~A and Flambaum V~V 2003 {\em Phys. Rev. A\/} {\bf 68}(2) 022502 \urlprefix\url{https://link.aps.org/doi/10.1103/PhysRevA.68.022502}

\bibitem{2015-Roy}
Roy S and Majumder S 2015 {\em Phys. Rev. A\/} {\bf 92}(1) 012508 \urlprefix\url{https://link.aps.org/doi/10.1103/PhysRevA.92.012508}

\bibitem{2022-Na}
Ohayon B, Ruiz R, Sun Z~H, Hagen G, Papenbrock T and Sahoo B~K 2022 {\em Phys. Rev. C\/} {\bf 105}(3) L031305 \urlprefix\url{https://link.aps.org/doi/10.1103/PhysRevC.105.L031305}

\bibitem{2023-Zn}
Sahoo B~K and Ohayon B 2023 {\em Phys. Rev. Res.\/} {\bf 5}(4) 043142 \urlprefix\url{https://link.aps.org/doi/10.1103/PhysRevResearch.5.043142}

\bibitem{2022-Cd}
Ohayon B, Hofsäss S, Padilla-Castillo J~E, Wright S~C, Meijer G, Truppe S, Gibble K and Sahoo B~K 2022 {\em New Journal of Physics\/} {\bf 24} 123040 \urlprefix\url{https://dx.doi.org/10.1088/1367-2630/acacbb}

\bibitem{2016-AlI}
Filippin L, Beerwerth R, Ekman J, Fritzsche S, Godefroid M and J\"onsson P 2016 {\em Phys. Rev. A\/} {\bf 94}(6) 062508 \urlprefix\url{https://link.aps.org/doi/10.1103/PhysRevA.94.062508}

\bibitem{2021-Al}
Heylen H, Devlin C~S, Gins W, Bissell M~L, Blaum K, Cheal B, Filippin L, Ruiz R~F~G, Godefroid M, Gorges C, Holt J~D, Kanellakopoulos A, Kaufmann S, Koszor\'us A, K\"onig K, Malbrunot-Ettenauer S, Miyagi T, Neugart R, Neyens G, N\"ortersh\"auser W, S\'anchez R, Sommer F, Rodr\'{\i}guez L~V, Xie L, Xu Z~Y, Yang X~F and Yordanov D~T 2021 {\em Phys. Rev. C\/} {\bf 103}(1) 014318 \urlprefix\url{https://link.aps.org/doi/10.1103/PhysRevC.103.014318}

\bibitem{Cubiss:Au:23}
Cubiss J~G, Andreyev A~N, Barzakh A~E, {Van Duppen} P, Hilaire S, P\'eru S, Goriely S, {Al Monthery} M, Althubiti N~A, Andel B, Antalic S, Atanasov D, Blaum K, Cocolios T~E, {Day Goodacre} T, {de Roubin} A, {Farooq-Smith} G~J, Fedorov D~V, Fedosseev V~N, Fink D~A, Gaffney L~P, Ghys L, Harding R~D, Huyse M, Imai N, Joss D~T, Kreim S, Lunney D, Lynch K~M, Manea V, Marsh B~A, {Martinez Palenzuela} Y, Molkanov P~L, Neidherr D, {O'Neill} G~G, Page R~D, Prosnyak S~D, Rosenbusch M, Rossel R~E, Rothe S, Schweikhard L, Seliverstov M~D, Sels S, Skripnikov L~V, Stott A, {Van Beveren} C, Verstraelen E, Welker A, Wienholtz F, Wolf R~N and Zuber K 2023 {\em Phys. Rev. Lett.\/} {\bf 131}(20) 202501

\bibitem{Skripnikov:17a}
Skripnikov L~V, Maison D~E and Mosyagin N~S 2017 {\em Phys. Rev. A\/} {\bf 95}(2) 022507

\bibitem{maa19}
Maa{\ss} B, H{\"u}ther T, K{\"o}nig K, Kr{\"a}mer J, Krause J, Lovato A, M{\"u}ller P, Pachucki K, Puchalski M, Roth R {\em et~al.\/} 2019 {\em Physical Review Letters\/} {\bf 122} 182501

\bibitem{Img23}
Imgram P, K\"onig K, Maa\ss{} B, M\"uller P and N\"ortersh\"auser W 2023 {\em Phys. Rev. Lett.\/} {\bf 131}(24) 243001 \urlprefix\url{https://link.aps.org/doi/10.1103/PhysRevLett.131.243001}

\bibitem{Gar16a}
{Garcia Ruiz, RF \it{et al}} 2016 \urlprefix\url{https://cds.cern.ch/record/2157183/files/INTC-I-171.pdf}

\bibitem{2020-MIRACLES}
Vil{\'e}n M and Malbrunot-Ettenauer S 2020 Miracls at isolde: The charge radii of exotic magnesium isotopes Tech. rep.

\bibitem{Pla23}
Plattner P, Wood E, Al~Ayoubi L, Beliuskina O, Bissell M~L, Blaum K, Campbell P, Cheal B, de~Groote R~P, Devlin C~S, Eronen T, Filippin L, Garcia~Ruiz R~F, Ge Z, Geldhof S, Gins W, Godefroid M, Heylen H, Hukkanen M, Imgram P, Jaries A, Jokinen A, Kanellakopoulos A, Kankainen A, Kaufmann S, K\"onig K, Koszor\'us A, Kujanp\"a\"a S, Lechner S, Malbrunot-Ettenauer S, M\"uller P, Mathieson R, Moore I, N\"ortersh\"auser W, Nesterenko D, Neugart R, Neyens G, Ortiz-Cortes A, Penttil\"a H, Pohjalainen I, Raggio A, Reponen M, Rinta-Antila S, Rodr\'{\i}guez L~V, Romero J, S\'anchez R, Sommer F, Stryjczyk M, Virtanen V, Xie L, Xu Z~Y, Yang X~F and Yordanov D~T 2023 {\em Phys. Rev. Lett.\/} {\bf 131}(22) 222502 \urlprefix\url{https://link.aps.org/doi/10.1103/PhysRevLett.131.222502}

\bibitem{Kre14}
Kreim K, Bissell M, Papuga J, Blaum K, {De Rydt} M, {Garcia Ruiz} R, Goriely S, Heylen H, Kowalska M, Neugart R, Neyens G, Nörtershäuser W, Rajabali M, {Sánchez Alarcón} R, Stroke H and Yordanov D 2014 {\em Physics Letters B\/} {\bf 731} 97--102 ISSN 0370-2693 \urlprefix\url{https://www.sciencedirect.com/science/article/pii/S0370269314001038}

\bibitem{2021-K}
Koszor{\'u}s {\'A}, Yang X, Jiang W, Novario S, Bai S, Billowes J, Binnersley C, Bissell M, Cocolios T~E, Cooper B {\em et~al.\/} 2021 {\em Nature Physics\/} {\bf 17} 439--443

\bibitem{Kos23}
Koszor\'us A, Block M, Campbell P, Cheal B, de~Groote R~P, Gins W, Moore I~D, Ortiz-Cortes A, Raggio A and Warbinek J 2023 {\em Sci. Rep.\/} {\bf 13} 4783

\bibitem{Mal22}
Malbrunot-Ettenauer S, Kaufmann S, Bacca S, Barbieri C, Billowes J, Bissell M~L, Blaum K, Cheal B, Duguet T, Ruiz R~F~G, Gins W, Gorges C, Hagen G, Heylen H, Holt J~D, Jansen G~R, Kanellakopoulos A, Kortelainen M, Miyagi T, Navr\'atil P, Nazarewicz W, Neugart R, Neyens G, N\"ortersh\"auser W, Novario S~J, Papenbrock T, Ratajczyk T, Reinhard P~G, Rodr\'{\i}guez L~V, S\'anchez R, Sailer S, Schwenk A, Simonis J, Som\`a V, Stroberg S~R, Wehner L, Wraith C, Xie L, Xu Z~Y, Yang X~F and Yordanov D~T 2022 {\em Phys. Rev. Lett.\/} {\bf 128}(2) 022502 \urlprefix\url{https://link.aps.org/doi/10.1103/PhysRevLett.128.022502}

\bibitem{2024-wang}
Wang S {\em et~al.\/} 2024 {\em arXiv\/} (\textit{Preprint} \eprint{2404.06046}) \urlprefix\url{https://arxiv.org/abs/2404.06046}

\bibitem{Hey16}
Heylen H, Babcock C, Beerwerth R, Billowes J, Bissell M~L, Blaum K, Bonnard J, Campbell P, Cheal B, Day~Goodacre T, Fedorov D, Fritzsche S, Garcia~Ruiz R~F, Geithner W, Geppert C, Gins W, Grob L~K, Kowalska M, Kreim K, Lenzi S~M, Moore I~D, Maass B, Malbrunot-Ettenauer S, Marsh B, Neugart R, Neyens G, N\"ortersh\"auser W, Otsuka T, Papuga J, Rossel R, Rothe S, S\'anchez R, Tsunoda Y, Wraith C, Xie L, Yang X~F and Yordanov D~T 2016 {\em Phys. Rev. C\/} {\bf 94}(5) 054321 \urlprefix\url{https://link.aps.org/doi/10.1103/PhysRevC.94.054321}

\bibitem{Roca-Maza.2018}
Roca-Maza X and Paar N 2018 {\em Prog. Part. Nucl. Phys.\/} {\bf 101} 96--176 ISSN 0146-6410 \urlprefix\url{https://www.sciencedirect.com/science/article/pii/S0146641018300334}

\bibitem{Adh21}
Adhikari D, Albataineh H, Androic D, Aniol K, Armstrong D~S, Averett T, Ayerbe~Gayoso C, Barcus S, Bellini V, Beminiwattha R~S, Benesch J~F, Bhatt H, Bhatta~Pathak D, Bhetuwal D, Blaikie B, Campagna Q, Camsonne A, Cates G~D, Chen Y, Clarke C, Cornejo J~C, Covrig~Dusa S, Datta P, Deshpande A, Dutta D, Feldman C, Fuchey E, Gal C, Gaskell D, Gautam T, Gericke M, Ghosh C, Halilovic I, Hansen J~O, Hauenstein F, Henry W, Horowitz C~J, Jantzi C, Jian S, Johnston S, Jones D~C, Karki B, Katugampola S, Keppel C, King P~M, King D~E, Knauss M, Kumar K~S, Kutz T, Lashley-Colthirst N, Leverick G, Liu H, Liyange N, Malace S, Mammei R, Mammei J, McCaughan M, McNulty D, Meekins D, Metts C, Michaels R, Mondal M~M, Napolitano J, Narayan A, Nikolaev D, Rashad M~N~H, Owen V, Palatchi C, Pan J, Pandey B, Park S, Paschke K~D, Petrusky M, Pitt M~L, Premathilake S, Puckett A~J~R, Quinn B, Radloff R, Rahman S, Rathnayake A, Reed B~T, Reimer P~E, Richards R, Riordan S, Roblin Y, Seeds S, Shahinyan A, Souder P, Tang L, Thiel M, Tian Y,
  Urciuoli G~M, Wertz E~W, Wojtsekhowski B, Yale B, Ye T, Yoon A, Zec A, Zhang W, Zhang J and Zheng X (PREX Collaboration) 2021 {\em Phys. Rev. Lett.\/} {\bf 126}(17) 172502 \urlprefix\url{https://link.aps.org/doi/10.1103/PhysRevLett.126.172502}

\bibitem{Wang.2013}
Wang N and Li T 2013 {\em Phys. Rev. C\/} {\bf 88}(1) 011301 \urlprefix\url{https://link.aps.org/doi/10.1103/PhysRevC.88.011301}

\bibitem{Brown.2017}
Brown B~A 2017 {\em Phys. Rev. Lett.\/} {\bf 119}(12) 122502 \urlprefix\url{https://link.aps.org/doi/10.1103/PhysRevLett.119.122502}

\bibitem{Yang.2018}
Yang J and Piekarewicz J 2018 {\em Phys. Rev. C\/} {\bf 97}(1) 014314 \urlprefix\url{https://link.aps.org/doi/10.1103/PhysRevC.97.014314}

\bibitem{Reinhard.2022b}
Reinhard P~G and Nazarewicz W 2022 {\em Phys. Rev. C\/} {\bf 105}(2) L021301 \urlprefix\url{https://link.aps.org/doi/10.1103/PhysRevC.105.L021301}

\bibitem{2023-weak}
Seng C~Y 2023 {\em Phys. Rev. Lett.\/} {\bf 130}(15) 152501 \urlprefix\url{https://link.aps.org/doi/10.1103/PhysRevLett.130.152501}

\bibitem{2023-EW}
Seng C~Y and Gorchtein M 2023 {\em Physics Letters B\/} {\bf 838} 137654 ISSN 0370-2693 \urlprefix\url{https://www.sciencedirect.com/science/article/pii/S0370269322007882}

\bibitem{2024-ft}
Seng C~Y and Gorchtein M 2024 {\em Phys. Rev. C\/} {\bf 109}(4) 045501 \urlprefix\url{https://link.aps.org/doi/10.1103/PhysRevC.109.045501}

\bibitem{2020-HT}
Hardy J~C and Towner I~S 2020 {\em Phys. Rev. C\/} {\bf 102}(4) 045501 \urlprefix\url{https://link.aps.org/doi/10.1103/PhysRevC.102.045501}

\bibitem{2024-Super}
Gorchtein M and Seng C~Y 2024 {\em Annual Review of Nuclear and Particle Science\/} ISSN 0163-8998 \urlprefix\url{https://www.annualreviews.org/content/journals/10.1146/annurev-nucl-102622-020726}

\bibitem{wil22}
{Wilkins, SG \it{et al}} 2022 \urlprefix\url{https://cds.cern.ch/record/2834696/files/INTC-P-646.pdf}

\bibitem{lyc23}
{Lynch, KM \it{et al}} 2023 \urlprefix\url{https://cds.cern.ch/record/2845948/files/INTC-P-657.pdf}

\bibitem{che22}
{Cheal, B \it{et al}} 2022 \urlprefix\url{https://cds.cern.ch/record/2834596/files/INTC-I-245.pdf}

\bibitem{lyc24}
{Lynch, KM \it{et al}} 2024 \urlprefix\url{https://cds.cern.ch/record/2894970/files/INTC-I-278.pdf}

\bibitem{Gaffney2013}
{Gaffney, L P \emph{et al}} 2013 {\em Nature\/} {\bf 497} 199--204 ISSN 0028-0836 \urlprefix\url{http://www.nature.com/articles/nature12073}

\bibitem{Lyn18}
Lynch K~M, Wilkins S~G, Billowes J, Binnersley C~L, Bissell M~L, Chrysalidis K, Cocolios T~E, Goodacre T~D, de~Groote R~P, Farooq-Smith G~J, Fedorov D~V, Fedosseev V~N, Flanagan K~T, Franchoo S, Garcia~Ruiz R~F, Gins W, Heinke R, Koszor\'us A, Marsh B~A, Molkanov P~L, Naubereit P, Neyens G, Ricketts C~M, Rothe S, Seiffert C, Seliverstov M~D, Stroke H~H, Studer D, Vernon A~R, Wendt K~D~A and Yang X~F 2018 {\em Phys. Rev. C\/} {\bf 97}(2) 024309 \urlprefix\url{https://link.aps.org/doi/10.1103/PhysRevC.97.024309}

\bibitem{behr2022nuclei}
Behr J 2022 {\em arXiv preprint arXiv:2203.06758\/}

\bibitem{Arr24}
Arrowsmith-Kron G, Athanasakis-Kaklamanakis M, Au M, Ballof J, Dobaczewski J~J, Garcia~Ruiz R~F, Hutzler N~R, Jayich A, Nazarewicz W and Singh J 2024 {\em Reports on Progress in Physics\/} \urlprefix\url{http://iopscience.iop.org/article/10.1088/1361-6633/ad1e39}

\bibitem{Dal23}
Dalton F, Flambaum V~V and Mansour A~J 2023 {\em Phys. Rev. C\/} {\bf 107}(3) 035502 \urlprefix\url{https://link.aps.org/doi/10.1103/PhysRevC.107.035502}

\bibitem{Wil24}
Wilkins S {\em et~al.\/} 2024 Observation of the distribution of nuclear magnetization in a molecule (\textit{Preprint} \eprint{2311.04121})

\bibitem{Butler2020}
{Butler, P A \emph{et al}} 2020 {\em Physical Review Letters\/} {\bf 124}(4) 042503 \urlprefix\url{https://link.aps.org/doi/10.1103/PhysRevLett.124.042503}

\bibitem{AHMAD1988244}
Ahmad S, Klempt W, Neugart R, Otten E, Reinhard P~G, Ulm G and Wendt K 1988 {\em Nuclear Physics A\/} {\bf 483} 244--268 ISSN 0375-9474 \urlprefix\url{https://www.sciencedirect.com/science/article/pii/0375947488905349}

\bibitem{1989-Otten}
W O~E 1989  {\bf 8} 517

\bibitem{2014-Fr}
Budin\ifmmode \check{c}\else \v{c}\fi{}evi\ifmmode~\acute{c}\else \'{c}\fi{} I, Billowes J, Bissell M~L, Cocolios T~E, de~Groote R~P, De~Schepper S, Fedosseev V~N, Flanagan K~T, Franchoo S, Garcia~Ruiz R~F, Heylen H, Lynch K~M, Marsh B~A, Neyens G, Procter T~J, Rossel R~E, Rothe S, Strashnov I, Stroke H~H and Wendt K~D~A 2014 {\em Phys. Rev. C\/} {\bf 90}(1) 014317 \urlprefix\url{https://link.aps.org/doi/10.1103/PhysRevC.90.014317}

\bibitem{2019-Ac}
Verstraelen E, Teigelh\"ofer A, Ryssens W, Ames F, Barzakh A, Bender M, Ferrer R, Goriely S, Heenen P~H, Huyse M, Kunz P, Lassen J, Manea V, Raeder S and Van~Duppen P 2019 {\em Phys. Rev. C\/} {\bf 100}(4) 044321 \urlprefix\url{https://link.aps.org/doi/10.1103/PhysRevC.100.044321}

\bibitem{bar19}
Barzakh A~E, Cubiss J~G, Andreyev A~N, Seliverstov M~D, Andel B, Antalic S, Ascher P, Atanasov D, Beck D, Biero\ifmmode~\acute{n}\else \'{n}\fi{} J, Blaum K, Borgmann C, Breitenfeldt M, Capponi L, Cocolios T~E, Day~Goodacre T, Derkx X, De~Witte H, Elseviers J, Fedorov D~V, Fedosseev V~N, Fritzsche S, Gaffney L~P, George S, Ghys L, He\ss{}berger F~P, Huyse M, Imai N, Kalaninov\'a Z, Kisler D, K\"oster U, Kowalska M, Kreim S, Lane J~F~W, Liberati V, Lunney D, Lynch K~M, Manea V, Marsh B~A, Mitsuoka S, Molkanov P~L, Nagame Y, Neidherr D, Nishio K, Ota S, Pauwels D, Popescu L, Radulov D, Rapisarda E, Revill J~P, Rosenbusch M, Rossel R~E, Rothe S, Sandhu K, Schweikhard L, Sels S, Truesdale V~L, Van~Beveren C, Van~den Bergh P, Van~Duppen P, Wakabayashi Y, Wendt K~D~A, Wienholtz F, Whitmore B~W, Wilson G~L, Wolf R~N and Zuber K 2019 {\em Phys. Rev. C\/} {\bf 99}(5) 054317 \urlprefix\url{https://link.aps.org/doi/10.1103/PhysRevC.99.054317}

\bibitem{2021-DFT}
Perera U~C, Afanasjev A~V and Ring P 2021 {\em Phys. Rev. C\/} {\bf 104}(6) 064313 \urlprefix\url{https://link.aps.org/doi/10.1103/PhysRevC.104.064313}

\bibitem{Web23}
Weber F, Albrecht-Sch\"onzart T~E, Allehabi S~O, Berndt S, Block M, Dorrer H, D\"ullmann C~E, Dzuba V~A, Ezold J~G, Flambaum V~V, Gadelshin V, Goriely S, Harvey A, Heinke R, Hilaire S, Kaja M, Kieck T, Kneip N, K\"oster U, Lantis J, Mokry C, M\"unzberg D, Nothhelfer S, Oberstedt S, P\'eru S, Raeder S, Runke J, Sonnenschein V, Stemmler M, Studer D, Th\"orle-Pospiech P, Tomita H, Trautmann N, Van~Cleve S, Warbinek J and Wendt K 2023 {\em Phys. Rev. C\/} {\bf 107}(3) 034313 \urlprefix\url{https://link.aps.org/doi/10.1103/PhysRevC.107.034313}

\bibitem{Not22}
Nothhelfer S, Albrecht-Sch\"onzart T~E, Block M, Chhetri P, D\"ullmann C~E, Ezold J~G, Gadelshin V, Gaiser A, Giacoppo F, Heinke R, Kieck T, Kneip N, Laatiaoui M, Mokry C, Raeder S, Runke J, Schneider F, Sperling J~M, Studer D, Th\"orle-Pospiech P, Trautmann N, Weber F and Wendt K 2022 {\em Phys. Rev. C\/} {\bf 105}(2) L021302 \urlprefix\url{https://link.aps.org/doi/10.1103/PhysRevC.105.L021302}

\bibitem{Blo21}
Block M, Laatiaoui M and Raeder S 2021 {\em Progress in Particle and Nuclear Physics\/} {\bf 116} 103834 ISSN 0146-6410 \urlprefix\url{https://www.sciencedirect.com/science/article/pii/S0146641020300818}

\bibitem{2024-YbKP2}
Kawasaki A, Kobayashi T, Nishiyama A, Tanabe T and Yasuda M 2024 {\em Phys. Rev. A\/} {\bf 109}(6) 062806 \urlprefix\url{https://link.aps.org/doi/10.1103/PhysRevA.109.062806}

\bibitem{All21}
Allehabi S~O, Dzuba V~A, Flambaum V~V and Afanasjev A~V 2021 {\em Phys. Rev. A\/} {\bf 103}(3) L030801 \urlprefix\url{https://link.aps.org/doi/10.1103/PhysRevA.103.L030801}

\bibitem{2022-NPOL}
Munro-Laylim P, Dzuba V~A and Flambaum V~V 2022 {\em Phys. Rev. A\/} {\bf 105}(4) 042814 \urlprefix\url{https://link.aps.org/doi/10.1103/PhysRevA.105.042814}

\bibitem{Ros13}
Rossi D~M {\em et~al.\/} 2013 {\em Phys. Rev. Lett.\/} {\bf 111}(24) 242503 \urlprefix\url{https://link.aps.org/doi/10.1103/PhysRevLett.111.242503}

\bibitem{Roc18b}
Roca-Maza X, Col\`o G and Sagawa H 2018 {\em Phys. Rev. Lett.\/} {\bf 120}(20) 202501 \urlprefix\url{https://link.aps.org/doi/10.1103/PhysRevLett.120.202501}

\bibitem{2021-LiKP}
Drake G~W~F, Dhindsa H~S and Marton V~J 2021 {\em Phys. Rev. A\/} {\bf 104}(6) L060801 \urlprefix\url{https://link.aps.org/doi/10.1103/PhysRevA.104.L060801}

\bibitem{2020-g}
Debierre V, Keitel C and Harman Z 2020 {\em Physics Letters B\/} {\bf 807} 135527 ISSN 0370-2693 \urlprefix\url{https://www.sciencedirect.com/science/article/pii/S0370269320303312}

\bibitem{2021-CaHCI}
Rehbehn N~H, Rosner M~K, Bekker H, Berengut J~C, Schmidt P~O, King S~A, Micke P, Gu M~F, M\"uller R, Surzhykov A and L\'opez-Urrutia J~R~C 2021 {\em Phys. Rev. A\/} {\bf 103}(4) L040801 \urlprefix\url{https://link.aps.org/doi/10.1103/PhysRevA.103.L040801}

\bibitem{2023-CaNL}
Viatkina A~V, Yerokhin V~A and Surzhykov A 2023 {\em Phys. Rev. A\/} {\bf 108}(2) 022802 \urlprefix\url{https://link.aps.org/doi/10.1103/PhysRevA.108.022802}

\bibitem{2024-simon}
R{\"o}ser D, M{\"o}ller L, Ke{\ss}ler H and Stellmer S 2024 {\em arXiv preprint arXiv:2406.06806\/}

\bibitem{2024-SrM}
Ge Z, Bai S, Eronen T, Jokinen A, Kankainen A, Kujanp{\"a}{\"a} S, Moore I, Nesterenko D and Reponen M 2024 {\em arXiv preprint arXiv:2404.02025\/}

\bibitem{2022-Wang}
Han J~Z, Pan C, Zhang K~Y, Yang X~F, Zhang S~Q, Berengut J~C, Goriely S, Wang H, Yu Y~M, Meng J, Zhang J~W and Wang L~J 2022 {\em Phys. Rev. Res.\/} {\bf 4}(3) 033049 \urlprefix\url{https://link.aps.org/doi/10.1103/PhysRevResearch.4.033049}

\bibitem{2023-Cd2}
Hofs\"ass S, Padilla-Castillo J~E, Wright S~C, Kray S, Thomas R, Sartakov B~G, Ohayon B, Meijer G and Truppe S 2023 {\em Phys. Rev. Res.\/} {\bf 5}(1) 013043 \urlprefix\url{https://link.aps.org/doi/10.1103/PhysRevResearch.5.013043}

\bibitem{2024-Sn}
Leibrandt D~R, Porsev S~G, Cheung C and Safronova M~S 2024 {\em Nature Communications\/} {\bf 15} 5663

\bibitem{2023-XeKP}
Rehbehn N~H, Rosner M~K, Berengut J~C, Schmidt P~O, Pfeifer T, Gu M~F and L\'opez-Urrutia J~R~C 2023 {\em Phys. Rev. Lett.\/} {\bf 131}(16) 161803 \urlprefix\url{https://link.aps.org/doi/10.1103/PhysRevLett.131.161803}

\bibitem{2022-Hg}
Witkowski M, Ciuryło R, Gogyan A, Linek A, Rodriguez R~M, Tecmer P and Zawada M 2022 New physics searches with isotope shifts of two hg clock transitions {\em 2022 Joint Conference of the European Frequency and Time Forum and IEEE International Frequency Control Symposium (EFTF/IFCS)\/} pp 1--2

\bibitem{2019-Ra}
Fan M, Holliman C~A, Wang A~L and Jayich A~M 2019 {\em Phys. Rev. Lett.\/} {\bf 122}(22) 223001

\bibitem{2022-Ra}
Holliman C~A, Fan M, Contractor A, Brewer S~M and Jayich A~M 2022 {\em Phys. Rev. Lett.\/} {\bf 128}(3) 033202 \urlprefix\url{https://link.aps.org/doi/10.1103/PhysRevLett.128.033202}

\bibitem{2023-Ra}
Fan M, Ready R~A, Li H, Kofford S, Kwapisz R, Holliman C~A, Ladabaum M~S, Gaiser A~N, Griswold J~R and Jayich A~M 2023 {\em Phys. Rev. Res.\/} {\bf 5}(4) 043201 \urlprefix\url{https://link.aps.org/doi/10.1103/PhysRevResearch.5.043201}

\bibitem{1969-muonic}
Wu C~S and Wilets L 1969 {\em Annual Review of Nuclear and Particle Science\/} {\bf 19} 527--606 ISSN 1545-4134 \urlprefix\url{https://www.annualreviews.org/content/journals/10.1146/annurev.ns.19.120169.002523}

\bibitem{2022-muonic}
Antognini A, Bacca S, Fleischmann A, Gastaldo L, Hagelstein F, Indelicato P, Knecht A, Lensky V, Ohayon B, Pascalutsa V {\em et~al.\/} 2022 {\em arXiv preprint arXiv:2210.16929\/}

\bibitem{2024-QUARTET}
Ohayon B, Abeln A, Bara S, Cocolios T~E, Eizenberg O, Fleischmann A, Gastaldo L, Godinho C, Heines M, Hengstler D, Hupin G, Indelicato P, Kirch K, Knecht A, Kreuzberger D, Machado J, Navratil P, Paul N, Pohl R, Unger D, Vogiatzi S~M, Schoeler K~v and Wauters F 2024 {\em Physics\/} {\bf 6} 206--215 ISSN 2624-8174 \urlprefix\url{https://www.mdpi.com/2624-8174/6/1/15}

\bibitem{2018-MuX}
Adamczak A, Antognini A, Berger N, Cocolios T~E, Dressler R, Eggenberger A, Eichler R, Indelicato P, Jungmann K, Kirch K {\em et~al.\/} 2018 Nuclear structure with radioactive muonic atoms {\em EPJ Web of Conferences\/} vol 193 (EDP Sciences) p 04014

\bibitem{2019-MuX}
Skawran A, Adamczak A, Antognini A, Berger N, Cocolios T~E, Dressler R, Duellmann C~E, Eichler R, Indelicato P, Jungmann K~P {\em et~al.\/} 2019 {\em Il Nuovo Cimento Della Societa Italiana di Fisica. C: Geophysics and Space Physics\/} {\bf 42} 125

\bibitem{2020-Abs}
Cocolios T~E 2020 Absolute charge radii of radioactive isotopes measured by muonic x-ray spectroscopy at psi Tech. rep.

\bibitem{2023-MUX}
Adamczak A, Antognini A, Berger N, Cocolios T~E, Deokar N, D{\"u}llmann C~E, Eggenberger A, Eichler R, Heines M, Hess H {\em et~al.\/} 2023 {\em The European Physical Journal A\/} {\bf 59} 15

\bibitem{2023-MUX2}
Heines M, Antwis L, Bara S, Caerts B, Cocolios T~E, Eisenwinder S, Fletcher J, Kieck T, Knecht A, Niikura M, Ritjoho N, Pereira L~M, Pohl R, Vantomme A, Vogiatzi S~M, {von Schoeler} K, Wauters F, Webb R, Zhao Q and Zweidler S 2023 {\em Nuclear Instruments and Methods in Physics Research Section B: Beam Interactions with Materials and Atoms\/} {\bf 541} 173--175 ISSN 0168-583X \urlprefix\url{https://www.sciencedirect.com/science/article/pii/S0168583X23002410}

\bibitem{2024-Ra}
Cocolios T~E and Knecht A 2024 Implantation of $^{226}$ra for the measurement of its absolute nuclear charge radius Tech. rep.

\bibitem{1997-bound}
Eides M~I and Grotch H 1997 {\em Annals of Physics\/} {\bf 260} 191--200 ISSN 0003-4916 \urlprefix\url{https://www.sciencedirect.com/science/article/pii/S0003491697957250}

\bibitem{2002-bound}
Shabaev V~M and Yerokhin V~A 2002 {\em Phys. Rev. Lett.\/} {\bf 88}(9) 091801 \urlprefix\url{https://link.aps.org/doi/10.1103/PhysRevLett.88.091801}

\bibitem{2019-bound}
Michel N, Zatorski J, Oreshkina N~S and Keitel C~H 2019 {\em Phys. Rev. A\/} {\bf 99}(1) 012505 \urlprefix\url{https://link.aps.org/doi/10.1103/PhysRevA.99.012505}

\bibitem{2020-bound}
Malyshev A~V, Glazov D~A and Shabaev V~M 2020 {\em Phys. Rev. A\/} {\bf 101}(1) 012513 \urlprefix\url{https://link.aps.org/doi/10.1103/PhysRevA.101.012513}

\bibitem{2022-g}
Sailer T, Debierre V, Harman Z, Hei{\ss}e F, K{\"o}nig C, Morgner J, Tu B, Volotka A~V, Keitel C~H, Blaum K {\em et~al.\/} 2022 {\em Nature\/} {\bf 606} 479--483

\bibitem{1992-Ne}
Fricke G, Herberz J, Hennemann T, Mallot G, Schaller L~A, Schellenberg L, Piller C and Jacot-Guillarmod R 1992 {\em Phys. Rev. C\/} {\bf 45}(1) 80--89 \urlprefix\url{https://link.aps.org/doi/10.1103/PhysRevC.45.80}

\bibitem{2019-Ne}
Ohayon B, Rahangdale H, Geddes A~J, Berengut J~C and Ron G 2019 {\em Phys. Rev. A\/} {\bf 99}(4) 042503 \urlprefix\url{https://link.aps.org/doi/10.1103/PhysRevA.99.042503}

\bibitem{Skripnikov2023_Po}
Skripnikov L~V and Barzakh A~E 2023 Revisited nuclear magnetic dipole and electric quadrupole moments of polonium isotopes (\textit{Preprint} \eprint{2311.00621})

\bibitem{Lyakh:11}
Lyakh D~I, Musia\l{} M, Lotrich V~F and Bartlett R~J 2011 {\em Chem.~Rev.\/} {\bf 112} 182–243

\bibitem{Kohn:12}
K\"{o}hn A, Hanauer M, M\"{u}ck L~A, Jagau T and Gauss J 2012 {\em Wiley Interdiscip. Rev. Comput. Mol. Sci.\/} {\bf 3} 176–197

\bibitem{Saue:99}
Saue T and Jensen H~J~A 1999 {\em J. Chem. Phys.\/} {\bf 111} 6211–6222

\bibitem{Jeziorski:81}
Jeziorski B and Monkhorst H~J 1981 {\em Phys. Rev. A\/} {\bf 24}(4) 1668--1681

\bibitem{Kucharski:91}
Kucharski S~A and Bartlett R~J 1991 {\em J. Chem. Phys.\/} {\bf 95} 8227–8238

\bibitem{Evangelista:11}
Evangelista F~A and Gauss J 2011 {\em J. Chem. Phys.\/} {\bf 134} 114102

\bibitem{2023-QEDRec}
Pachucki K and Yerokhin V~A 2023 {\em Phys. Rev. Lett.\/} {\bf 130}(5) 053002 \urlprefix\url{https://link.aps.org/doi/10.1103/PhysRevLett.130.053002}

\bibitem{1998-Spurious}
Shabaev V~M, Artemyev A~N, Beier T, Plunien G, Yerokhin V~A and Soff G 1998 {\em Phys. Rev. A\/} {\bf 57}(6) 4235--4239 \urlprefix\url{https://link.aps.org/doi/10.1103/PhysRevA.57.4235}

\bibitem{2023-recQED}
Yerokhin V~A and Oreshkina N~S 2023 {\em Phys. Rev. A\/} {\bf 108}(5) 052824 \urlprefix\url{https://link.aps.org/doi/10.1103/PhysRevA.108.052824}

\bibitem{pachucki2024heavy}
Pachucki K and Yerokhin V~A 2024 {\em arXiv preprint arXiv:2407.20703\/}

\end{thebibliography}

\end{document}